\documentclass[sigconf]{acmart}
\usepackage{xcolor,colortbl}

\definecolor{Gray}{gray}{0.85}
\definecolor{LightCyan}{rgb}{0.88,1,1}
\newcolumntype{a}{>{\columncolor{Gray}}c}
\newcolumntype{b}{>{\columncolor{white}}c}

\usepackage{amssymb}
\usepackage{hyperref}
\usepackage{epstopdf}

\usepackage{graphicx}
\usepackage[caption=false]{subfig}

\usepackage{enumerate}
\usepackage{slashbox}
\usepackage{amsmath}
\usepackage{verbatim}

\usepackage{microtype}
\DisableLigatures{encoding = *, family = * }

\usepackage[nolist]{acronym}
\begin{acronym}
	\acro{AAA}{Authentication, Authorization and Accounting}
	\acro{ACL}{Access Control List}
	\acro{AKI}{Accountable Key Infrastructure}
	\acro{API}{Application Programming Interface}
    \acro{BSM}{Basic Safety Message}
    \acro{BYOD}{Bring Your Own Device}
    \acro{BF}{Bloom Filter}
	\acro{C2C-CC}{Car2Car Communication Consortium}
	\acro{C2I}{Car-to-Infrastructure}
	\acro{C$^2$RL}{Compressed \ac{CRL}}
	\acro{CA}{Certification Authority}
	\acro{CN}{Common Name}
	\acro{CAM}{Cooperative Awareness Message}
	\acro{CAMP VSC3}{Crash Avoidance Metrics Partnership Vehicle Safety Consortium}
	\acro{CIA}{Confidentiality, Integrity and Availability}
	\acro{CRL}{Certificate Revocation List}
	\acro{CDN}{Content Delivery Network}
	\acro{COCA}{Cornell OnLine Certification Authority}
	\acro{CSR}{Certificate Signing Request}
	\acro{DAA}{Direct Anonymous Attestation}
	\acro{DDoS}{Distributed DoS}
	\acro{DDH}{Decisional Diffie-Helman}
	\acro{DENM}{Decentralized Environmental Notification Message}
	\acro{DHT}{Distributed Hash Table}
	\acro{DL/ECIES}{Discrete Logarithm and Elliptic Curve Integrated Encryption Scheme}
	\acro{DoS}{Denial of Service}
	\acro{DoT}{Department of Transportation}
	\acro{DPA}{Data Protection Agency}
	\acro{DSRC}{Dedicated Short Range Communication}
	\acro{DSS}{Digital Signature Standard}
	\acro{DTLS}{Datagram \ac{TLS}}
	\acro{ECU}{Electronic Control Unit}
	\acro{EDR}{Event Data Recorder}
	\acro{ETSI}{European Telecommunications Standards Institute}
	\acro{ECDSA}{Elliptic Curve Digital Signature Algorithm}
	\acro{ECC}{Elliptic Curve Cryptography}
	\acro{EVITA}{E-safety Vehicle Intrusion protected Applications}
	\acro{FOT}{Field Operational Testing}
	\acro{FPGA}{Field-Programmable Gate Array}
	\acro{GPA}{Global Passive Adversary}
	\acro{GN}{GeoNetworking}
	\acro{GS-VLR}{Group Signatures with Verifier Local Revocation}
	\acro{GS}{Group Signatures}
	\acro{GM}{Group Manager}
	\acro{GBA}{Generic Bootstrapping Architecture}
	\acro{GUI}{Graphic User Interface}
	\acro{HSM}{Hardware Security Module}
	\acro{HTTP}{Hypertext Transfer Protocol}
	\acro{IEEE}{Institute of Electrical and Electronics Engineers}
	\acro{IETF}{Internet Engineering Task Force}
	\acro{IoT}{Internet of Things}
	\acro{ITS}{Intelligent Transport System}
	\acro{IT}{Information Technologies}
	\acro{IMSI}{International Mobile Subscriber Identity}
	\acro{IMEI}{International Mobile Station Equipment Identity}
	\acro{IdP}{Identity Provider}
	\acro{IDS}{Intrusion Detection System}
	\acro{ISP}{Internet Service Provider}
	\acro{LEA}{Law Enforcement Agency}
	\acro{LCPP}{Lightweight Conditional Privacy Preservation}
	\acro{LTC}{Long Term Certificate}
	\acro{LTCA}{Long Term CA}
	\acro{H-LTCA}{Home-LTCA}
	\acro{F-LTCA}{Foreign-LTCA}
	\acro{LDAP}{Lightweight Directory Access Protocol}
	\acro{LBS}{Location Based Service}
	\acro{LTE}{Long Term Evolution}
	\acro{LuST}{Luxembourg SUMO Traffic}
	\acro{MAC}{Message Authentication Code}
	\acro{MCA}{Message \ac{CA}}
	\acro{MEA}{Misbehavior Evaluation Authority}
	\acro{MANET}{Mobile Ad-hoc Network}
	\acro{MPB}{Most Pieces Broadcast}
	\acro{NoW}{Network on Wheel}
	\acro{OBU}{On-Board Unit}
	\acro{OEM}{Original Equipment Manufacturer}
	\acro{OCSP}{Online Certificate Status Protocol}
	\acro{PCA}{Pseudonym CA}
	\acro{PDP}{Policy Decision Point}
	\acro{PEP}{Policy Enforcement Point}
	\acro{PIR}{Private Information Retrieval}
	\acro{PKC}{Public Key Cryptography}
	\acro{PKCS}{Public Key Cryptosystem}
	\acro{PKI}{Public-Key Infrastructure}
	\acro{PRECIOSA}{Privacy Enabled Capability in Co-operative Systems and Safety Applications}
	\acro{PRESERVE}{Preparing Secure Vehicle-to-X Communication Systems}
	\acro{P2P}{peer-to-peer}
	\acro{PS}{Participatory Sensing}
	\acro{RA}{Resolution Authority}
	\acro{REST}{Representational State Transfer}
	\acro{RBAC}{Role Based Access Control}
	\acro{RCA}{Root \acs{CA}}
	\acro{RSU}{Roadside Unit}
	\acro{SAML}{Security Assertion Markup Language}
	\acro{SAS}{Sample Aggregation Service}
	\acro{SCMS}{Security Credential Management System}
	\acro{SCORE@F}{Système COopératif Routier Expérimental Français}
	\acro{SDSI}{Simple Distributed Security Infrastructure}
	\acro{SRAAC}{Secure Revocable Anonymous Authenticated Inter-Vehicle Communication}
	\acro{SeVeCom}{Secure Vehicle Communication}
	\acro{SIT}{Sichere Informationstechnologie}
	\acro{SLC}{Short-Lived Certificate}
	\acro{SoA}{Service-oriented-Approach}
	\acro{SIFS}{Short Inter Frame Space}
	\acro{SSO}{Single-Sign-On}
	\acro{SSL}{Secure Sockets Layer}
	\acro{SOAP}{Simple Object Access Protocol}
	\acro{TACK}{Temporary Anonymous Certified Key}
	\acro{TS}{Task Service}
	\acro{TLS}{Transport Layer Security}
	\acro{TPM}{Trusted Platform Module}
	\acro{TTP}{Trusted Third Party}
	\acro{TVR}{Ticket Validation Repository}
	\acro{URI}{Uniform Resource Identifier}
	\acro{VANET}{Vehicular Ad-hoc Network}
	\acro{V2I}{Vehicle-to-Infrastructure}
	\acro{V2V}{Vehicle-to-Vehicle}
	\acro{V2X}{\ac{V2V}/\ac{V2I}}
	\acro{VC}{Vehicular Communication}
	\acro{VM}{Virtual Machine}
	\acro{VSS}{\ac{VC} Security Subsystem}
	\acro{WAVE}{Wireless Access in Vehicular Environments}
	\acro{WSDL}{Web Services Discovery Language}
	\acro{W3C}{World Wide Web Consortium}
	\acro{V}{Vehicle}
	\acro{VANET}{Vehicular Ad-hoc Network}
	\acro{VLR}{Verifier-Local Revocation}
	\acro{VPKI}{Vehicular Public-Key Infrastructure}
	\acro{VM}{Virtual Machine}
	\acro{WS}{Web Service}
	\acro{WoT}{Web of Trust}
	\acro{WSACA}{\ac{WAVE} Service Advertisement \ac{CA}}
	\acro{XML}{Extensible Markup Language}
	\acro{XACML}{eXtensible Access Control Markup Language}
	\acro{3G}{3rd Generation}
\end{acronym}

\usepackage{makecell}

\usepackage{arydshln}

\usepackage{xcolor}

\usepackage{algpseudocode}

\usepackage{setspace}

\errorcontextlines\maxdimen

\makeatletter
\newcommand*{\algrule}[1][\algorithmicindent]{\makebox[#1][l]{\hspace*{.5em}\thealgruleextra\vrule height \thealgruleheight depth \thealgruledepth}}%
\newcommand*{\thealgruleextra}{}
\newcommand*{\thealgruleheight}{.75\baselineskip}
\newcommand*{\thealgruledepth}{.25\baselineskip}

\newcount\ALG@printindent@tempcnta
\def\ALG@printindent{%
	\ifnum \theALG@nested>0
	\ifx\ALG@text\ALG@x@notext
	\else
	\unskip
	\addvspace{-1pt}
	\ALG@printindent@tempcnta=1
	\loop
	\algrule[\csname ALG@ind@\the\ALG@printindent@tempcnta\endcsname]%
	\advance \ALG@printindent@tempcnta 1
	\ifnum \ALG@printindent@tempcnta<\numexpr\theALG@nested+1\relax
	\repeat
	\fi
	\fi
}%
\usepackage{etoolbox}
\patchcmd{\ALG@doentity}{\noindent\hskip\ALG@tlm}{\ALG@printindent}{}{\errmessage{failed to patch}}
\makeatother

\newbox\statebox
\newcommand{\myState}[1]{%
	\setbox\statebox=\vbox{#1}%
	\edef\thealgruleheight{\dimexpr \the\ht\statebox+1pt\relax}%
	\edef\thealgruledepth{\dimexpr \the\dp\statebox+1pt\relax}%
	\ifdim\thealgruleheight<.75\baselineskip
	\def\thealgruleheight{\dimexpr .75\baselineskip+1pt\relax}%
	\fi
	\ifdim\thealgruledepth<.25\baselineskip
	\def\thealgruledepth{\dimexpr .25\baselineskip+1pt\relax}%
	\fi
	\State #1%
	\def\thealgruleheight{\dimexpr .75\baselineskip+1pt\relax}%
	\def\thealgruledepth{\dimexpr .25\baselineskip+1pt\relax}%
}

\usepackage{footnote}

\usepackage{rotating}

\usepackage{soul}

\usepackage{algorithm}

\usepackage{tikz}
\usetikzlibrary{shadows,positioning,calc}
\tikzset{multiple/.style = {double copy shadow={shadow xshift=1ex,shadow
			yshift=-1.5ex,draw=black!30},fill=white,draw=black,thick,minimum height = 1cm,minimum
		width=2cm},
	ordinary/.style = {rectangle,draw,thick,minimum height = 1cm,minimum width=2cm}}

\usetikzlibrary{decorations.pathreplacing}

\usepackage{booktabs}
\usepackage{multirow}
\usepackage{siunitx}

\usepackage{algpseudocode}

\makeatletter
\renewcommand{\ALG@beginalgorithmic}{\footnotesize} 
\makeatother

\usepackage{afterpage}
\newlength{\oldtextfloatsep}

\renewcommand\footnotetextcopyrightpermission[1]{} 
\usepackage[utf8]{inputenc}

\usepackage[normalem]{ulem}

\usepackage{booktabs} 

\begin{document}
\hyphenation{harmon-ization vulnera-ble invest-igated comm-uni-cations Con-sor-tium data-base ano-ny-mi-ty a-no-ny-mi-za-tion e-quip-ped pse-udo-ny-mi-ty co-lo-gne in-fra-struc-ture pseu-do-nyms pseu-do-nym informat-ion mecha-nism cryptogra-phic necessita-tes}

\title[A Vehicle-Centric Approach for Certificate Revocation List Distribution in \acsp{VANET}]{Efficient, Scalable, and Resilient Vehicle-Centric \\ Certificate Revocation List Distribution in \acsp{VANET}}

\author{Mohammad Khodaei}
\orcid{}
\affiliation{%
  \institution{Networked Systems Security Group}
  \streetaddress{KTH Royal Institute of Technology}
  \city{Stockholm} 
  \state{Sweden} 
}
\email{khodaei@kth.se}

\author{Panos Papadimitratos}
\affiliation{%
  \institution{Networked Systems Security Group}
  \streetaddress{KTH Royal Institute of Technology}
  \city{Stockholm} 
  \state{Sweden} 
}
\email{papadim@kth.se}


\begin{abstract}

In spite of progress in securing \ac{VC} systems, there is no consensus on how to distribute \acp{CRL}. The main challenges lie exactly in (i) crafting an efficient and timely distribution of \acp{CRL} for numerous anonymous credentials, \emph{pseudonyms}, (ii) maintaining strong privacy for vehicles prior to revocation events, even with \emph{honest-but-curious} system entities, (iii) and catering to computation and communication constraints of on-board units with intermittent connectivity to the infrastructure. Relying on peers to distribute the \acp{CRL} is a double-edged sword: \emph{abusive peers} could ``pollute'' the process, thus degrading the timely \acp{CRL} distribution. In this paper, we propose a \emph{vehicle-centric} solution that addresses all these challenges and thus closes a gap in the literature. Our scheme radically reduces \ac{CRL} distribution overhead: each vehicle receives \acp{CRL} corresponding only to its region of operation and its actual trip duration. Moreover, a ``fingerprint'' of \ac{CRL} `pieces' is attached to a subset of (verifiable) pseudonyms for fast \ac{CRL} `piece' validation (while mitigating resource depletion attacks abusing the \ac{CRL} distribution). Our experimental evaluation shows that our scheme is efficient, scalable, dependable, and practical: with no more than 25 KB/s of traffic load, the latest \ac{CRL} can be delivered to 95\% of the vehicles in a region (50$\times$50 KM) within 15s, i.e., more than 40 times faster than the state-of-the-art. Overall, our scheme is a comprehensive solution that complements standards and can catalyze the deployment of secure and privacy-protecting \ac{VC} systems. 

\end{abstract}

\maketitle

\section{Introduction}
\label{sec:crl-dis-introduction}

\acresetall

\ac{V2V} and \ac{V2I} communications seek to enhance transportation safety and efficiency. It has been well-understood that \ac{VC} systems are vulnerable to attacks and that the privacy of their users is at stake. As a result, security and privacy solutions have been developed by standardization bodies (IEEE 1609.2 WG~\cite{1609-2016} and \acs{ETSI}~\cite{ETSI-102-638}), harmonization efforts (C2C-CC~\cite{c2c}), and projects (\acs{SeVeCom}~\cite{papadimitratos2007architecture}, \acs{PRESERVE}~\cite{preserve-url}, and CAMP~\cite{whyte2013security}). A consensus towards using \ac{PKC} to protect \ac{V2X} communication is reached: a set of Certification Authorities (CAs) constitutes the \ac{VPKI}, providing multiple anonymous credentials, termed \emph{pseudonyms}, to legitimate vehicles. Vehicles switch from one pseudonym to a non-previously used one towards unlinkability of digitally signed messages, and improved sender privacy for \ac{V2V}/\ac{V2I} messages. Pseudonymity is conditional in the sense that the corresponding long-term vehicle identity (\ac{LTC}) can be retrieved by the \ac{VPKI} entities if deviating from system policies.

In fact, vehicles can be compromised or faulty and disseminate erroneous information across the \ac{V2X} network~\cite{Papadi:C:08, raya2006certificaterevocation}. They should be held \emph{accountable} for such actions and credentials (their \acp{LTC} and their pseudonyms) can be revoked. To efficiently revoke a set of pseudonyms, one can disclose a single entry for all (revoked) pseudonyms of the vehicle~\cite{fischer2006secure, stumpf2007trust, laberteaux2008security, haas2009design}. However, upon a revocation event, all non-revoked (but expired) pseudonyms belonging to the ``misbehaving'' vehicle would also be linked. Linking pseudonyms with lifetimes prior to a revocation event implies that all the corresponding digitally signed messages will be trivially linked. Even if revocation is justified, this does not imply that a user \emph{``deserves''} to abolish privacy prior to the revocation event. Avoiding such a situation, i.e., achieving \emph{perfect-forward-privacy}, can be guaranteed if the \ac{VPKI} entities are \emph{fully-trustworthy}~\cite{haas2011efficient}. However, we need to guarantee strong user privacy even in the presence of \emph{honest-but-curious} \ac{VPKI} entity; recent revelations of mass surveillance, e.g.,~\cite{nsa, era2015cryptography}, show that assuming service providers are fully-trustworthy is no longer a viable approach.

A main concern, relevant to all proposals in the literature~\cite{papadimitratos2008certificate, haas2011efficient, laberteaux2008security, haas2009design, nowatkowski2010certificate, nowatkowski2009cooperative} is efficiency and scalability, essentially low communication and computation overhead even as system dimension grows. Consider first typical operational constraints: the average daily commute time is less than an hour (on average 29.2 miles and 46 minutes per day)~\cite{whyte2013security, acs-survey, newsroom-how-much-motorists-drive} while the latencies for the dissemination of a full \ac{CRL} can exceed the actual trip duration~\cite{DOTHS812014}. One can compress \ac{CRL} using a \acf{BF}~\cite{raya2006certificaterevocation, raya2007eviction, rigazzi2017optimized}; however, the size of a \ac{CRL} grows linearly with the number of revoked pseudonyms, thus necessitates larger \acp{BF}. More so, a sizable portion of the \ac{CRL} information is irrelevant to a receiving vehicle and can be left unused. This, at the system level, constitutes waste of computation, communication (bandwidth), and storage resources. In turn, it leads to higher latency for all vehicles to reconstruct the \ac{CRL}, i.e., a degradation of timely distribution.

Alternatively, vehicles can only validate revocation status of (their neighbors') pseudonyms through an \ac{OCSP}~\cite{myers1999x}. Even if a \ac{VPKI} system can comfortably handle such a demanding load~\cite{khodaei2014ScalableRobustVPKI}, \ac{OCSP} cannot be used as a standalone solution in \ac{VC} systems: it requires continuous connectivity and significant bandwidth dedicated to revocation traffic, thus impractical due to the network volatility and scale~\cite{raya2006certificaterevocation}. Moreover, what would be the course of action if the \ac{VPKI} were not reachable for other reasons, e.g., during a \ac{DoS} attack? So, the challenge is \emph{how can one distribute the most relevant revocation information to a given vehicle, per trip, and ensure timely revocation even without uninterrupted connectivity to the \ac{VPKI}?}

The computation overhead for the verification of the \ac{CRL} could interfere with safety- and time-critical operations especially if one considers typical \ac{VC} rates of 10 safety beacons per second, and thus processing of possibly hundreds of messages from neighboring vehicles per second. Simply put, with existing computation and communication overhead and given the time critical nature of safety applications in \ac{VC} systems, minimizing the overhead for \ac{CRL} verification and distribution is paramount.

From a different viewpoint, we need to allocate as little bandwidth as possible for the \ac{CRL} distribution in order not to interfere with safety critical operations or enable an attacker to broadcast a fake \ac{CRL} at a high rate. However, this should be hand in hand with timely \ac{CRL} distribution. This can be achieved with the use of \acp{RSU}~\cite{papadimitratos2008certificate}; however, dense deployment of \acp{RSU} in a large-scale environment is costly. If the deployment is sparse, a significant delay could be introduced. Alternatively, the \ac{CRL} can be distributed in a peer-to-peer, epidemic manner~\cite{laberteaux2008security, haas2011efficient, haas2009design}. This is a double-edged sword: \emph{abusive peers}, seeking to compromise the trustworthiness of the system, could pollute the \ac{CRL} distribution and mount a clogging \ac{DoS} attack.

Despite the plethora of research efforts, none addresses all challenges at hand. In this paper, we show how to \emph{efficiently revoke a very large volume of pseudonyms while providing strong user privacy protection, even in the presence of honest-but-curious \ac{VPKI} entities. Our system effectively, resiliently, and in a timely manner disseminate the authentic \ac{CRL} throughout a large-scale (multi-domain) \ac{VC} system.} Moreover, \emph{we ensure that the \ac{CRL} distribution incurs low overhead and prevents abuse of the distribution mechanism.}

\emph{Contributions:} Our comprehensive security and privacy-preserving solution systematically addresses all key aspects of \ac{CRL}-based revocation, i.e., security, privacy, and efficiency. This is based on few simple yet powerful, as it turns out, ideas. We propose making the \ac{CRL} acquisition process \emph{vehicle-centric}: each vehicle only receives the pieces of \acp{CRL} corresponding to its targeted region and its actual trip duration, i.e., obtaining only region- and time-relevant revocation information. Moreover, randomly chosen pseudonyms issued by the \ac{VPKI} are selected to piggyback a notification about new \ac{CRL}-update events and an authenticator for efficiently validating pieces of the latest \ac{CRL}; in other words, validation of the \ac{CRL} pieces \emph{almost for free}. These novel features dramatically reduce the \ac{CRL} size and \ac{CRL} validation overhead, while they significantly increase its resiliency against resource depletion attacks.

In the rest of the paper, we critically survey the state-of-the-art research efforts (Sec.~\ref{sec:crl-dis-related-work}) and describe the system model (Sec.~\ref{sec:crl-dis-model-requirements}). We present system design (Sec.~\ref{sec:crl-dis-design}), followed by qualitative and quantitative analysis (Sec.~\ref{sec:crl-dis-scheme-analysis-evaluation}). We then conclude the paper (Sec.~\ref{sec:crl-dis-conclusions}).

\section{Related Work}
\label{sec:crl-dis-related-work}

The need to evict misbehaving or compromised~\cite{Papadi:C:08} vehicles from a \ac{VC} system is commonly accepted, because such vehicles can threaten the safety of vehicles and users and degrade transportation efficiency. \ac{CRL} distribution is of central importance and it is the final and definitive line of defense~\cite{1609-2016, ETSI-102-638, gerlach2007security, papadimitratos2007architecture, papadimitratos2007architecture, raya2007eviction}: only the \acs{VPKI} can \emph{``ultimately''} revoke a vehicle by including its unexpired certificates' serial numbers in a \ac{CRL}.

The literature proposes distribution of the \ac{CRL} via \acp{RSU}~\cite{papadimitratos2008certificate} and car-to-car epidemic communication~\cite{laberteaux2008security, haas2009design, haas2011efficient}, with enhancements on the distribution of pieces~\cite{nowatkowski2010certificate, nowatkowski2009cooperative} evaluated in~\cite{nowatkowski2010scalable, amoozadeh2012certificate}. A na\"ive solution would be to digitally sign the entire \ac{CRL} and broadcast it; however, it imposes difficulties in downloading a large \ac{CRL} file and exchanging it over short contact period (with an \ac{RSU} or a peer). Splitting the digitally signed \ac{CRL} into multiple pieces is vulnerable to \emph{pollution} attacks: in the absence of fine-grained authentication, per \ac{CRL} piece, an adversary can delay or even prevent reception by injecting fake pieces. Thus, the straightforward solution is to have the \ac{VPKI} prepare the \ac{CRL}, split it into multiple pieces, sign each piece, and distribute all of them across the \ac{VC} system. \acp{RSU} can broadcast \ac{CRL} pieces randomly or in a round-robin fashion~\cite{papadimitratos2008certificate}, and vehicles can relay pieces until all vehicles receive all pieces necessary to reconstruct the \ac{CRL}~\cite{laberteaux2008security}. Erasure codes can be used to enhance the fault-tolerance of the \ac{CRL} piece distribution in the highly volatile \ac{VC} environment~\cite{papadimitratos2008certificate, ardelean2009CRLImplementation}.

Signing each \ac{CRL} piece so that it is self-verifiable, incurs significant computation overhead, which grows linearly with the number of \ac{CRL} pieces, both for the \ac{VPKI} and for the receiving vehicles. Furthermore, an attacker could aggressively forge \ac{CRL} pieces for a \ac{DoS} attack leveraging signature verification delays~\cite{hsiao2011flooding} that can prevent vehicles from obtaining the genuine \ac{CRL} pieces. A \emph{``precode-and-hash''} scheme~\cite{nguyen2016secure} proposes to calculate a hash value of each pre-coded piece, sign it, and disseminate it with higher priority. Each relaying node can apply a different precode to the original \ac{CRL} and act as a secondary source. However, by applying different encodings to the original \ac{CRL} file, another receiver cannot reconstruct the entire \ac{CRL} from the pieces, encoded differently by various relaying nodes. To mitigate pollution and \ac{DoS} attacks, we propose to piggyback a fingerprint (a \ac{BF}~\cite{bloom1970space, mitzenmacher2002compressed}) for \ac{CRL} pieces into a subset of pseudonyms to validating \ac{CRL} pieces ``for free''.

To efficiently revoke an ensemble of pseudonyms, one can enable revocation of multiple pseudonyms with a single \ac{CRL} entry, to reduce the \ac{CRL} size, e.g.,~\cite{fischer2006secure, stumpf2007trust, laberteaux2008security, haas2009design}. Despite a huge reduction in size, such schemes do not provide \emph{perfect-forward-privacy}: upon a revocation event and \ac{CRL} release, all the ``non-revoked'' but previously expired pseudonyms belonging to the evicted entity would be linked as well. Although perfect-forward-privacy can be achieved by leveraging a hash chain~\cite{haas2011efficient}, the pseudonyms' issuer can trivially link all pseudonyms belonging to a vehicle, and thus the pseudonymously authenticated messages, towards tracking it for the entire duration of its presence in the system~\cite{fischer2006secure, stumpf2007trust, laberteaux2008security, haas2009design, haas2011efficient}.

Compressing \acp{CRL} using a \ac{BF} was proposed for compact storage of revocation entries~\cite{raya2007eviction}, or to efficiently distribute them across the network~\cite{raya2006certificaterevocation, raya2007eviction, rigazzi2017optimized}. However, the challenge is twofold: scalability and efficiency. The size of a \ac{CRL} linearly grows with the number of revoked pseudonyms, but also a substantial portion of the ``compressed'' \ac{CRL} can be irrelevant to a receiving vehicle and be left unused. Moreover, as it becomes clear in Sec.~\ref{subsec:crl-dis-qualitative-analysis}, compressing \acp{CRL} using a \ac{BF} does not necessarily reduce the size of a \ac{CRL} as vehicles can be provided with possibly hundreds of pseudonyms~\cite{1609-2016}. Unlike such schemes~\cite{raya2006certificaterevocation, raya2007eviction, rigazzi2017optimized}, we do not compress the \ac{CRL}: our scheme disseminates only trip-relevant revocation information to vehicles and it utilizes a \ac{BF} to provide a condensed authenticator for the \ac{CRL} pieces. Our scheme leverages and \emph{enhances} the functionality of the state-of-the-art \ac{VPKI} system~\cite{khodaei2018Secmace} towards efficiently revoking a batch of pseudonyms without compromising user privacy backwards: upon a revocation event, all pseudonyms prior to the revocation event remain unlinkable (a detailed description in Sec.~\ref{subsec:crl-dis-security-protocols}).

Alternatively, vehicles could validate pseudonym status (revocation) information through \ac{OCSP}~\cite{myers1999x}. However, due to intermittent \ac{VC} network connectivity, significant usage of the bandwidth by time- and safety-critical operations, and substantial overhead for the \ac{VPKI} (assuming the server is reachable), \ac{OCSP} cannot really be used as a standalone solution~\cite{raya2006certificaterevocation}. A hybrid solution could rely on distributing certificate status information to other mobile nodes~\cite{marias2005adopt, forne2009certificate, ganan2012toward, ganan2013coach, ganan2013becsi}; however, the system would be subject to the reachability (of sufficiently many cooperative) and the trustworthiness of such nodes. In our scheme, we ensure that the latest \ac{CRL} is efficiently, effectively, and timely distributed among all vehicles without any assumption on persistent reachability and trustworthiness of specific mobile nodes.

Research efforts also focused on how to protect the \ac{VC} systems from misbehaving nodes, by temporarily ``revoking'' (isolating) them from further access to the system~\cite{raya2006certificaterevocation, raya2007eviction, moore2008fast, wasef2009edr, bissmeyer2014misbehavior} until connection to the \ac{VPKI} is established and they are fully evicted from the system. Before the \ac{VPKI} performs the ``actual'' eviction and \ac{CRL} distribution, these protocols build evidence, in fact local agreement, that a given wrongdoer is present. This can serve towards isolating misbehaving vehicles before the corresponding \ac{VPKI} entity takes the \emph{``ultimate''} decision and commences the latest \ac{CRL} distribution.

\acs{C2C-CC}~\cite{c2c} and V-token~\cite{schaub2010v} propose to revoke only the \ac{LTC} of vehicles and let the pseudonyms expire. PUCA~\cite{puca2014} requires the owner of the pseudonym to trigger revocation, i.e., the system cannot evict a misbehaving entity from the system. Clearly, leaving it up to the misbehaving entity, or allowing it to act for a significant period till pseudonyms expire, creates an unacceptable vulnerability window. Another line of studies proposes geo-casting a \emph{``self-revocation''} message, by the \ac{VPKI}, across a region, to wipe out the credentials from the \ac{HSM} of a misbehaving vehicle~\cite{raya2006certificaterevocation, raya2007eviction, papadimitratos2008secure, forster2015rewire}. However, an adversary could control incoming messages, and prevent the \emph{``self-revocation''} instruction from reaching the \ac{HSM}, i.e., such schemes alone cannot guarantee the trustworthiness of the system against misbehavior unless the \ac{VPKI} distributes the \ac{CRL} enabling legitimate vehicles to defend themselves against misbehavior or faulty peers. 

Alternatively, the \ac{VPKI} could provide vehicles for a long period, e.g., 25 years, worth of pseudonyms with a decryption key for, e.g., a weekly batch of pseudonyms, delivered periodically~\cite{kumar2017binary}. This would eliminate the need for bidirectional connectivity to the \ac{VPKI} to obtain pseudonyms. To evict a vehicle, the \ac{VPKI} can stop delivering the corresponding decryption key to the vehicle \ac{HSM}. Still, it is imperative to distribute the \ac{CRL} and cover the (weekly) period and the corresponding revoked pseudonyms. Furthermore, having released a \ac{CRL} towards the end of a week, signed messages with the private keys corresponding to the recently revoked pseudonyms (included in the \ac{CRL}) can be linked, i.e., backwards-trackable for a week (no \emph{perfect-forward-privacy} for that period)~\cite{DOTHS812014}. 

Outside the \ac{VC} realm, a recent comparative evaluation of classic Internet schemes is available~\cite{clark2013sok}. Such schemes, e.g.,~\cite{micali2002scalable, solworth2008instant, iliadis2003towards, cooper2000more, micali1996efficient, charitonccsp, larisch2017crlite}, cannot be leveraged due to the nature of \ac{VC} systems, i.e., short-lived pseudonyms, highly dynamic intermittent connectivity, and resource constraints. For example, CRLite~\cite{larisch2017crlite} stores \acp{CRL} in a \emph{filter-cascade} \ac{BF} without any false positive or false negative; however, this necessitates little change in the set of revoked and non-revoked certificates. Obviously, this contradicts \emph{on-demand} pseudonym acquisition strategies for \ac{VC} systems, e.g.,~\cite{khodaei2018Secmace, khodaei2016evaluating, khodaei2014ScalableRobustVPKI, fischer2006secure, schaub2010v, puca2014, bibmeyer2013copra, ma2008pseudonym, khodaei2017RHyTHM}, which are more efficient (than preloading for a long duration, e.g.,~\cite{kumar2017binary}) in terms of pseudonym utilization and revocation, thus more effective in fending off misbehavior.

\section{Model and Requirements}
\label{sec:crl-dis-model-requirements}

\subsection{System Model and Assumptions}
\label{subsec:crl-dis-system-model-assumptions}

A \ac{VPKI} consists of a set of Certification Authorities (CAs) with distinct roles: the \ac{RCA}, the highest-level authority, certifies other lower-level authorities; the \ac{LTCA} is responsible for the vehicle registration and the \acf{LTC} issuance, and the \ac{PCA} issues pseudonyms for the registered vehicles. Pseudonyms have a lifetime (a validity period), typically ranging from minutes to hours; in principle, the shorter the pseudonym lifetime is, the higher the unlinkability and thus the higher privacy protection can be achieved. We assume that each vehicle is registered only with its \emph{\ac{H-LTCA}}, the \emph{policy decision and enforcement point}, reachable by the registered vehicles. Without loss of generality, a \emph{domain} can be defined as a set of vehicles in a region, registered with the \ac{H-LTCA}, subject to the same administrative regulations and policies~\cite{khodaei2015VTMagazine}. There can be several \acp{PCA}, each active in one or more domains. Each vehicle can cross in to \emph{foreign} domains and communicate with the \emph{\ac{F-LTCA}} towards obtaining pseudonyms, i.e., a new set of pseudonyms when entering a new domain, to operate as a native vehicle in that region. Trust between two domains can be established with the help of the \ac{RCA}, or through cross certification. Moreover, the certificates of higher-level authorities are installed in the \acp{OBU}, which are loosely synchronized with the \ac{VPKI} servers. The \acp{RSU} could be deployed by other authorities than the \ac{VPKI} ones, thus they only expose minimal information, e.g., IP address and location, to the corresponding \acp{PCA}.

All vehicles (\acp{OBU}) registered in the system are provided with \acp{HSM}, ensuring that private keys never leave the \ac{HSM}. Moreover, we assume that there is a misbehavior detection system, e.g.,~\cite{bissmeyer2014misbehavior}, that triggers the revocation\footnote{The faulty behavior detection depends on, e.g., data-centric plausibility and consistency checks, and it is orthogonal to this investigation.}. The \acf{RA} can initiate a process to resolve and revoke all pseudonyms of a misbehaving vehicle: it interacts with the corresponding \acp{PCA} and \ac{LTCA} (a detailed protocol description, e.g., in~\cite{khodaei2014ScalableRobustVPKI, khodaei2018Secmace}) to resolve and revoke all credentials issued for a misbehaving vehicle. Consequently, the misbehaving vehicle can no longer obtain credentials from the \ac{VPKI}. The \ac{VPKI} is responsible for distributing the \acp{CRL} and notifying all legitimate entities about the revocation; this implies a new \ac{CRL}-update event.\footnote{The revocation information of other system entities, e.g., \ac{VPKI} entities, need to be distributed as well. Here, we only focus on the distribution of revoked pseudonyms.}

\subsection{Adversarial Model}
\label{subsec:crl-dis-adversarial-model}

We extend the general adversary model in secure vehicular communications~\cite{papadimitratos2006securing} to include \ac{VPKI} entities that are \emph{honest-but-curious}, i.e., entities complying with security protocols and policies, but motivated to profile users. In a multi-domain \ac{VC} environment, internal adversaries, i.e., malicious, compromised, or non-cooperative clients, and external adversaries, i.e., unauthorized entities, raise four challenges. More specifically in the context of this work, adversaries can try to (i) exclude revoked pseudonym serial numbers from a \ac{CRL}, (ii) add valid pseudonyms by forging a fake \ac{CRL} (piece), or (iii) prevent legitimate entities from obtaining genuine and the most up-to-date \ac{CRL} (pieces), or delay the \ac{CRL} distribution by replaying old, spreading fake \ac{CRL} (pieces), or performing a \ac{DoS} attack. This allows wrong-doers to remain operational in the \ac{VC} system using their current revoked pseudonym sets. Moreover, they might be simply non-cooperative or malicious, tempted to prevent other vehicles from receiving a notification on a new \ac{CRL}-update event, thus preventing them from requesting to download the \acp{CRL}. Lastly, (iv) \ac{VPKI} entities (in collusion with vehicle communication observers) could potentially link messages signed under (non-revoked but expired) pseudonyms prior to the revocation events, e.g., inferring sensitive information from the \acp{CRL} towards linking pseudonyms, and thus tracking vehicles backwards. The \acp{PCA} operating in a domain (or across domains) could also collude, i.e., share information that each of them individually has, to harm user privacy.\footnote{Note that \emph{``malicious''} \ac{VPKI} entities could attempt to influence the distribution of \acp{CRL}, e.g., manipulating the \ac{CRL} entries unlawfully; this is out of the scope of our honest-but-curious adversarial model.} 

\subsection{Requirements}
\label{subsec:crl-dis-requirements}

Security and privacy requirements for \ac{V2X} communications have been specified in the literature, e.g., as early as~\cite{papadimitratos2006securing}, and additional requirements specifically for \ac{VPKI} entities in~\cite{khodaei2018Secmace}. Next, we compile security and privacy, as well as functional and performance, requirements for the \ac{CRL} distribution problem. 

\emph{R1. Fine-grained authentication, integrity, and non-repudiation:} Each \ac{CRL} (piece) should be authenticated and its integrity be protected, i.e., preventing alternation or replays. Moreover, each \ac{CRL} (piece) should be non-repudiably connected to its originator (the \ac{VPKI} entity). 

\emph{R2. Unlinkability (perfect-forward-privacy):} \acp{CRL} should not enable any observer (even in collusion with a single \ac{VPKI} entity) to link pseudonyms (and thus the corresponding signed messages) prior to their revocation. In fact, upon a revocation event, all non-revoked previously expired pseudonyms of an evicted vehicle should remain unlinkable. 

\emph{R3. Availability:} The system should ensure any legitimate vehicle can obtain the latest \ac{CRL} within a reasonable time interval despite of benign failures, e.g., system faults or crashes, or network outages, e.g., intermittent connectivity. Moreover, the system should be resilient to active disruptions, including resource depletion attacks. 

\emph{R4. Efficiency:} Generating, validating, and disseminating the \ac{CRL} (pieces) and revocation event notification should be efficient and scalable even if the number of vehicles and credentials grow, i.e., incurring low computation and communication overhead. Moreover, a small fraction of bandwidth should be used for \ac{CRL} distribution, in order not to interfere with transportation safety- and time-critical operations. However, allocation of a small amount of bandwidth in a timely fashion should be sufficient to distribute \acp{CRL} to all legitimate vehicles. 

\emph{R5. Explicit and/or implicit notification on revocation events:} The system should notify, explicitly or implicitly, every legitimate vehicle within the system (domain) regarding revocation events and then \ac{CRL}-updates (availability of new revocation information).

\section{Design}
\label{sec:crl-dis-design}

\subsection{Motivation and Overview} 
\label{subsec:crl-dis-motivation-and-overview}

\textbf{Preliminary assumptions:} We leverage the state-of-the-art \ac{VPKI} system~\cite{khodaei2018Secmace} that provides pseudonyms in an \emph{on-demand} fashion: each vehicle \emph{``decides''} when to trigger the pseudonym acquisition process based on various factors~\cite{khodaei2016evaluating}. Such a scheme requires sparse connectivity to the \ac{VPKI}, but it facilitates an \ac{OBU} to be \emph{preloaded} with pseudonyms proactively, covering a longer period, e.g., a week or a month, should the connectivity be expected heavily intermittent. The efficiency, scalability and robustness of the \ac{VPKI} system is systematically investigated~\cite{khodaei2016evaluating, khodaei2018Secmace} with the \ac{VPKI} handles a large workload. Moreover, it enhances user privacy, notably preventing linking pseudonyms based on \emph{timing information} (the instance of issuance and the pseudonym lifetime) as well as offers strong user privacy protection even in the presence of \emph{honest-but-curious} \ac{VPKI} entities. More precisely, a universally fixed interval, $\Gamma$, is specified by the \ac{H-LTCA} and all pseudonyms in that domain are issued with the lifetime ($\tau_{P}$) aligned with the \ac{VPKI} clock. Vehicles obtain pseudonyms on-the-fly as they operate, and the number of pseudonyms in a request is $\frac{\Gamma}{\tau_{p}}$, i.e., no prior calculation needed. As a result of this policy, at any point in time, all the vehicles transmit using pseudonyms that are indistinguishable thanks to this time alignment, i.e., eliminating any distinction among pseudonym sets of different vehicles, thus enhancing user privacy. We leverage and \emph{enhance} the functionality of this \ac{VPKI} system; in particular, our solution necessitates two modifications during pseudonym acquisition process, notably (i) implicitly binding pseudonyms issued to a given requester per $\Gamma$, and (ii) integrating a fingerprint into a subset of the pseudonyms for efficient \ac{CRL} validation.

\textbf{High-level overview:} The default policy is to distribute all revocation information to all vehicles. Nonetheless, this approach ignores the locality, the temporal nature of pseudonyms, and other constraints, e.g., the average daily commute time. Locality could be geographical, i.e., credentials relative to the corresponding region, and temporal, i.e., relevance to the lifetime of pseudonyms with respect to the trip duration of a vehicle. To efficiently, effectively, and timely distribute the \acp{CRL} across the \ac{V2X} network, we propose making the \ac{CRL} acquisition process \emph{vehicle-centric}, i.e., through \emph{a content-based and context-sensitive ``publish-subscribe''} scheme~\cite{eugster2003many, huang2004publish}.

\begin{figure} [!t]
	\vspace{-3em}
	\begin{center}
		\centering
		\includegraphics[trim=0cm 0.25cm 0cm 0cm, clip=true, width=0.42\textwidth,height=0.42\textheight,keepaspectratio]{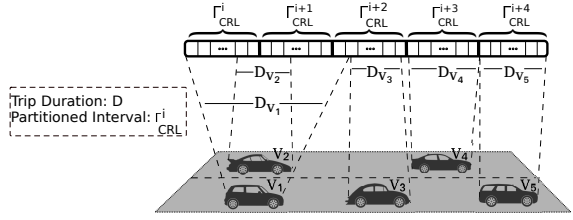}
		\vspace{-0.5em}
		\caption{\ac{CRL} as a Stream: {\small $V_{1}$} subscribes to {\small $\{\Gamma_{CRL}^{i}, \Gamma_{CRL}^{i+1}, \Gamma_{CRL}^{i+2}\}$}; {\small $V_{2}: \{\Gamma_{CRL}^{i}, \Gamma_{CRL}^{i+1}\}$}; {\small $V_{3}: \{\Gamma_{CRL}^{i+2}\}$}; {\small $V_{4}: \{\Gamma_{CRL}^{i+3}\}$}; and {\small $V_{5}: \{\Gamma_{CRL}^{i+4}\}$}.}
		\label{fig:crl-dis-crl-stream}
	\end{center}
	\vspace{-1.5em}
\end{figure}

Fig.~\ref{fig:crl-dis-crl-stream} shows that by starting a new trip, each vehicle only subscribes to receive the pieces of \acp{CRL}, i.e., the content, corresponding to its actual trip duration and its targeted region, i.e., the context. To reap the benefits of the ephemeral nature pseudonyms and the timely-aligned pseudonym provisioning policy, towards an effective, efficient, and scalable \ac{CRL} distribution, a fixed interval, $\Gamma_{CRL}$, is predetermined by the \acp{PCA} in the domain. They publicize revoked pseudonyms whose lifetimes fall within $\Gamma_{CRL}$, i.e., distributing only the serial number of these pseudonyms rather than publishing the entire \ac{CRL}. Note that $\Gamma$, the universally fixed interval to obtain pseudonyms~\cite{khodaei2018Secmace}, and $\Gamma_{CRL}$ are not necessarily aligned due to the unpredictable nature of revocation events. 

When a vehicle reliably connects to the \ac{VPKI}, it can obtain the \emph{``necessary''} \ac{CRL} pieces corresponding to its trip duration during the pseudonym acquisition phase. However, if reliable connectivity is not guaranteed, or if a vehicle obtained (possibly preloaded with enough) pseudonyms in advance, or a new revocation event happens, one can be notified about a new \ac{CRL}-update (revocation) event: a signed fingerprint (a \acf{BF}~\cite{bloom1970space, mitzenmacher2002compressed}) of \ac{CRL} pieces is broadcasted by \acp{RSU} and it is integrated in a subset of recently issued pseudonyms, this way readily broadcasted by vehicles (termed \emph{fingerprint-carrier} nodes) along with their \acp{CAM}. This essentially piggybacks a notification about the latest \ac{CRL}-update event and an authenticator for validating \ac{CRL} pieces. This provides \ac{CRL} validation for free: pseudonyms are readily validated by the receiving vehicles since each vehicle verifies the signature on a pseudonym before validating the content of a \ac{CAM}, i.e., the verification of \ac{CRL} pieces does not incur extra computation overhead. This eliminates the need for signature verification, but a \ac{BF} membership test, for each \ac{CRL} piece as the fingerprint is signed with the private key of the \ac{PCA}. 

Our scheme does not require prior knowledge on trip duration in order to obtain \acp{CRL}, i.e., a vehicle can be oblivious to the trip duration. In fact, such information would not be relevant to the \ac{CRL} dissemination: due to the unpredictable nature of revocation events, the \acp{PCA} disseminate at each point revoked pseudonyms whose lifetimes fall within a $\Gamma_{CRL}$ interval. As long as a vehicle moves inside a domain, it does not need to receive \acp{CRL} from other domains: all vehicles in the domain are issued pseudonyms by the \acp{PCA} in that domain. In other words, our scheme does not require any communication and cooperation between \acp{RSU} and \acp{PCA} from different domains on \ac{CRL} construction and distribution tasks; only \acp{PCA}-\acp{RSU} collaboration within a domain. The \acp{PCA} operating in a domain construct the \acp{CRL} and push the \ac{CRL} pieces to the \acp{RSU} so that the \acp{RSU} broadcast the \ac{CRL} pieces for the current $\Gamma_{CRL}$.

\begin{figure} [!t]
	\vspace{-3em}
	\begin{center}
		\centering
		\includegraphics[width=0.42\textwidth,height=0.42\textheight,keepaspectratio]{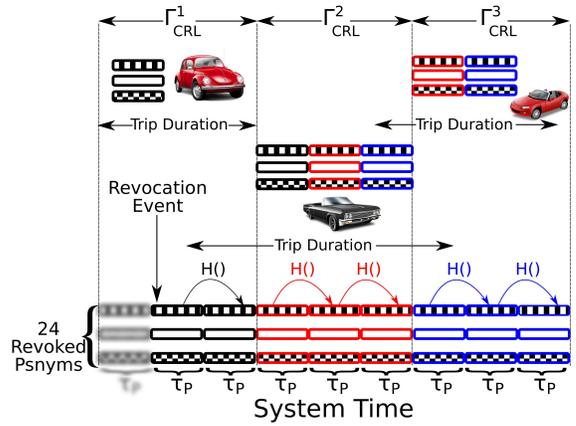}
		\vspace{-0.5em}
		\caption{A vehicle-centric approach: each vehicle only subscribes for pieces of \acp{CRL} corresponding to its trip duration.}
		\label{fig:crl-dis-vehicle-centric-overview}
	\end{center}
	\vspace{-1.5em}
\end{figure}

Fig.~\ref{fig:crl-dis-vehicle-centric-overview} illustrates an example of 24 revoked pseudonyms to be distributed. A vehicle traveling within $\Gamma^{1}_{CRL}$ would possibly only face revoked pseudonyms with a lifetime falling in that interval, 6 pseudonyms, shown in black, instead of all 24 entries (the blurred pseudonyms are expired, thus not included in the \ac{CRL}). These 6 revoked pseudonyms within $\Gamma^{1}_{CRL}$ can be implicitly bound without compromising their unlinkability prior to the revocation event, in a way that one can simply derive subsequent pseudonyms from an anchor (the blurred pseudonyms are non-revoked but expired and they cannot be linked to the revoked ones; this becomes clear later). Thus, in this example, distributing 3 entries for that vehicle is sufficient. Another vehicle, however, traveling for a longer duration, e.g., from the middle of $\Gamma^{1}_{CRL}$ till the beginning of $\Gamma^{3}_{CRL}$, would need to be provided with all 24 revocation entries, i.e., requiring 9 entries to derive all 24 revoked pseudonyms. 

In a more realistic example, assume there are 1 million vehicles in the system, each has 6 hours worth of pseudonyms (72 pseudonyms per day with $\Gamma=30$ min and $\tau_{P}=5$ min, i.e., 6 pseudonyms per $\Gamma$), all are issued timely aligned with the rest with non-overlapping intervals~\cite{khodaei2018Secmace}. Suppose 1 percent of them are compromised or their sensors became faulty and thus evicted from the system. As a result, the revocation information to be disseminated for a day contains 720,000 entires, thus a \ac{CRL} of around 22 MB (with 256-bit long serial numbers per pseudonym). By implicitly binding pseudonyms belonging to each \ac{OBU}, one can distribute 1 entry for a batch of revoked pseudonyms per $\Gamma$ (with some additional information), in total, 12 entries per revoked vehicle instead of 72 entries. Thus, the size of the \ac{CRL} for that day becomes 7.3 MB, with 120,000 entries (with 256-bit serial numbers and 256-bit of complementary information for each entry). This already shows a significant reduction of the \ac{CRL} size. However, distributing all that revocation information ignores the temporal nature of pseudonyms and the vehicle trip duration; it is more effective to distribute revocation information for a protocol-selectable period in the near future. Therefore, when a vehicle is to travel approximately within a $\Gamma_{CRL}$ interval, assumed for example to be 30 min, it will only receive pieces of information for that $\Gamma_{CRL}$, i.e., around 10,000 entries and thus a \ac{CRL} size of 625 KB instead of 22 MB, i.e., 3 orders of magnitude reduction of the \ac{CRL} size distributed at any point in time.

\subsection{Security Protocols}
\label{subsec:crl-dis-security-protocols}

In a nutshell, the \acp{PCA} operating in a domain construct the \acp{CRL} by sorting the revoked pseudonyms based on their validity periods in a $\Gamma_{CRL}$ interval and push them to the \acp{RSU} (Sec.~\ref{subsubsec:crl-dis-pca-operation-for-crl-construction}). For ease of exposition, we assume there is one \ac{PCA}, even though the extension of our scheme with multiple \acp{PCA} within a given domain is straightforward. \acp{RSU} and fingerprint-carrier peers publish the \ac{CRL}-update notification and the \ac{CRL} pieces (Sec.~\ref{subsubsec:crl-dis-operations-for-publishing-crl}). Upon receiving a new revocation event, each vehicle broadcasts a query to its neighbors to fetch the (missing) pieces of the \ac{CRL}, e.g., similarly to~\cite{das2004spawn}, corresponding to its actual trip duration (Sec.~\ref{subsubsec:crl-dis-operations-for-crl-subscription}). Finally, it parses recovered \ac{CRL} pieces and stores them locally (Sec.~\ref{subsubsec:crl-dis-vehcile-operation-for-parsing-crl-pieces}).

Beyond \ac{CRL} distribution protocols, we provide a modified pseudonym acquisition process (Sec.~\ref{subsubsec:crl-dis-pseudonym-acquisition-process}): all pseudonyms belonging to a requester in a $\Gamma$ are issued in a way that does not link them, unless the \ac{PCA} reveals only the first revoked pseudonym serial number in a $\Gamma$ interval. Moreover, a fraction of pseudonyms is equipped with a fingerprint of \ac{CRL} pieces in a $\Gamma$ interval, to facilitate fast validation of \ac{CRL} pieces. The notation is given in Table~\ref{table:crl-dis-protocols-notation}. 

\subsubsection{\textbf{Pseudonym Acquisition Process (Protocol~\ref{protocol:crl-dis-issuing-psnyms})}}
\label{subsubsec:crl-dis-pseudonym-acquisition-process}

A vehicle first requests an anonymous ticket~\cite{khodaei2016evaluating, khodaei2014ScalableRobustVPKI} from its \ac{H-LTCA}, using it to interact with the desired \ac{PCA} to obtain pseudonyms. Upon reception of a valid ticket, it generates \ac{ECDSA} public/private key pairs~\cite{1609-2016, ETSI-102-638} and sends the request to the \ac{PCA}~\cite{khodaei2016evaluating, khodaei2014ScalableRobustVPKI}. Vehicle-\ac{LTCA} is over mutually authenticated \ac{TLS}~\cite{dierks2008transport} tunnels (or \ac{DTLS}~\cite{rescorla2012datagram}) and the vehicle-\ac{PCA} communication is over a unidirectional (server-only) authenticated \ac{TLS} (or \ac{DTLS}). 

\begin{table}
	\vspace{-1.5em}
	\centering
	\caption{Notation Used in the Protocols.}
	\vspace{-1.25em}
	\label{table:crl-dis-protocols-notation}
	\hspace{-0.95em}
	\resizebox{0.499\textwidth}{!}
	{
		\renewcommand{\arraystretch}{1.0001}
		\begin{tabular}{ | c | c || c | c | }
			\hline
			\textbf{Notation} & \textbf{Description} & \textbf{Notation} & \textbf{Description} \\\hline\hline
			$(P^{i}_{v})_{pca}$, $P^{i}_{v}$ & a valid psnym signed by the \acs{PCA} & $Append()$ & appending a revoked psnym SN to \acp{CRL} \\\hline 

			\shortstack{$(K^i_v, k^i_v)$} & \shortstack{psnym pub./priv. key pairs} & BFTest() & \ac{BF} membership test \\\hline
			
			$(K_{pca}; Lk_{pca})$ & long-term pub./priv. key pairs & $p$, $K$ & false positive rate, optimal hash functions \\\hline 
			
			\shortstack{$(msg)_{\sigma_{v}}$} & \shortstack{signed msg with vehicle's priv. key} & \shortstack{$\Gamma$} & \shortstack{interval to issue time-aligned psnyms} \\\hline 
			
			$LTC$ & \acl{LTC} & $\Gamma_{CRL}$ & interval to release \acp{CRL} \\\hline 
			
			$t_{now}, t_s, t_e$ & a fresh, starting, ending timestamp & $RIK$ & revocation identifiable key \\\hline 
			
			$T_{timeout}$ & response reception timeout & $\mathbb{B}$ & max. bandwidth for \ac{CRL} distribution \\\hline 
			
			$n\textnormal{-}tkt$, $(n\textnormal{-}tkt)_{ltca}$ & a native ticket & $\mathbb{R}$ & revocation rate \\\hline 
			
			$Id_{req}, Id_{res}$ & request/response identifiers & N & total number of \ac{CRL} pieces in each $\Gamma_{CRL}$ \\\hline 
			
			$SN$ & psnym serial number & n & number of remaining psnyms in each batch \\\hline 
			
			$Sign(Lk_{ca}, msg)$ & signing a msg with \acs{CA}'s priv. key & k & index of the first revoked psnym \\\hline 
			
			$Verify(LTC_{ca}, msg)$ & verifying with the \acs{CA}'s pub. key & $CRL_{v}$ & \ac{CRL} version \\\hline
			
			$GenRnd(), rand(0,*)$ & GEN. a random number, or in range & $\emptyset$ & Null or empty vector \\\hline 
			
			$H^{k}(), H$ & hash function ($k$ times), hash value & k, j, m, $\zeta$ & temporary variables \\\hline 
		\end{tabular} 
	}
	\vspace{0.5em}	
\end{table}

\setlength{\textfloatsep}{0pt}
\begin{algorithm}[t!]
	\floatname{algorithm}{Protocol}
	\caption{Issuing Pseudonyms (by the \acs{PCA})}
	\label{protocol:crl-dis-issuing-psnyms}
	\algloop{For}{}

	\algblock{Begin}{End}
	\begin{algorithmic}[1]
		\Procedure{IssuePsnyms}{$Req$}
		\State {\scriptsize $Req\to{(Id_{req}, t_{s}, t_{e}, (tkt)_{\sigma_{ltca}}, \{(K^1_v)_{\sigma_{k^1_v}},\cdots,(K^n_v)_{\sigma_{k^n_v}}\},nonce,t_{now})}$} 
		\State $\text{Verify}(\ac{LTC}_{ltca}, (tkt)_{\sigma_{ltca}})$ 
		\State $Rnd_{v} \gets GenRnd()$
		\For{i:=1 to \textbf{n}}{}
		\Begin
		\State $\text{Verify}(K^{i}_{v}, (K^i_v)_{\sigma_{k^i_v}})$ 
		\State ${RIK_{P^i_v} \gets H(IK_{tkt} || K^i_v || t_{s}^i || t_{e}^i|| H^{i}(Rnd_{v})})$
		\If {$i = 1$} 
		\State $SN^i \gets H(RIK_{P^i_v} || H^{i}(Rnd_{v}))$
		\Else
		\State $SN^i \gets H(SN^{i-1} || H^{i}(Rnd_{v}))$
		\EndIf
		\State ${\zeta \leftarrow (SN^i, K^i_v, CRL_{v}, BF_{\Gamma_{CRL}^{i}}, RIK_{P^i_v}, t_{s}^i, t_{e}^i)}$ 

		\State $(P^i_v)_{\sigma_{pca}} \leftarrow Sign(Lk_{pca}, \zeta)$
		\End
		\State \textbf{return} $(Id_{res}, \{(P^1_v)_{\sigma_{pca}}, \dots, (P^n_v)_{\sigma_{pca}}\}, Rnd_{v}, nonce\textnormal{+}1, t_{now})$
		\EndProcedure
	\end{algorithmic}
\end{algorithm}

Having received a request, the \ac{PCA} verifies the ticket signed by the \ac{H-LTCA} (assuming trust is established between the two) (steps~\ref{protocol:crl-dis-issuing-psnyms}.2\textendash\ref{protocol:crl-dis-issuing-psnyms}.3). Then, the \ac{PCA} generates a random number (step~\ref{protocol:crl-dis-issuing-psnyms}.4) and initiates a proof-of-possession protocol to verify the ownership of the corresponding private keys by the vehicle (step \ref{protocol:crl-dis-issuing-psnyms}.7). Then, it calculates $H(IK_{tkt} || K^i_v || t_{s}^i || t_{e}^i || H^{i}(Rnd_{v}))$\footnote{$IK_{tkt}$ in a ticket prevents even a compromised \ac{H-LTCA} from mapping the ticket to a different \ac{LTC} during resolution process~\cite{khodaei2018Secmace}.}, the \emph{``revocation identifiable key''} ($RIK$). This essentially prevents a compromised \ac{PCA} from mapping a different ticket during resolution process (step~\ref{protocol:crl-dis-issuing-psnyms}.8). The \ac{PCA} implicitly correlates a batch of pseudonyms belonging to each requester (steps~\ref{protocol:crl-dis-issuing-psnyms}.9\textendash\ref{protocol:crl-dis-issuing-psnyms}.13). This essentially enables efficient distribution of the \ac{CRL}: the \ac{PCA} only needs to include one entry per batch of pseudonyms without compromising their unlinkability. Finally, the \ac{PCA} issues the pseudonyms (steps~\ref{protocol:crl-dis-issuing-psnyms}.14\textendash\ref{protocol:crl-dis-issuing-psnyms}.15) and delivers the response (step~\ref{protocol:crl-dis-issuing-psnyms}.17). Note that a \ac{PCA} randomly selects some of the pseudonyms to be fingerprint-carriers by integrating a \ac{BF} of all \ac{CRL} pieces within a $\Gamma_{CRL}$ ({\small $BF_{\Gamma_{CRL}^{i}}$}) (step~\ref{protocol:crl-dis-issuing-psnyms}.14). This parameter (fraction of fingerprint-carriers) can be set based on different factors, e.g., frequency of revocation events and coverage of deployed \acp{RSU}, which are beyond the scope of this work.

\begin{algorithm}[!t]
	\floatname{algorithm}{Protocol}
	\caption{\ac{CRL} Construction (by the \acs{PCA})} 
	\label{protocol:crl-dis-algo-construction}
	\begin{algorithmic}[1]
		\vspace{-0.1em}
		\Procedure{GenCRL}{$\Gamma_{CRL}^{i}, \mathbb{B}$}
		\myState{$Piece_{\Gamma_{CRL}^i} \gets \emptyset$}
		
		\vspace{-0.1em}
		\Repeat 
		\vspace{-0.1em}
		\myState{$\{SN^{k}_{P}, H^{k}_{Rnd_{v}}, n\} \gets fetchRevokedPsnyms(\Gamma_{CRL}^{i})$}
		\If {$SN^{k}_{P} \not= Null$}
		\vspace{-0.1em}
		\myState{$Piece_{\Gamma_{CRL}^{i}} \gets Append(\{SN^{k}_{P}, H^{k}_{Rnd_{v}}, n\})$}
		\vspace{-0.1em}
		\EndIf
		\Until{$SN^{k}_{P} == Null$}
		
		\myState{$N \gets {\Bigg \lceil} \dfrac{size(Piece_{\Gamma_{CRL}^{i}})}{\mathbb{B}} {\Bigg \rceil} $} \Comment{calculating number of pieces with a given $\mathbb{B}$}
		\vspace{-0.1em}
		\For{$j\gets 0, N$} \Comment{N: number of pieces in $\Gamma_{CRL}^{i}$}
		\myState{$Piece^{j}_{\Gamma_{CRL}^{i}} \gets Split(Piece_{\Gamma_{CRL}^{i}}, \mathbb{B}, N)$} \Comment{splitting into $N$ pieces}
		\vspace{-0.1em}
		\EndFor
		\vspace{-0.1em}
		\myState{\textbf{return} $\{(Piece^{1}_{\Gamma_{CRL}^i}),\dots,(Piece^{N}_{\Gamma_{CRL}^i})\}$}
		\vspace{-0.1em}
		\EndProcedure
	\end{algorithmic}
\end{algorithm}

\subsubsection{\textbf{\ac{PCA} Operation for \ac{CRL} Construction (Protocol~\ref{protocol:crl-dis-algo-construction})}}
\label{subsubsec:crl-dis-pca-operation-for-crl-construction}

When a vehicle is to be evicted, the \ac{PCA} sorts revoked pseudonyms based on the pseudonyms validity intervals in each $\Gamma_{CRL}$. It then appends the following data for each batch of pseudonyms: (i) the serial number of the first revoked pseudonym in the chain ($SN^i$), (ii) a hash value ($H^{i}_{Rnd_{v}}$), and (iii) the number of remaining pseudonyms in this batch ($n$) (steps~\ref{protocol:crl-dis-algo-construction}.2\textendash~\ref{protocol:crl-dis-algo-construction}.8). It then splits the \ac{CRL} into multiple pieces according to the maximum allocated bandwidth, i.e., system parameter $\mathbb{B}$, for \ac{CRL} distribution (steps~\ref{protocol:crl-dis-algo-construction}.9\textendash~\ref{protocol:crl-dis-algo-construction}.13). The number of revocation entries is proportional to the number of pseudonyms and vehicles, and revocation events, e.g., due to vehicle-compromising malware propagation, evaluated in Sec.~\ref{sec:crl-dis-scheme-analysis-evaluation}.

\begin{algorithm}[!t] 
	\floatname{algorithm}{Protocol}
	\caption{Publishing \acp{CRL} (by the \acsp{OBU})} 
	\label{crl-dis-algo-publishing-crl}
	\begin{algorithmic}[1]
		\Procedure{PublishCRL()}{} 
		\myState{$\{(Id_{req}, \Gamma_{CRL}^i, [indexes])\} = receiveQuery((\zeta)_{\sigma_{P^{i}_{v}}})$}
		\myState{$Verify (P^{i}_{v}, (\zeta)_{\sigma_{P^{i}_{v}}})$} 
		\myState{$CRL^*_{\Gamma_{CRL}^i} = search_{local}(\Gamma_{CRL}^i)$} \Comment{search local repository}
		\myState{$j \gets rand(0, *)$} \Comment{randomly select one of the available pieces}
		\If {$CRL^j_{\Gamma_{CRL}^i} \not= \emptyset$}
		\myState{$broadcast(\{Id_{res}, CRL^{j}_{\Gamma_{CRL}^i}\})$}
		\EndIf
		\EndProcedure
	\end{algorithmic}
\end{algorithm}

\subsubsection{\textbf{Operations for Publishing the \ac{CRL} (Protocol~\ref{crl-dis-algo-publishing-crl})}}
\label{subsubsec:crl-dis-operations-for-publishing-crl}

Each \ac{RSU} continuously broadcasts the signed fingerprint of \ac{CRL} pieces, to notify vehicles in a region about any new revocation event. The transmission rate of the signed fingerprint corresponding to the current $\Gamma^{i}_{CRL}$ can gradually decrease towards the end of $\Gamma^{i}_{CRL}$; instead, the transmission rate of the signed fingerprint for $\Gamma^{i+1}_{CRL}$ can moderately increase. This ``ensures'' that all legitimate vehicles are notified about a new revocation event, thus being capable to request and efficiently validate \ac{CRL} pieces (evaluated in Fig.~\ref{fig:crl-dis-average-end-to-end-latency-to-fetch-CRL-lust-dataset-fingerprint-dissemination}.b). Upon reception and validation of a query, an \ac{RSU} commences transmission across the wireless data link with a low-rate transmission (without any acknowledgment from peers). 

Upon receiving an authentic query for the missing \ac{CRL} pieces (steps~\ref{crl-dis-algo-publishing-crl}.2\textendash\ref{crl-dis-algo-publishing-crl}.3) by a neighboring vehicle, a vehicle searches its local repository and randomly chooses one of the requested pieces and broadcasts it (steps~\ref{crl-dis-algo-publishing-crl}.4\textendash\ref{crl-dis-algo-publishing-crl}.8). The maximum allocated bandwidth for \ac{CRL} distribution is $\mathbb{B}$, chosen to be much smaller than $C$, the bandwidth the data link support ($\mathbb{B}\ll C$). Such a rate limiting mechanism ensures that a compromised insider cannot abuse the allocated bandwidth towards performing a \ac{DoS} attack, thus \ac{CRL} distribution does not interfere with other safety-critical operations. 

\begin{algorithm}[!t]
	\floatname{algorithm}{Protocol}
	\caption{Subscribing to \ac{CRL} Pieces (by the \acsp{OBU})} 
	\label{crl-dis-algo-subscribing-to-crl-pieces-from-rus-obu}
	\begin{algorithmic}[1]
		\Procedure{SubscribeCRL}{$\Gamma_{CRL}^{i}, N$} 
		\myState{$resp_{final} \gets \emptyset, j \gets 0, t \gets t_{now} + T_{timeout}$} 
		\Repeat 
		\myState{$ \zeta \leftarrow (Id_{req}, \Gamma_{CRL}^i, [missing \: pieces \: indexes])$}
		\myState{$(\zeta)_{\sigma_{v}} \leftarrow Sign(k^{i}_{v}, \zeta)$}
		\myState{$broadcast((\zeta)_{\sigma_{P^{i}_{v}}}, P^{i}_{v})$} 
		\myState{$Piece^{j}_{\Gamma_{CRL}^{i}} \gets receiveBefore(t)$} 
		\If {$BFTest(Piece^{j}_{\Gamma_{CRL}^{i}}, BF_{\Gamma_{CRL}^{i}})$} 
		\myState{$resp_{final} \gets Store(Piece^{j}_{\Gamma_{CRL}^{i}})$} \Comment{storing in local repository} 
		\EndIf
		\myState{$j \gets j + 1$}
		\Until{$j > N$}
		\myState{\textbf{return} $resp_{final}$}
		\EndProcedure
	\end{algorithmic}
\end{algorithm}

\subsubsection{\textbf{Operations for \ac{CRL} Subscription (Protocol~\ref{crl-dis-algo-subscribing-to-crl-pieces-from-rus-obu})}}
\label{subsubsec:crl-dis-operations-for-crl-subscription}

Each vehicle can receive necessary \ac{CRL} pieces corresponding to its actual trip duration from nearby \acp{RSU} or neighboring vehicles. A vehicle broadcasts a signed query to its neighbors, to receive the missing pieces of the revocation information of $\Gamma^{i}_{CRL}$ during which the vehicle wishes to travel (steps~\ref{crl-dis-algo-subscribing-to-crl-pieces-from-rus-obu}.2\textendash~\ref{crl-dis-algo-subscribing-to-crl-pieces-from-rus-obu}.6). Having received a \ac{CRL} piece, it simply validates the piece by testing against the signed fingerprint (already obtained from \acp{RSU} in vicinity or integrated in a subset of recently issued pseudonyms broadcasted in the network). If the \ac{BF} test is successful, it accepts that piece and keeps requesting until successfully receiving all remaining pieces (steps~\ref{crl-dis-algo-subscribing-to-crl-pieces-from-rus-obu}.7\textendash~\ref{crl-dis-algo-subscribing-to-crl-pieces-from-rus-obu}.12).

\subsubsection{\textbf{Operations for Parsing \ac{CRL}}}
\label{subsubsec:crl-dis-vehcile-operation-for-parsing-crl-pieces}

Upon reception and validation of a \ac{CRL} piece, each vehicle derives the revoked pseudonym serial numbers from the obtained hash anchors, by calculating a hash value $n$ times: {\small $H(SN^{i}||H(H^{i}_{Rnd_v}))$}. Revocation entries can be stored in local storage, e.g.,~\cite{ganan2012toward}, and searched with $O(log(n))$ time complexity. To enhance revocation status validation, a vehicle could generate a \ac{BF} locally~\cite{haas2011efficient} with constant computational cost ($O(1)$) for insertions and search operations but at a cost of a false positive rate. Note that the revocation entries are stored for the period they are valid for, i.e., within a $\Gamma_{CRL}^{i}$ interval.

\section{Scheme Analysis and Evaluation}
\label{sec:crl-dis-scheme-analysis-evaluation}

We first discuss how our scheme satisfies the security and privacy requirements, as well as operational requirements defined in Sec.~\ref{subsec:crl-dis-requirements} and then demonstrate quantitatively its efficiency, scalability, and resiliency through an extensive experimental evaluation. 

\subsection{Qualitative Analysis}
\label{subsec:crl-dis-qualitative-analysis}

\emph{Fine-grained authentication, integrity, and non-repudiation:} The authenticity and integrity of each \ac{CRL} piece is validated by testing each piece against the fingerprint, periodically broadcasted by \acp{RSU} and integrated in a subset of recently issued pseudonyms (R1). Moreover, no \ac{PCA} can deny the inclusion of pseudonym serial number as the fingerprint of \ac{CRL} pieces is signed with the \ac{PCA}'s private key (R1). Furthermore, each query to obtain \ac{CRL} pieces is authenticated, in fact signed with the current valid pseudonym of the vehicle, thus preventing from abusing mechanism. If a \emph{legitimate-looking} node aggressively requests \ac{CRL} pieces, responding to such requests can be of the lowest priority and they are reported as potential misbehavior.

\begin{figure} [!t]
	\vspace{-1.75em}
	\begin{center}
		\centering
		\subfloat[Vehicle-centric scheme]{
			\hspace{-1em} 
			\includegraphics[trim=0cm 0.15cm 0.5cm 1.45cm, clip=true, width=0.265\textwidth,height=0.265\textheight,keepaspectratio]{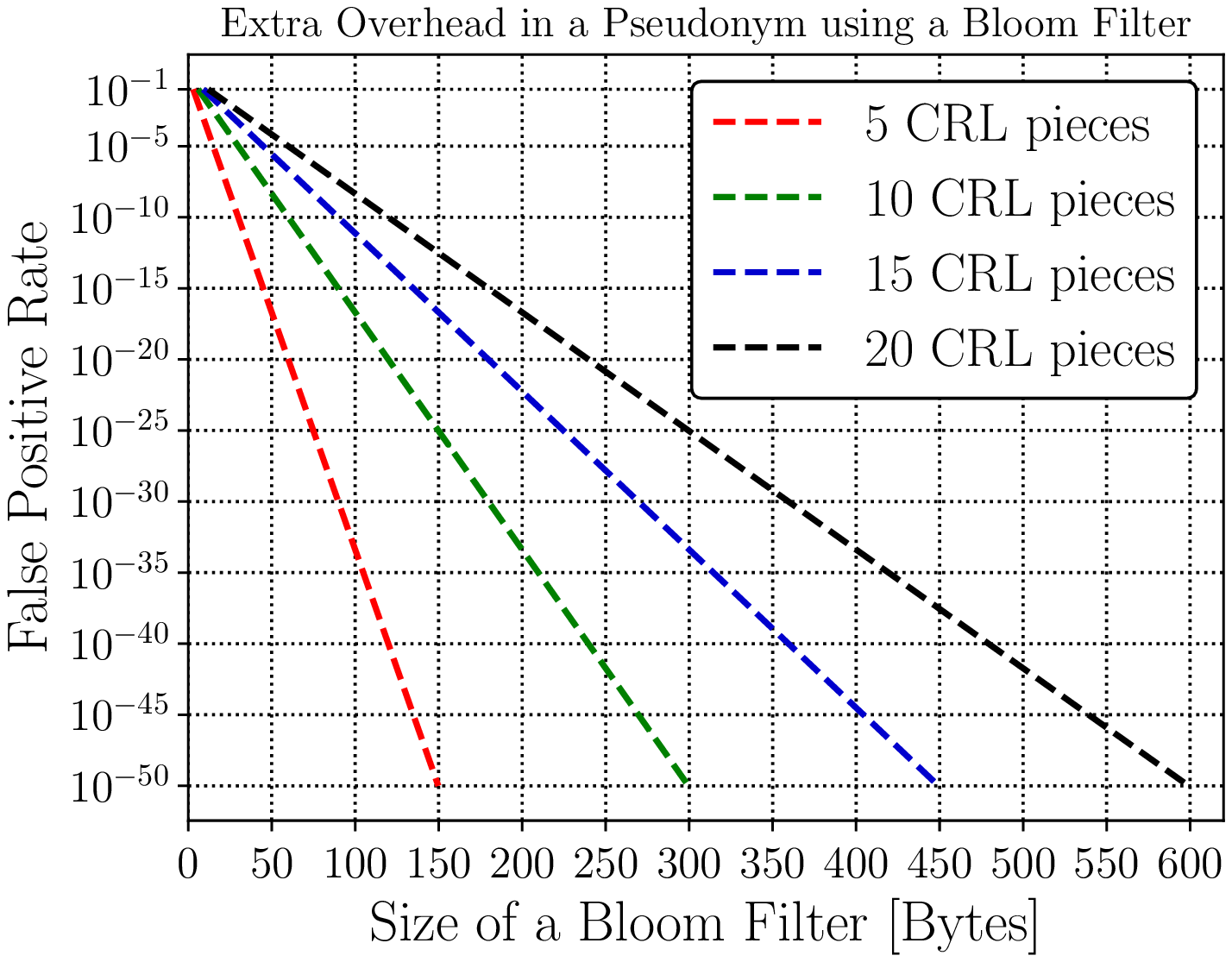}}
		\subfloat[Precode-and-hash scheme~\cite{nguyen2016secure}]{
			\hspace{-1.35em} 
			\includegraphics[trim=0cm 0.15cm 0.5cm 1.45cm, clip=true, width=0.265\textwidth,height=0.265\textheight,keepaspectratio]{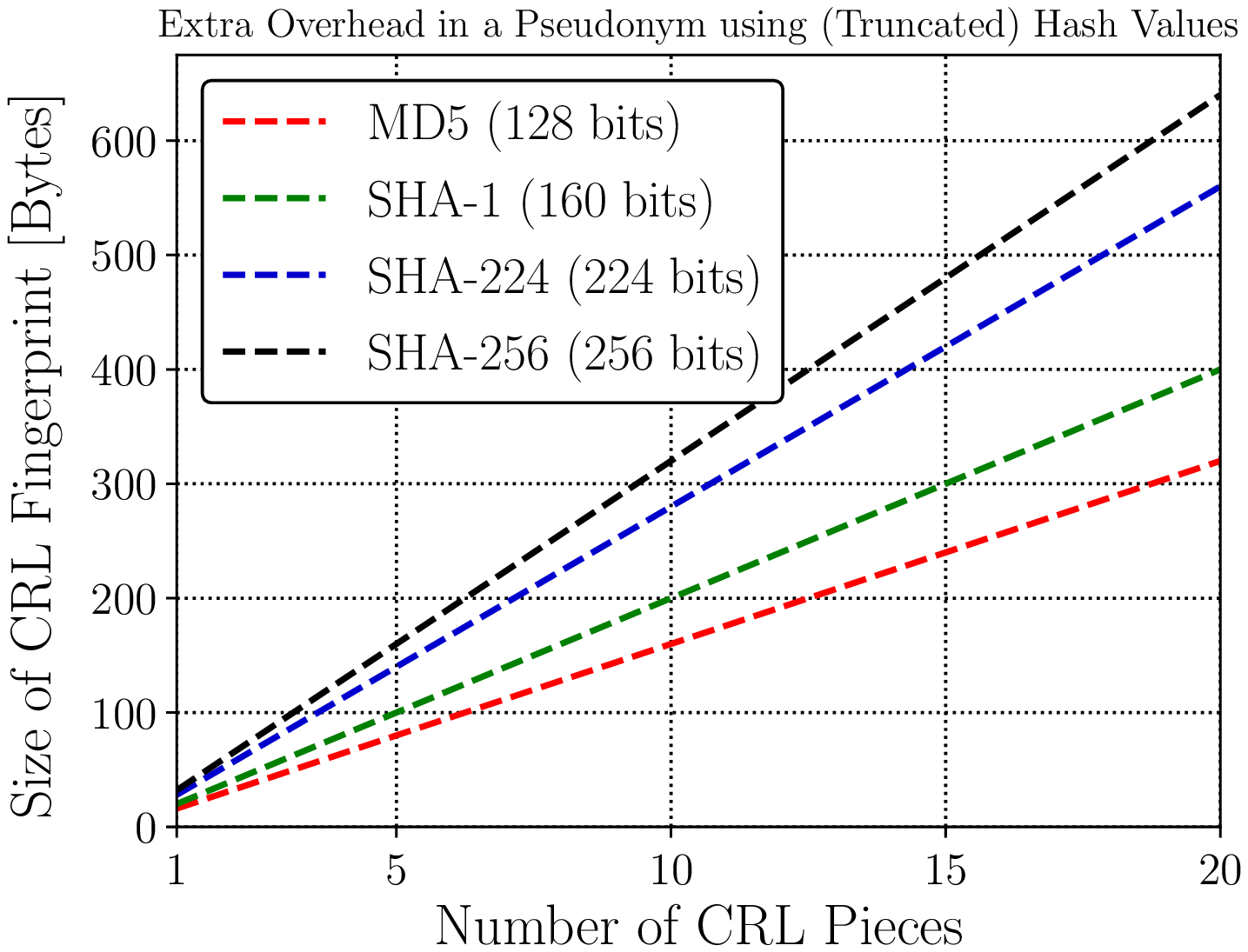}}
		\vspace{-0.75em}
		\caption{Extra overhead for \ac{CRL} fingerprints.}
		\label{fig:crl-dis-extra-overhead-for-crl-fingerprints}
	\end{center}
	\vspace{-0em}
\end{figure}

Representing \ac{CRL} pieces in a space-efficient \ac{BF} trades off communication overhead for a false positive rate ($p$). Fig.~\ref{fig:crl-dis-extra-overhead-for-crl-fingerprints}.a shows that the \ac{BF} size linearly increases as the false positive rate decreases. For example, for 10 \ac{CRL} pieces covering one $\Gamma_{CRL}$, and $p = 10^{-20}$ (with the optimal number of hash functions), the \ac{BF} size and thus the overhead for each pseudonym is 120 bytes. This eliminates the need to sign each \ac{CRL} piece. However, one might target the false positive rate of a \ac{BF} towards generating a fake piece of \ac{CRL} to be accepted as legitimate. This is different from a pollution or a \ac{DDoS} attack: not only would it prevent a legitimate vehicle from obtaining a genuine \ac{CRL} piece, but also disseminate an \emph{authentic-looking} piece that passes the \ac{BF} test; in fact, such attacks can rely on sheer computational power.

Our scheme resists such attacks that attempt to exclude revoked pseudonym serial numbers or add valid ones by forging a fake \ac{CRL} piece that passes the BF test.\footnote{Generating a fake \ac{BF} with completely different valid pseudonyms serial number necessitates accessing at least, e.g., $10^{20}$, valid pseudonyms, i.e., a more powerful adversary (\emph{malicious \ac{VPKI} entities}), and is beyond the scope of our adversarial model.} An adversary could buy top-notch bitcoin-mining hardware, Antminer-S9~\cite{antminerS9Review} (14TH/s, \$3,000). If $\Gamma_{CRL}=1$ hour and $p=10^{-20}$, and the optimal number of hash functions, $K=67$, the adversary needs 132,936 Antminer-S9 (\$400M) to generate a bogus piece within $\Gamma_{CRL}$ ({\small $\frac{10^{20}\times67}{14\times10^{12}}$}). Alternatively, he could join AntPool~\cite{antpoolchina}, one of the largest Bitcoin mining pools, ($1,604,608 \: TH/s$) to generate a fake piece in 70 min, which might seem to be practical. However, if $p=10^{-22}$ (with $K=73$) or even $p=10^{-23}$ (with $K=76$), the adversary would need 5 or 55 days, respectively ({\small $\frac{10^{22}\times73}{1.6\times10^{18}}=126h$, $\frac{10^{23}\times76}{1.6\times10^{18}}=1,319h$}). With inherently short $\tau_{P}$ (important for unlinkability and thus privacy) and $\Gamma_{CRL}$, proper choice of $p$ makes attacks infeasible; in other words, irrelevant, as forged pieces refer to already expired credentials. Upon receiving conflicting pieces, vehicles report misbehavior to the \ac{VPKI} to take appropriate actions, e.g., adjusting $p$. The results of our experiments in Sec.~\ref{subsec:crl-dis-quantitative-analysis} rely on $p=10^{-30}$ and $K=100$.

The \ac{PCA} can concatenate the hash values for each \ac{CRL} piece~\cite{nguyen2016secure}, or alternatively truncate the output of hash functions. Fig.~\ref{fig:crl-dis-extra-overhead-for-crl-fingerprints}.b shows the size of a \ac{CRL} fingerprint with different hash functions. For instance, by employing \emph{precode-and-hash} with SHA1 (20 bytes output size)~\cite{nguyen2016secure}, the size of a fingerprint for 20 \ac{CRL} pieces becomes 400 bytes; whereas employing our scheme results in an extra overhead of 311 bytes ($p=10^{-25}$) or 371 bytes for the extremely low false positive rate ($p=10^{-30}$). 

\emph{Unlinkability (perfect-forward-privacy):} Upon a revocation event and \ac{CRL} release, an external observer can try to link the revoked pseudonyms backwards (towards the beginning of the $\Gamma$ interval). However, it is infeasible to link the previously non-revoked (but expired) pseudonyms belonging to a misbehaving vehicle due to the utilization of a hash-chain during pseudonym issuance process ({\small $SN^i \gets H(RIK_{P^i_v} || H^{i}_{Rnd_{v}})$} or {\small $SN^{i} \gets H(SN^{i-1} || H(H^{i}_{Rnd_{v}})$}), i.e., strong user privacy protection for a period, during which the vehicle was not compromised (R2).

In collusion with \ac{V2X} observers, honest-but-curious \acp{PCA} operating in a given domain might be tempted to infer sensitive information from the pseudonyms, e.g., timing information, or, in our context, the \acp{CRL}, towards linking pseudonym sets and tracking a vehicle. However, all the issued pseudonyms are aligned with global system time (\ac{PCA} clock), thus, there is no distinction among pseudonyms based on pseudonym timing information. Moreover, the \acp{CRL} do not disclose extra information to harm user privacy\footnote{Each \ac{PCA} can trivially link the issued pseudonyms for the same vehicle as a response to a single request. However, one can configure the system to achieve \emph{full unlinkability}, i.e., $\Gamma$ is set equal to $\tau_{P}$ and force obtaining each single pseudonym with a different ticket. This implies that even honest-but-curious \acp{PCA} cannot link any two pseudonyms issued for a single vehicle, but it would be impractical in most setting.}. Moreover, \acp{PCA} randomly select a subset of pseudonyms to be fingerprint-carries; thus, correlating any of these pseudonyms does not imply that they belong to the same vehicle (R2).

\emph{Availability:} We leverage \acp{RSU} and car-to-car epidemic distribution to disseminate \ac{CRL} pieces and signed fingerprints for increased availability or intermittent connectivity (R3). The resilience to pollution and \ac{DDoS} attacks stems from three factors: (i) a huge reduction of the \ac{CRL} size, notably because of distributing \ac{CRL} information only for relevant periods of time, (ii) very efficient verification of \ac{CRL} pieces, i.e., testing against a \ac{BF} with hash and not signature validation, and (iii) integrating the fingerprint of \ac{CRL} pieces in a subset of pseudonyms (R3).

\begin{figure} [!t]
	\vspace{-1.75em}
	\begin{center}
		\centering
		\subfloat[\ac{CRL} size comparison]{
			\hspace{-2em} 
			\includegraphics[trim=0.1cm 0.35cm 0.4cm 1.25cm, clip=true, width=0.266\textwidth,height=0.266\textheight,keepaspectratio]{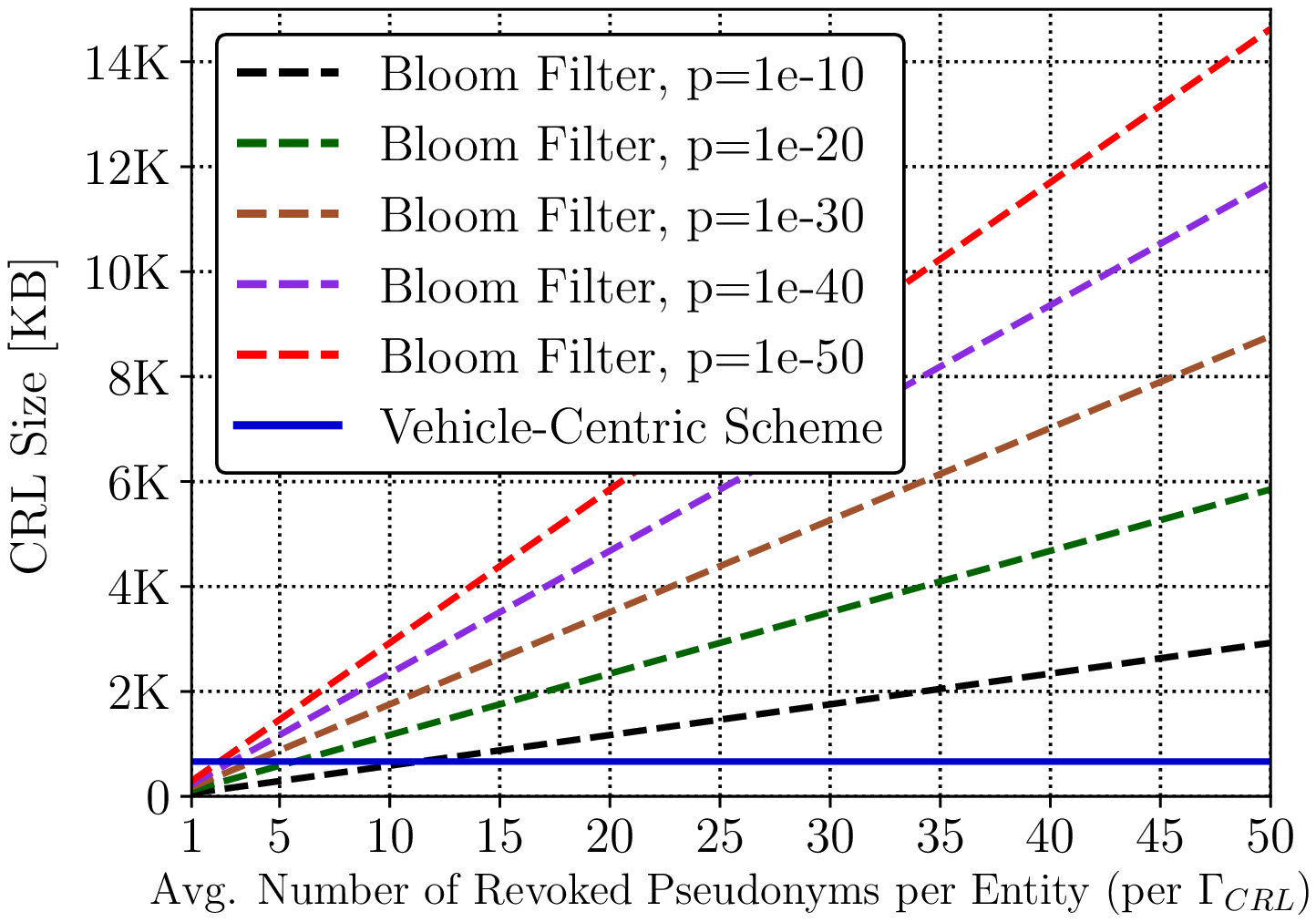}}
		\centering
		\subfloat[\ac{C$^2$RL} as a factor of false positive rate]{ 
			\hspace{-1.25em} 
			\includegraphics[trim=0cm 0.35cm 0.4cm 1.25cm, clip=true, width=0.266\textwidth,height=0.266\textheight,keepaspectratio]{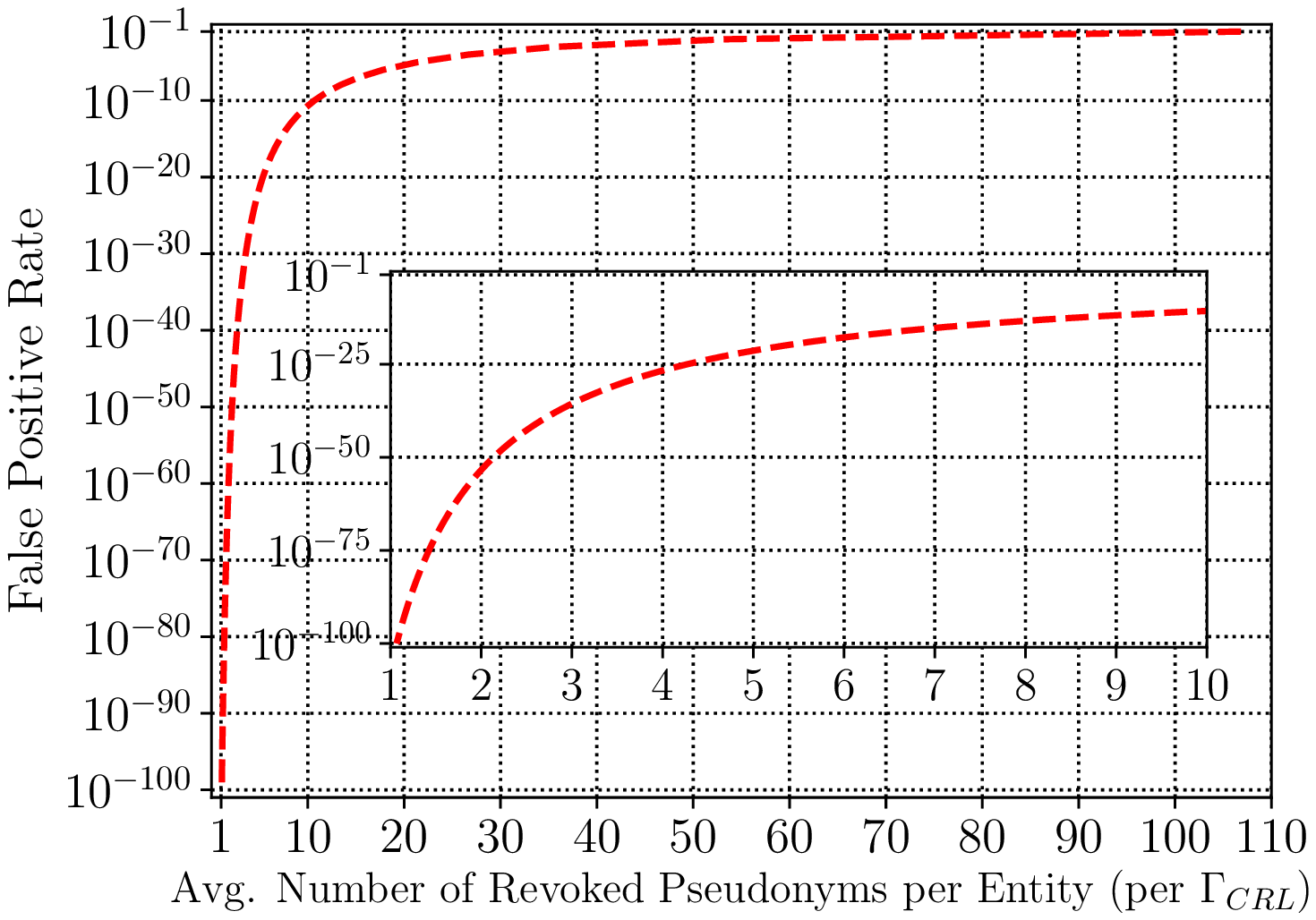}}
		\vspace{-0.75em}
		\caption{(a) \ac{CRL} size comparison for \ac{C$^2$RL} and vehicle-centric scheme (10,000 revoked vehicles). (b) Achieving vehicle-centric comparable \ac{CRL} size for the \ac{C$^2$RL} scheme.}
		\label{fig:crl-dis-compression-size-comparison}
	\end{center}
	\vspace{0.25em}
\end{figure}

\emph{Efficiency:} The efficiency stems from the efficient construction of an authenticator for \ac{CRL} pieces (minimal overhead on the \ac{PCA} side), fast verification of each piece (minimal overhead on the vehicle side), and implicit binding of a batch of pseudonyms. Moreover, leveraging recurrent interactions with the \ac{VPKI}, which issues time-aligned pseudonyms for all vehicles, and distributing \acp{CRL} with respect to locality, the ephemeral nature of credentials, and the average trip duration enhances efficiency (R4). We allocate a small fraction of bandwidth for \ac{CRL} distribution and we apply a rate limiting mechanism to prevent abuse of the mechanism (R3-R4). However, allocating a small amount of bandwidth is sufficient to timely distribute \acp{CRL} to practically all legitimate vehicles within the system (R4), as demonstrated in Sec.~\ref{subsubsec:crl-dis-experimental-setup}. Note that if pseudonyms were provided for a long period and vehicles had only unidirectional connectivity~\cite{kumar2017binary}, then the \ac{VPKI} cannot integrate new information into the pseudonyms for efficiency reasons. Thus, the signed fingerprint of \ac{CRL} pieces would need to be disseminated through \acp{RSU} on a weekly basis.

\emph{Explicit and/or implicit notification on revocation events:} Malicious entities might try to prevent other legitimate vehicles from receiving \ac{CRL}-update notifications, thus preventing them from requesting the latest \ac{CRL}, i.e., compromising availability and essentially harming the \ac{VC} system security (as evicted nodes would remain undetected). \acp{RSU} periodically broadcast the signed fingerprint, corresponding to all \ac{CRL} pieces of a given $\Gamma_{CRL}$, to ensure reception of the \ac{CRL} validation authenticator in a region. Moreover, the \acp{PCA} randomly choose a subset of recently issued pseudonyms to piggyback the \ac{CRL}-update notification. Vehicles beacon \acp{CAM} at a high rate, each signed with the private key of a pseudonym that possibly carries a notification about a \ac{CRL}-update event and attach the pseudonym to a significant fraction of \acp{CAM}, in fact free notification about a revocation event at any point in time in the system (R5). Further evidence to the availability, the resiliency, and the efficiency, is provided through the detailed experimental evaluation in Sec.~\ref{subsec:crl-dis-quantitative-analysis}.

\textbf{\ac{CRL} size comparison:} The size of a \ac{CRL} by compressing the revocation information into a \ac{BF}, i.e., \ac{C$^2$RL} scheme~\cite{raya2006certificaterevocation, raya2007eviction, rigazzi2017optimized}, is {\footnotesize $m_{BF}=- \dfrac{N \times M \times ln\,p}{(ln2)^2}$}~\cite{tarkoma2012theory}, where $N$ is the total number of compromised vehicles, $M$ is the average number of revoked pseudonyms per vehicle per $\Gamma_{CRL}$, and $p$ is the probability of false positive\footnote{Remark: the two false positive rates mentioned here are different in essence; one is for compressing the \ac{CRL} entries in \ac{C$^2$RL} scheme and the another one is for efficiently validating \ac{CRL} pieces in our vehicle-centric scheme.}. Fig.~\ref{fig:crl-dis-compression-size-comparison}.a illustrates that the size of a \ac{CRL} with \ac{C$^2$RL} grows linearly with $M$. Using our vehicle-centric scheme, it is sufficient to disclose one entry to revoke all pseudonyms of an evicted vehicle within a $\Gamma_{CRL}$ interval, i.e., the size of a \ac{CRL} in each $\Gamma_{CRL}$ is a constant value with respect to $M$: $(256+256) \times N$, with 256 bits for a pseudonym serial number and 256 bits for its corresponding hash value (excluding an extra byte, the number of remaining pseudonyms in each batch). Fig.~\ref{fig:crl-dis-compression-size-comparison}.b shows that compressing revocation information with a \ac{BF} could have comparable overhead, i.e., \ac{CRL} size, with our scheme only if the probability of false positive increases. For example, if $M=10$, the false positive rate for \ac{C$^2$RL} scheme should be $10^{-10}$ to achieve a \ac{CRL} size comparable to our scheme; otherwise, compressing a \ac{CRL} with a \ac{BF} is not as efficient as our scheme. Exactly because each \ac{PCA} issues multiple pseudonyms in each $\Gamma$ (for various reasons, e.g., \ac{VPKI} performance and connectivity)~\cite{khodaei2018Secmace}, we achieve a significant improvement over \ac{C$^2$RL}, e.g., 2.6 reduction in \ac{CRL} size when $M=10$ and $p=10^{-30}$.


\begin{table}
	\vspace{-0em}
	\centering
	\caption{Simulation Parameters (\acs{LuST} dataset).} 
	\vspace{-1em}
	\label{table:crl-dis-simulation-parameters}
	\resizebox{0.47\textwidth}{!}
	{
		\renewcommand{\arraystretch}{1.2}
		\begin{tabular}{ | c | c ||| c | c | }
			\hline
			\textbf{Parameters} & \textbf{Value} & \textbf{Parameters} & \textbf{Value} \\\hline\hline
			
			\ac{CRL}/Fingerprint TX interval & 0.5s/5s & Pseudonym lifetime & 30s-600s \\\hline 
			Carrier frequency & 5.89 GHz & Area size & 50 KM $\times$ 50 KM \\\hline 
			TX power & 20mW & Number of vehicles & 138,259 \\\hline
			Physical layer bit-rate & 18Mbps & Number of trips & 287,939 \\\hline
			Sensitivity & -89dBm & Average trip duration & 692.81s \\\hline
			Thermal noise & -110dBm & Duration of simulation & 4 hour (7-9, 17-19) \\\hline 
			\ac{CRL} dist. Bandwidth ($\mathbb{B}$) & 10, 25, 50 KB/s & $\Gamma$ & 1-60 min \\\hline 
			Number of \acp{RSU} & 100 & $\Gamma_{CRL}$ & 60 min \\\hline 
		\end{tabular}
	}
	\vspace{-0.5em}
\end{table}

\begin{table}
	\vspace{-0.35em}
	\caption{\acs{LuST} Revocation Information {\small ($\mathbb{R}=1\%$, $\mathbb{B}=10 KB/s$).}}
	\vspace{-1em}
	\label{table:crl-dis-lust-total-pseudonyms-based-on-lifetime}
	\resizebox{0.47\textwidth}{!}
	{
		\renewcommand{\arraystretch}{1.5}
		\begin{tabular}{| c ||| *{4}{c} |}
			\hline
			\shortstack{\\ {} \textbf{Pseudonym} \\ \textbf{Lifetime}} & \shortstack{\textbf{Number of} \\ \textbf{Psnyms}} & \shortstack{\textbf{Number of} \\ \textbf{Revoked Psnyms}} & \shortstack{\textbf{Average} \\ \textbf{Number per $\Gamma_{CRL}$}} & \shortstack{\textbf{Number of} \\ \textbf{Pieces}} \\\hline\hline 
			\textbf{$\tau_{P}$=30s} & \textbf{3,425,565} & \textbf{34,256} & \textbf{1,428} & \textbf{12} \\\hline 
			\textbf{$\tau_{P}$=60s} & \textbf{1,712,782} & \textbf{17,128} & \textbf{710} & \textbf{6} \\\hline 
			\textbf{$\tau_{P}$=300s} & \textbf{342,556} & \textbf{3,426} & \textbf{143} & \textbf{2} \\\hline 
			\textbf{$\tau_{P}$=600s} & \textbf{171,278} & \textbf{1,713} & \textbf{72} & \textbf{1} \\\hline 
		\end{tabular}
	}
	\vspace{-0.5em}
\end{table}

\begin{table}
	\centering
	\vspace{-0.35em}
	\caption{Simulation Parameters for \acs{LuST} Dataset ($\tau_{P}=60s$).}
	\vspace{-1em}
	\label{table:crl-dis-simulation-parameters-with-diff-revocation-rates}
	\resizebox{0.47\textwidth}{!}
	{
		\renewcommand{\arraystretch}{0.99}
		\hspace{-1em}
		\begin{tabular}{|c|c|c|c|c|c|c|c|c|c|c|c|c|}
			\hline
			\multirow{3}{*}{\LARGE {\textbf{\shortstack{Revocation \\ Rate ($\mathbb{R}$)}}}} & \multicolumn{4}{c|}{{\LARGE \textbf{\shortstack{{} \\ Baseline Scheme}}}} & \multicolumn{4}{c|}{{\LARGE \textbf{\shortstack{{} \\ Vehicle-Centric Scheme}}}} \\ 
			\cline{2-13}
			& \multirow{2}{*}{{\LARGE \textbf{\shortstack{\ac{CRL} \\ Entries}}}} & \multicolumn{1}{c|}{{\LARGE \textbf{\shortstack{{} \\ 10 KB/s}}}} & \multicolumn{1}{c|}{{\LARGE \textbf{\shortstack{{} \\ 25 KB/s}}}} & \multicolumn{1}{c|}{{\LARGE \textbf{\shortstack{{} \\ 50 KB/s}}}} & \multirow{2}{*}{{\LARGE \textbf{\shortstack{\ac{CRL} \\ Entries}}}} & \multicolumn{1}{c|}{{\LARGE \textbf{\shortstack{{} \\ 10 KB/s}}}} & \multicolumn{1}{c|}{{\LARGE \textbf{\shortstack{{} \\ 25 KB/s}}}} & \multicolumn{1}{c|}{{\LARGE \textbf{\shortstack{{} \\ 50 KB/s}}}} \\
			\cline{3-5}\cline{7-9}
			& & {\LARGE \textbf{\shortstack{{} \\ Pieces}}} & {\LARGE \textbf{\shortstack{{} \\ Pieces}}} & {\LARGE \textbf{\shortstack{{} \\ Pieces}}} & & {\LARGE \textbf{\shortstack{{} \\ Pieces}}} & {\LARGE \textbf{\shortstack{{} \\ Pieces}}} & {\LARGE \textbf{\shortstack{{} \\ Pieces}}} \\
			\hline
			{\LARGE \textbf{\shortstack{{} \\ 0.5\%}}} & {\LARGE \shortstack{{} \\ 8,500}} & {\LARGE \shortstack{{} \\ 70}} & {\LARGE \shortstack{{} \\ 30}} & {\LARGE \shortstack{{} \\ 15}} & {\LARGE \shortstack{{} \\ 355}} & {\LARGE \shortstack{{} \\ 3}} & {\LARGE \shortstack{{} \\ 2}} & {\LARGE \shortstack{{} \\ 1}} \\
			\hline
			{\LARGE \textbf{\shortstack{{} \\ 1\%}}} & {\LARGE 17,000} & {\LARGE 140} & {\LARGE 59} & {\LARGE 30} & {\LARGE 710} & {\LARGE 6} & {\LARGE 3} & {\LARGE 2} \\
			\hline
			{\LARGE \textbf{\shortstack{{} \\ 2\%}}} & {\LARGE 34,000} & {\LARGE 279} & {\LARGE 117} & {\LARGE 59} & {\LARGE 1,417} & {\LARGE 12} & {\LARGE 5} & {\LARGE 3} \\
			\hline
			{\LARGE \textbf{\shortstack{{} \\ 3\%}}} & {\LARGE 51,000} & {\LARGE 419} & {\LARGE 175} & {\LARGE 89} & {\LARGE 2,125} & {\LARGE 18} & {\LARGE 8} & {\LARGE 4} \\
			\hline
			{\LARGE \textbf{\shortstack{{} \\ 4\%}}} & {\LARGE 68,000} & {\LARGE 558} & {\LARGE 233} & {\LARGE 118} & {\LARGE 2,834} & {\LARGE 24} & {\LARGE 10} & {\LARGE 5} \\
			\hline
			{\LARGE \textbf{\shortstack{{} \\ 5\%}}} & {\LARGE 85,000} & {\LARGE 697} & {\LARGE 291} & {\LARGE 148} & {\LARGE 3,542} & {\LARGE 30} & {\LARGE 13} & {\LARGE 7} \\
			\hline
		\end{tabular}
	}
	\vspace{0.5em}
\end{table}

\subsection{Quantitative Analysis}
\label{subsec:crl-dis-quantitative-analysis}

\subsubsection{Experimental Setup}
\label{subsubsec:crl-dis-experimental-setup}

We use OMNET++~\cite{omnetpp} and the Veins framework to simulate a large-scale scenario using SUMO~\cite{behrisch2011sumo} with a realistic mobility trace, the \acs{LuST} dataset~\cite{codeca2015lust}. For the cryptographic protocols and primitives (\ac{ECDSA}-256 and SHA-256 as per IEEE 1609.2~\cite{1609-2016} and \acs{ETSI}~\cite{ETSI-102-638}), we use OpenSSL. \ac{V2I} communication is IEEE 802.11p\footnote{Our setup is in-line with the deployment of \ac{VC} systems, with sparse deployment of \acp{RSU} and IEEE 802.1p for safety critical applications~\cite{filippiieee802}. Furthermore, the US \acl{DoT} supports \ac{DSRC} to distribute \ac{CRL} updates ($\Delta$-\acp{CRL}), even though a full \ac{CRL} update cannot be supported as the download time might be longer than the average trip duration~\cite{DOTHS812014}. Although Cellular-V2X could be an alternative communication technology, it is not cost-effective (compared to deploying \ac{DSRC}+\ac{LTE})~~\cite{DOTHS812014, dsrc-cheaper-than-cellular} and it is far behind in the deployment phase~\cite{filippiieee802}. Our experiment is orthogonal to the choice of communication, even though it is envisioned to combine both technologies~\cite{filippiieee802, abboud2016interworking}.}~\cite{IEEE-WAVE-2016} and cryptographic protocols and primitives were executed on a virtual machine (dual-core 2.0 GHz). 

\textbf{Placement of the \acp{RSU}:} To effectively place the \acp{RSU}~\cite{liang2012optimal}, we sorted the intersections with the highest numbers of vehicles passing by. We then placed the \acp{RSU} based on these \emph{``highly-visited''} intersections (preferably with non-overlapping radio ranges of \acp{RSU}).

\textbf{Metrics:} We evaluate the latency to obtain the latest \ac{CRL} pieces, i.e., from the time a vehicle enters the system until it successfully downloads them (protocols~\ref{protocol:crl-dis-algo-construction} to~\ref{crl-dis-algo-subscribing-to-crl-pieces-from-rus-obu}). We choose a small amount of bandwidth ($\mathbb{B}$) for the distribution, e.g., 10-50 KB/s, in order not to interfere with safety-critical operations. Note that request-triggered \ac{CRL} piece broadcasts at 10-50 KB/s (80-400 Kbit/s) are practical because 802.11p supports data-rates up to 24 Mbit/s~\cite{IEEE-WAVE-2016}.

Table~\ref{table:crl-dis-simulation-parameters} shows the simulation parameters; Tables~\ref{table:crl-dis-lust-total-pseudonyms-based-on-lifetime} and~\ref{table:crl-dis-simulation-parameters-with-diff-revocation-rates} show the simulation information for the \acs{LuST} dataset with respect to different pseudonyms lifetimes ($\tau_{P}$), revocation rates ($\mathbb{R}$), and maximum bandwidth for distributing \ac{CRL} pieces ($\mathbb{B}$). We assume that the revocation events are uniformly distributed over a day. For example, if $\tau_{P}=60s$, the total number of pseudonyms for one day is around 1.7M. Assuming 1\% of the pseudonyms are revoked\footnote{To the best of our knowledge, no statistic is available for the expected percentage of revoked pseudonyms in \ac{VC} systems. However, ``Let's Encrypt'', as one of the largest \acsp{CA} in the Internet, reports around 0.2\% of revoked certificates~\cite{lets-encrypt}. Note that in \ac{VC} systems, vehicles are to be provided with multiple, possibly hundreds, of pseudonyms.} ($\mathbb{R}=$1\%), there will be around 17K revoked pseudonyms in a day. With our \emph{vehicle-centric} approach, each vehicle only needs to obtain pieces of information for the interval it travels. When $\Gamma_{CRL}=1$ hour, the average number of entries per $\Gamma_{CRL}$ is around 710. Assuming $\mathbb{B}$ is up to 10 KB/s, total number of pieces will be 6.\footnote{These numbers come from the actual implementation of encoded packets. Each \ac{CRL} piece contains different fields including version, index, total number of pieces in each $\Gamma_{CRL}$, and the entries, serialized with the C++ boost library.}

\subsubsection{Summary of Results}
\label{subsubsec:crl-dis-summary-of-results}

Our vehicle-centric scheme converges more than 40 times faster than the state-of-the-art~\cite{haas2011efficient, laberteaux2008security, haas2009design}, termed here the \emph{baseline} scheme, with a similar experimental set up (Fig.~\ref{fig:crl-dis-comparison-number-of-cognizant-nodes-in-the-system}.b). Moreover, with the baseline scheme, the number of vehicles that successfully downloaded the latest \ac{CRL}, referred to as \emph{cognizant vehicles}, is highly dependent on the revocation rate and it significantly drops when the revocation rate increases from 0.5\% to 5\%. However, the performance of our scheme is not affected by the revocation rate: the number of cognizant nodes remains almost intact even if the revocation rate increases up to 5\% (Fig.~\ref{fig:crl-dis-percentage-of-cognizant-nodes-with-diff-revocation-rate}). Furthermore, our scheme is more resilient to pollution and \ac{DoS}/\ac{DDoS} attacks: with 25\% of vehicles in the baseline system compromised, one could prevent almost all legitimate vehicles from obtaining the \acp{CRL}; however, with our scheme, the percentage of informed vehicles remains almost intact even if 50\% of the vehicles are compromised (Fig.~\ref{fig:crl-dis-dos-attack-resilience-comparison-percentage-of-cognizant-nodes-with-attackers}). Moreover, our scheme outperforms the baseline scheme in terms of computation overhead: signing and verifying 100 \ac{CRL} pieces for the baseline scheme require 51 ms and 39 ms, respectively; however, for our scheme, signature and verification delay for 100 \ac{CRL} pieces is 1 and 12 ms, respectively (Fig.~\ref{fig:crl-dis-computation-communication-latency-comparison}.a). Finally, our experiments confirm that our scheme outperforms the baseline scheme in terms of communication overhead, and notably security overhead (Fig.~\ref{fig:crl-dis-computation-communication-latency-comparison}.b).

\subsubsection{Vehicle-Centric Performance Evaluation}
\label{subsubsec:crl-dis-vehicle-centric-performance-evaluation}

\begin{figure} [!t]
	\vspace{-3.25em}
	\begin{center}
		\centering
		\subfloat[Vehicle-centric scheme ($\mathbb{B}=$10 KB/s)]{
			\hspace{-1.5em}
			\includegraphics[trim=0.15cm 0cm 0cm 1.35cm, clip=true, width=0.273\textwidth,height=0.273\textheight,keepaspectratio]{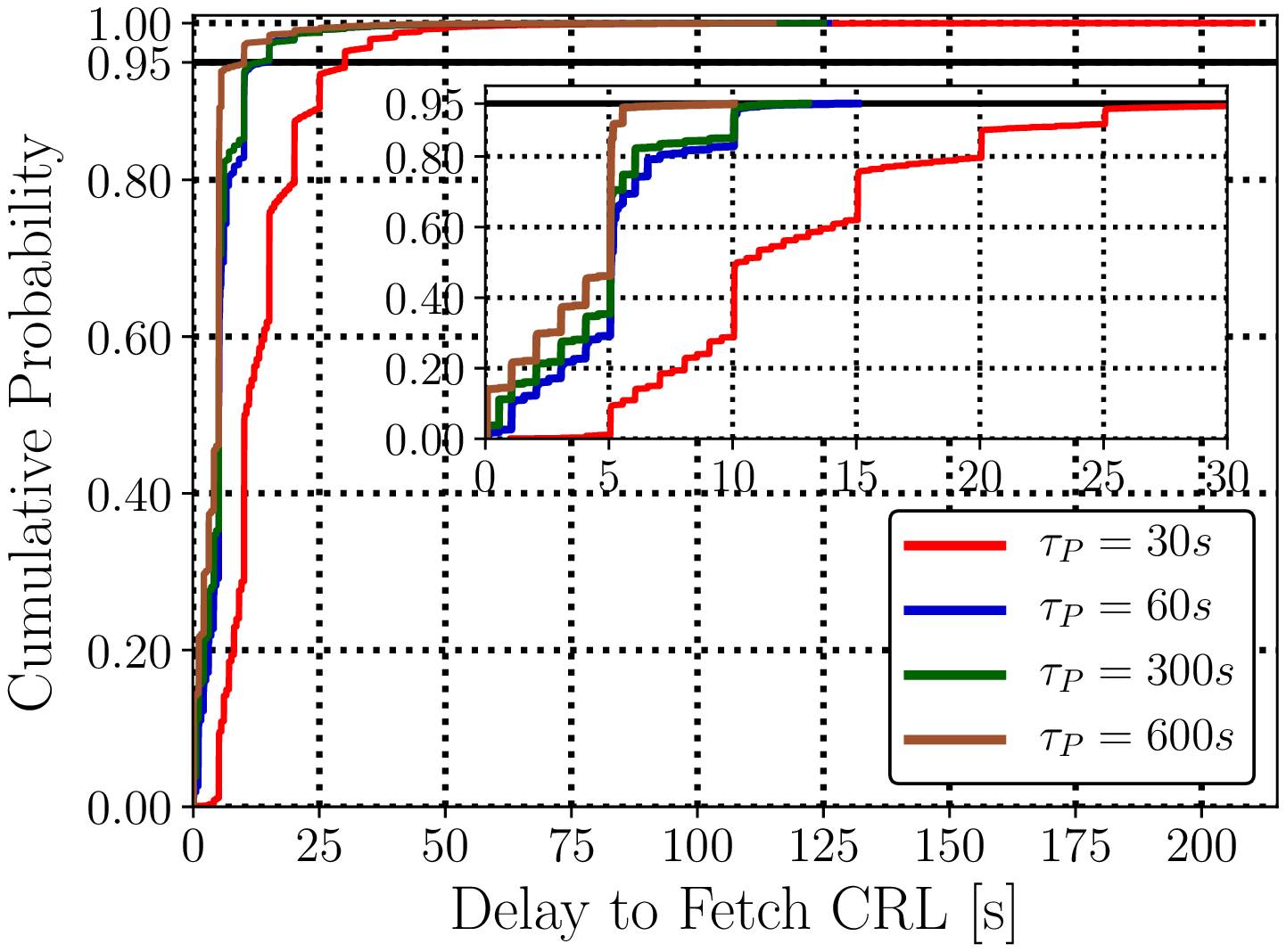}}
		\subfloat[Vehicle-centric scheme ($\mathbb{B}=$10 KB/s)]	{
			\hspace{-1.25em}
			\includegraphics[trim=0.15cm 0cm 0cm 1.35cm, clip=true, width=0.273\textwidth,height=0.273\textheight,keepaspectratio]{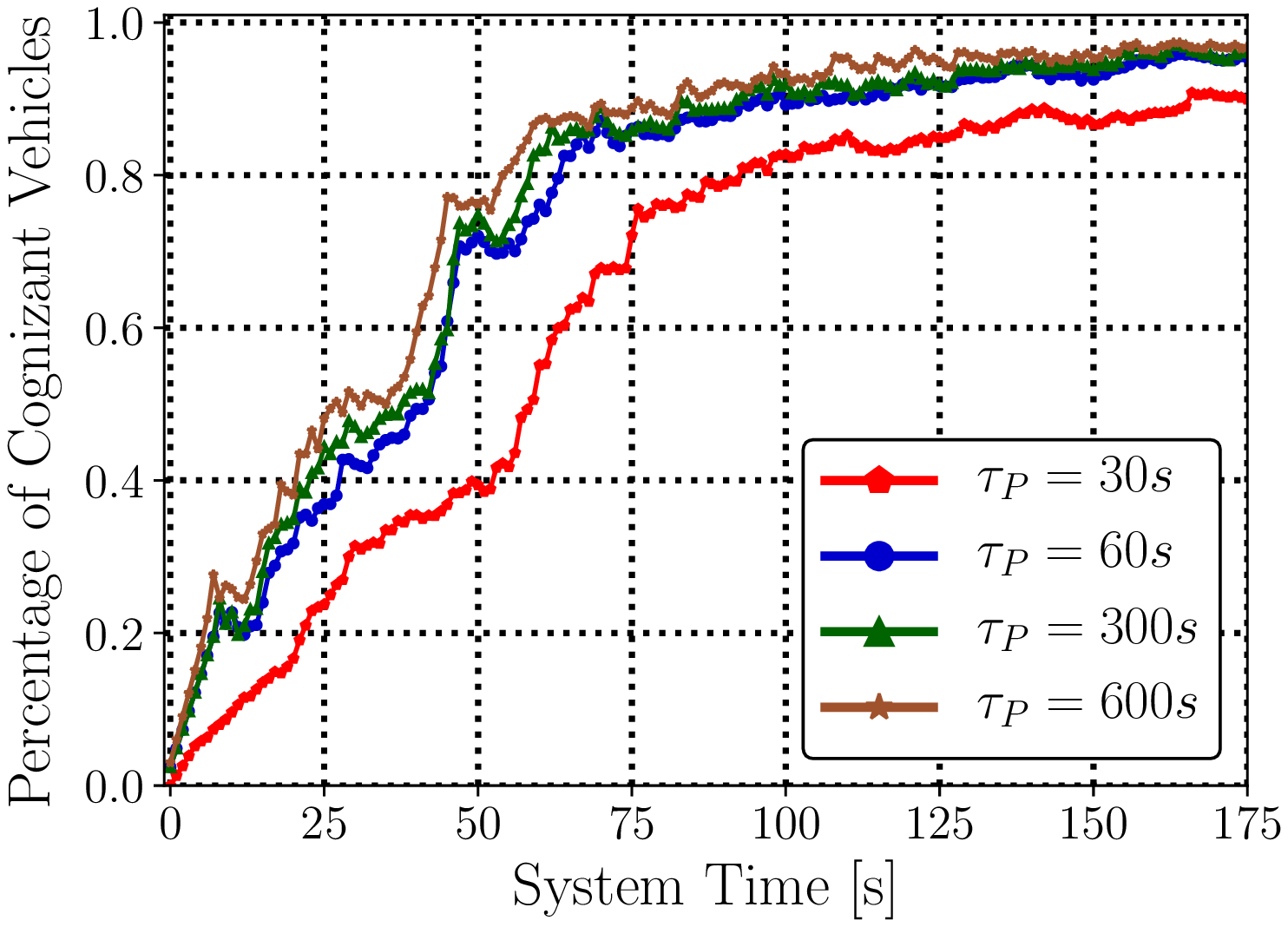}}
		\vspace{-0.75em}
		\caption{(a) End-to-end latency to fetch \ac{CRL} pieces. \\ (b) Percentage of cognizant vehicles over time.}
		\label{fig:crl-dis-e2e-latency-to-fetch-crl-cognizant-nodes-over-time}
	\end{center}
	\vspace{-1.25em}
\end{figure}

Fig.~\ref{fig:crl-dis-e2e-latency-to-fetch-crl-cognizant-nodes-over-time}.a shows the CDF of end-to-end latencies to obtain the needed \ac{CRL}. For example, with $\tau_{P}=60s$, 95\% of the vehicles received the needed pieces in 15s. Fig.~\ref{fig:crl-dis-e2e-latency-to-fetch-crl-cognizant-nodes-over-time}.b shows the percentage of cognizant vehicles over time, i.e., those that successfully obtained the \ac{CRL} pieces. Obviously, the longer the pseudonym lifetime is, the shorter the \ac{CRL} size is, thus the faster the convergence time becomes. For example, the percentage of cognizant nodes at system time 50 sec, with pseudonym lifetime 30s and 600s, is 39\% and 76\%, respectively.

\begin{figure} [!t]
	\vspace{-1.5em}
	\begin{center}
		\centering
		\subfloat[Vehicle-centric scheme ($\mathbb{B}=$25 KB/s)]{
			\hspace{-1.5em}
			\includegraphics[trim=0.25cm 0.1cm 0.5cm 1.25cm, clip=true, totalheight=0.147\textheight,angle=0,keepaspectratio]{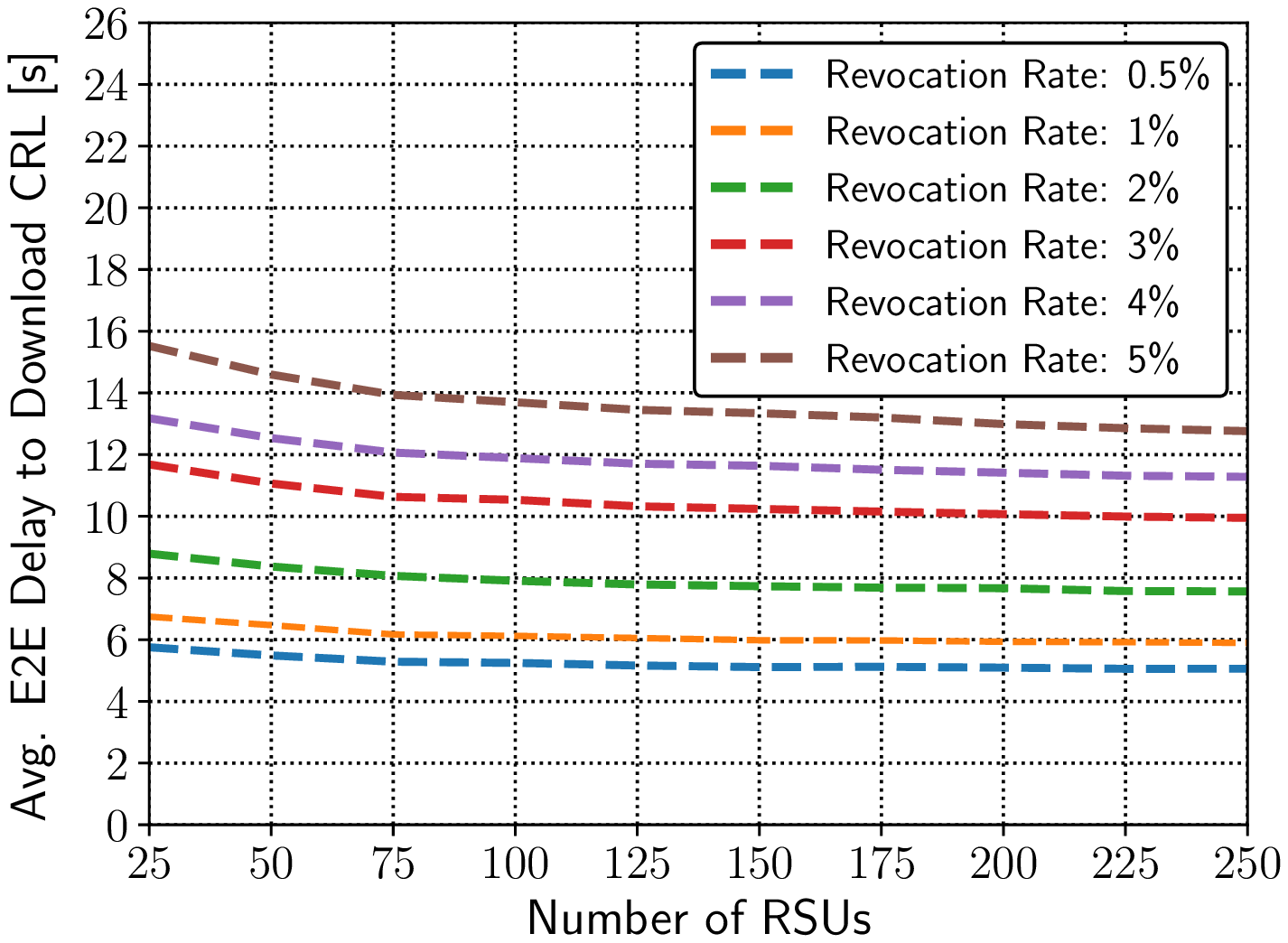}}
		\subfloat[Vehicle-centric scheme ($TX=$5s)]{
			\hspace{-1em}
			\includegraphics[trim=0.15cm 0.1cm 0.5cm 1.25cm, clip=true,totalheight=0.147\textheight,angle=0,keepaspectratio]{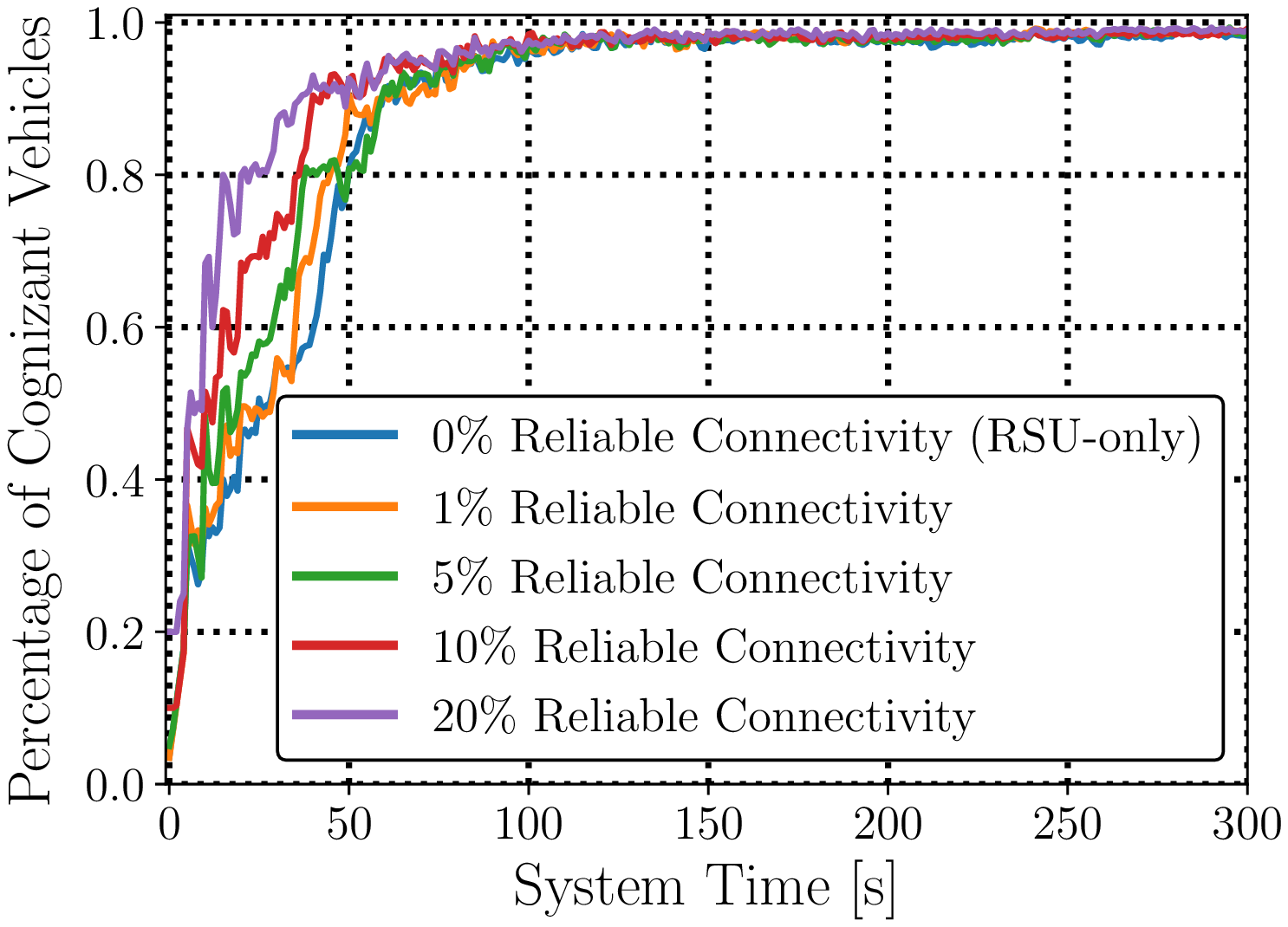}}
		\vspace{-0.75em}
		\caption{(a) Average end-to-end delay to download \acp{CRL}. (b) Dissemination of \ac{CRL} fingerprints.}
		\label{fig:crl-dis-average-end-to-end-latency-to-fetch-CRL-lust-dataset-fingerprint-dissemination}
	\end{center}
	\vspace{-0em}
\end{figure}

Fig.~\ref{fig:crl-dis-average-end-to-end-latency-to-fetch-CRL-lust-dataset-fingerprint-dissemination}.a shows the average end-to-end delay to download the \ac{CRL} as a function of the number of \acp{RSU} for our scheme. The delays were averaged over vehicles operating during the rush hours. The total number of pseudonyms is 1.7M ($\tau_{P}=60s$) and the maximum bandwidth to distribute \ac{CRL} pieces is 25 KB/s. In general, a higher number of \acp{RSU} and a lower revocation rate result in a lower average delay to obtain the \ac{CRL}. For example, the average latency, with $\mathbb{R}=$1\%, decreases from 6.91 to 6.23 as the number of \acp{RSU} increases from 25 to 100. As Fig.~\ref{fig:crl-dis-average-end-to-end-latency-to-fetch-CRL-lust-dataset-fingerprint-dissemination}.a shows, leveraging the car-to-car epidemic \ac{CRL} distribution makes the deployment of a large number of \acp{RSU} unnecessary. The optimal number of \acp{RSU} to be deployed for a given domain can be properly determined to achieve a certain level of quality of service. Further discussion is beyond the scope of our work. 

Fig.~\ref{fig:crl-dis-average-end-to-end-latency-to-fetch-CRL-lust-dataset-fingerprint-dissemination}.b shows how fast a \ac{CRL} fingerprint is distributed: the signed fingerprint of \ac{CRL} pieces is periodically broadcasted only by \acp{RSU}~\cite{nguyen2016secure}, or they are broadcasted by \acp{RSU} (approx. 365 bytes with $TX=5s$) and, in addition, integrated into a subset of pseudonyms with 36 bytes of extra overhead ($p=10^{-30}$, $\mathbb{R}=0.5\%$). Obviously, the distribution of \ac{CRL} fingerprints with our scheme is faster when there is a small fraction of vehicles with reliable connectivity. However, there is a time lag from the time a \ac{PCA} releases \ac{CRL} fingerprints until practically all vehicles are informed about a new \ac{CRL}-update event. Depending on the percentage of vehicles with reliable connectivity and the frequency of revocation events, the \ac{PCA} could \emph{``predict''} a suitable time to reveal the \ac{CRL} fingerprint to ensure that every legitimate vehicle operating within the system would receive the \ac{CRL} fingerprint. For example, the \ac{PCA} could integrate in a fraction of the recently issued pseudonyms the fingerprint of the current $\Gamma_{CRL}$ and integrate in another fraction of newly issued pseudonyms the fingerprint of the subsequent $\Gamma_{CRL}$. 

\subsubsection{Performance Comparison}
\label{subsubsec:crl-dis-performance-comparison}

We compare our scheme with the \emph{baseline} scheme~\cite{haas2011efficient, laberteaux2008security, haas2009design} that uses \acp{RSU} and car-to-car epidemic distribution, with the same assumptions, configuration, and system parameters. For the baseline scheme, the \acs{CA} signs each \ac{CRL} piece and can specify a \emph{``time interval''} so that each vehicle receives $\mathbb{D}$ pseudonyms during the pseudonym acquisition process. As a result, for each batch of revoked pseudonyms, a single $s_{i}$ (256 bit) is disclosed. Similarly, the \ac{PCA} in our scheme can be configured to issue $\mathbb{D}$ pseudonyms per $\Gamma$, i.e., {\footnotesize $\mathbb{D} = \dfrac{\Gamma}{\tau_{P}}$}. To revoke a batch of $\mathbb{D}$ pseudonyms, the serial number of the first revoked pseudonym in the chain and a random number, each 256 bits long, are disclosed. For both schemes, we assume a fully-unlinkable pseudonym provisioning policy~\cite{khodaei2018Secmace}, i.e., $\Gamma$ = $\tau_{P} = 1 min$.\footnote{We aim to stress the system with even an impractical configuration. The performance of the two schemes would improve if the system is configured with more conservative parameters, e.g., $\Gamma=10 \tau_{P}$ (10 pseudonyms per $\Gamma$). But we want to ensure that even under the most demanding condition our vehicle-centric scheme remains practical.}

\begin{figure} [!t]
	\vspace{-3.25em}
	\begin{center}
		\centering
		\subfloat[7:00-7:10 am ($\mathbb{B}=$25 KB/s)]{
			\hspace{-1.45em} \includegraphics[trim=0.15cm 0.25cm 0.5cm 1.3cm, clip=true, width=0.273\textwidth,height=0.273\textheight,keepaspectratio]{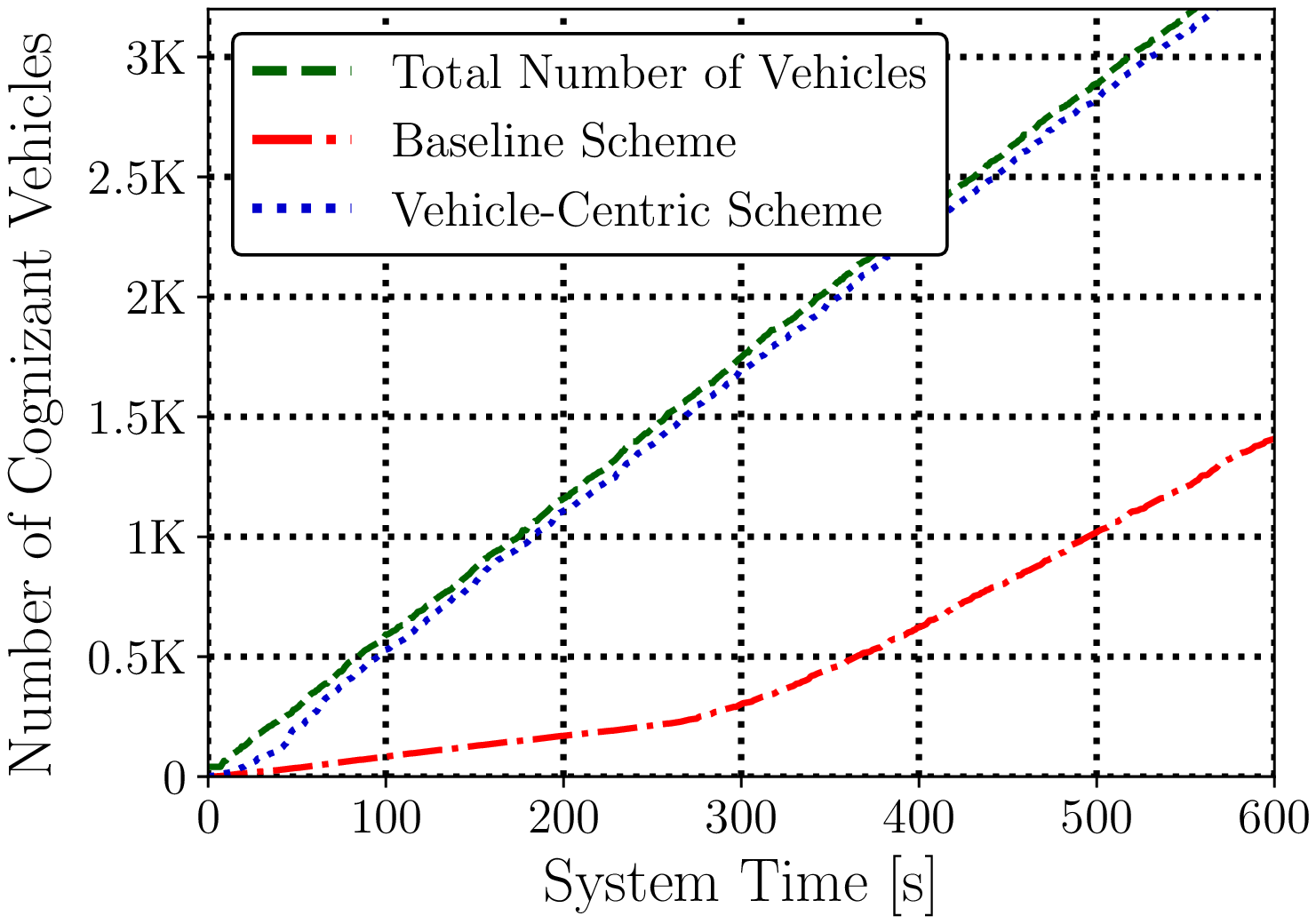}} 
		\subfloat[7-9 am, 5-7 pm ($\mathbb{B}=$25 KB/s)]{
			\hspace{-1.15em} \includegraphics[trim=0.15cm 0.25cm 0.5cm 1.3cm, clip=true, width=0.273\textwidth,height=0.273\textheight,keepaspectratio]{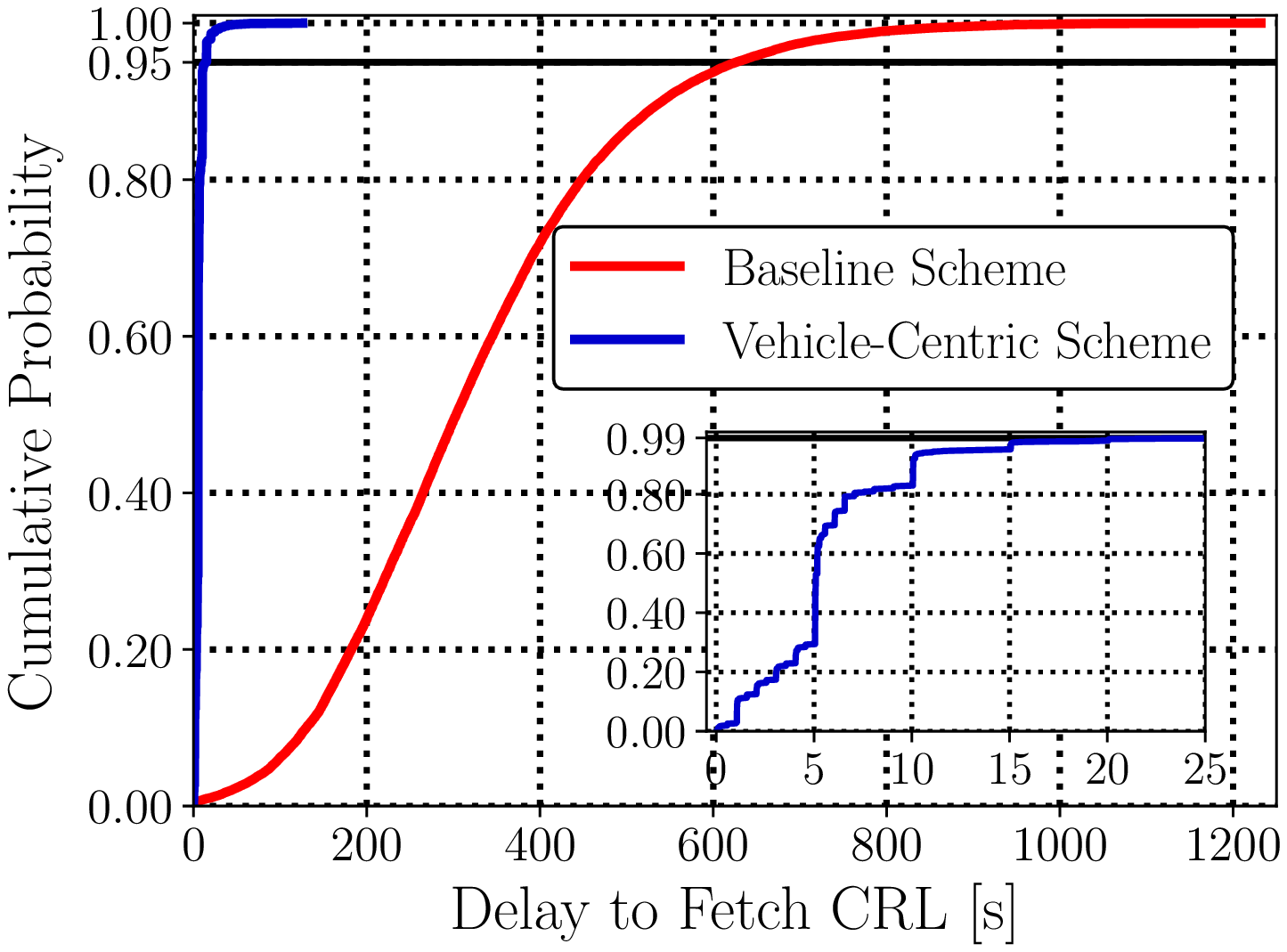}}
		\vspace{-0.75em}
		\caption{End-to-end delay to fetch \acp{CRL} {\small ($\mathbb{R}$ = 1\%, $\tau_{P}$ = 60s)}.}
		\label{fig:crl-dis-comparison-number-of-cognizant-nodes-in-the-system}
	\end{center}
	\vspace{-0em}
\end{figure}

We further assume that vehicles are provided with enough pseudonyms corresponding to their actual trips for a day. Upon a revocation event, information on all revoked pseudonyms for the day is disseminated for the baseline scheme. In contrast, with our scheme, the \ac{CRL} entries are distributed in a time prioritized manner, i.e., revoked pseudonyms whose validity intervals fall within the current $\Gamma_{CRL}$. Moreover, by disseminating signed \ac{BF} in advance, the verification cost is minimal compared to baseline signature verification, i.e., zero delay to verify the \ac{BF} integrated in \emph{fingerprint-carrier} pseudonyms or one signature verification for all \ac{CRL} pieces.

Fig.~\ref{fig:crl-dis-comparison-number-of-cognizant-nodes-in-the-system}.a shows the number of cognizant vehicles over time for the baseline and our scheme. Vehicle-centric distribution of the \ac{CRL} pieces converges faster: the number of cognizant vehicles is very close to the actual number of vehicles in the system. Fig.~\ref{fig:crl-dis-comparison-number-of-cognizant-nodes-in-the-system}.b shows the CDF of delays for the two schemes: for the baseline, $F_x(t=626s)=0.95$, whereas with our scheme, $F_x(t=15s)=0.95$, i.e., converging more than 40 times faster. The principal reasons for such significant improvements are the prioritization of the revocation entries based on their validity intervals, thus a huge reduction in size, as well as the efficient verification of \ac{CRL} pieces.

\begin{figure} [!t]
	\vspace{-3.25em}
	\begin{center}
		\centering
		\subfloat[Baseline scheme ($\mathbb{B}=$50 KB/s)]{
			\hspace{-0.95em} \includegraphics[trim=0.1cm 0.15cm 0.5cm 1.28cm, clip=true, width=0.26\textwidth,height=0.26\textheight,keepaspectratio]{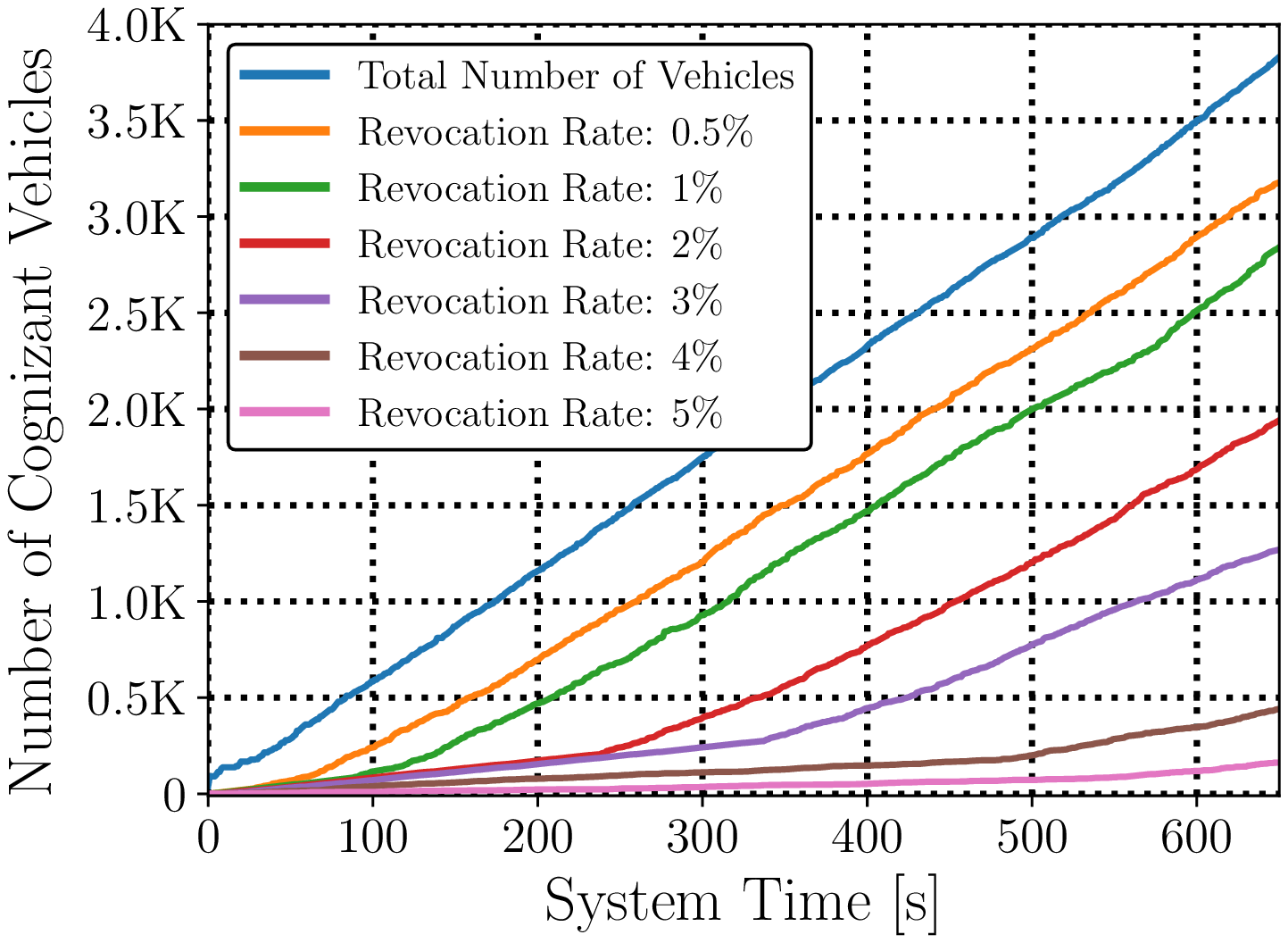}}
		\subfloat[Vehicle-centric scheme ($\mathbb{B}=$50 KB/s)] {
			\hspace{-0.95em} \includegraphics[trim=0.1cm 0.15cm 0.5cm 1.28cm, clip=true, width=0.26\textwidth,height=0.26\textheight,keepaspectratio]{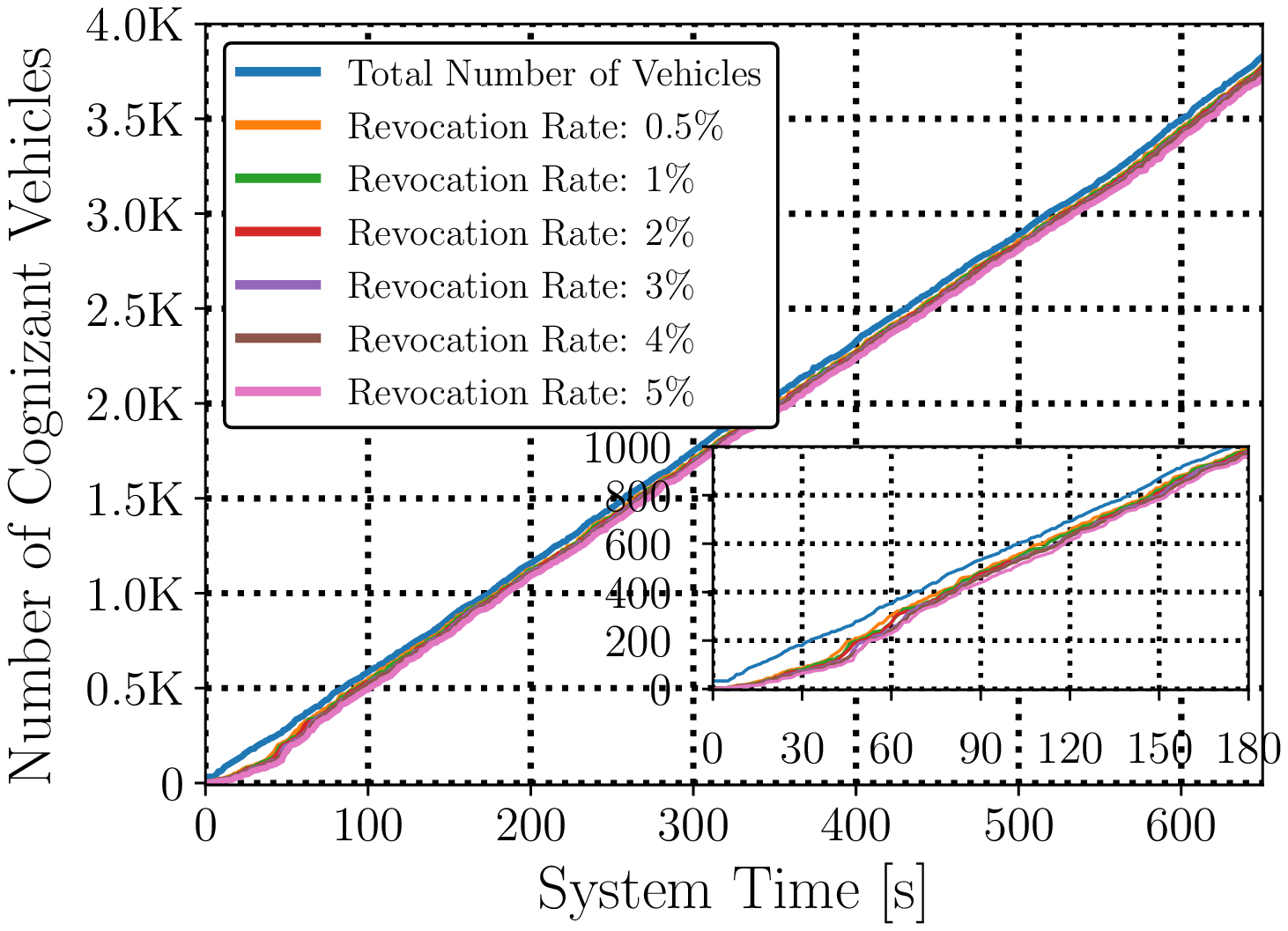}}
		\vspace{-1em}
		\caption{Cognizant vehicles with different revocation rates.}
		\label{fig:crl-dis-percentage-of-cognizant-nodes-with-diff-revocation-rate}
	\end{center}
	\vspace{-1em}
\end{figure}

Fig.~\ref{fig:crl-dis-percentage-of-cognizant-nodes-with-diff-revocation-rate}.a shows the number of informed vehicles with different revocation rates ($\mathbb{R}$) for the baseline scheme. The number of cognizant vehicles is highly affected by $\mathbb{R}$: the number of informed vehicles drops by half when $\mathbb{R}$ increases from 0.5\% to 3\%. Interestingly, the number of cognizant vehicles with $\mathbb{R}=5\%$ is practically negligible, i.e., the majority of vehicles cannot obtain the \ac{CRL} pieces within their trip duration because of the huge \ac{CRL} size. Assume that the total number of pseudonyms is $\mathbb{T}$ and all system configuration parameters are identical. For the baseline scheme, the size of the \ac{CRL}, {\small $\mathbb{T} \times \mathbb{R}$}, linearly increases with $\mathbb{R}$. On the contrary, Fig.~\ref{fig:crl-dis-percentage-of-cognizant-nodes-with-diff-revocation-rate}.b shows that our scheme is \emph{not} affected by $\mathbb{R}$: the number of cognizant vehicles grows as fast as the total number of vehicles in the system. The \ac{PCA} classifies the revocation entries based on $\Gamma_{CRL}$ intervals; thus, the size of an \emph{effective \ac{CRL}} is {\footnotesize $\dfrac{\mathbb{T} \times \mathbb{R}}{|\Gamma_{CRL}|}$}, where $|\Gamma_{CRL}|$ is the number of intervals in a day, e.g., $|\Gamma_{CRL}|$ is 24 when $\Gamma_{CRL}=1$ hour. This results in a huge reduction in \ac{CRL} size, thus ensuring much faster \acp{CRL} distribution. 

\begin{figure} [!t]
	\vspace{-1.75em}
	\begin{center}
		\centering
		\subfloat[Baseline scheme ($\mathbb{B}=$25 KB/s)]{
			\hspace{-0.95em} 
			\includegraphics[trim=0.35cm 0.25cm 1.1cm 1.36cm, clip=true, width=0.25\textwidth,height=0.25\textheight,keepaspectratio]{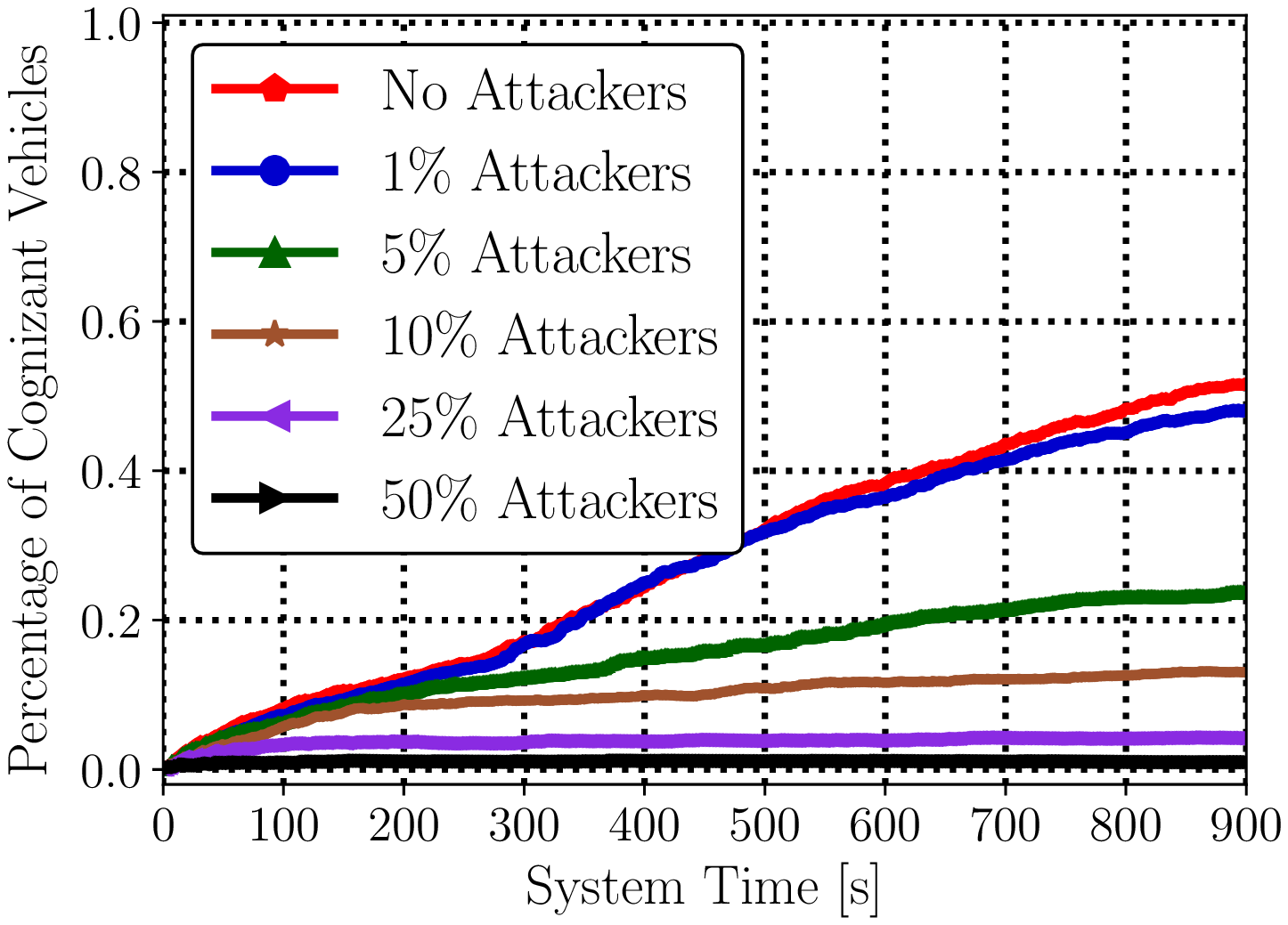}}
		\subfloat[Vehicle-centric scheme ($\mathbb{B}=$25 KB/s)]{
			\hspace{-0.5em} 
			\includegraphics[trim=0.25cm 0.25cm 1.1cm 1.36cm, clip=true, width=0.25\textwidth,height=0.25\textheight,keepaspectratio]{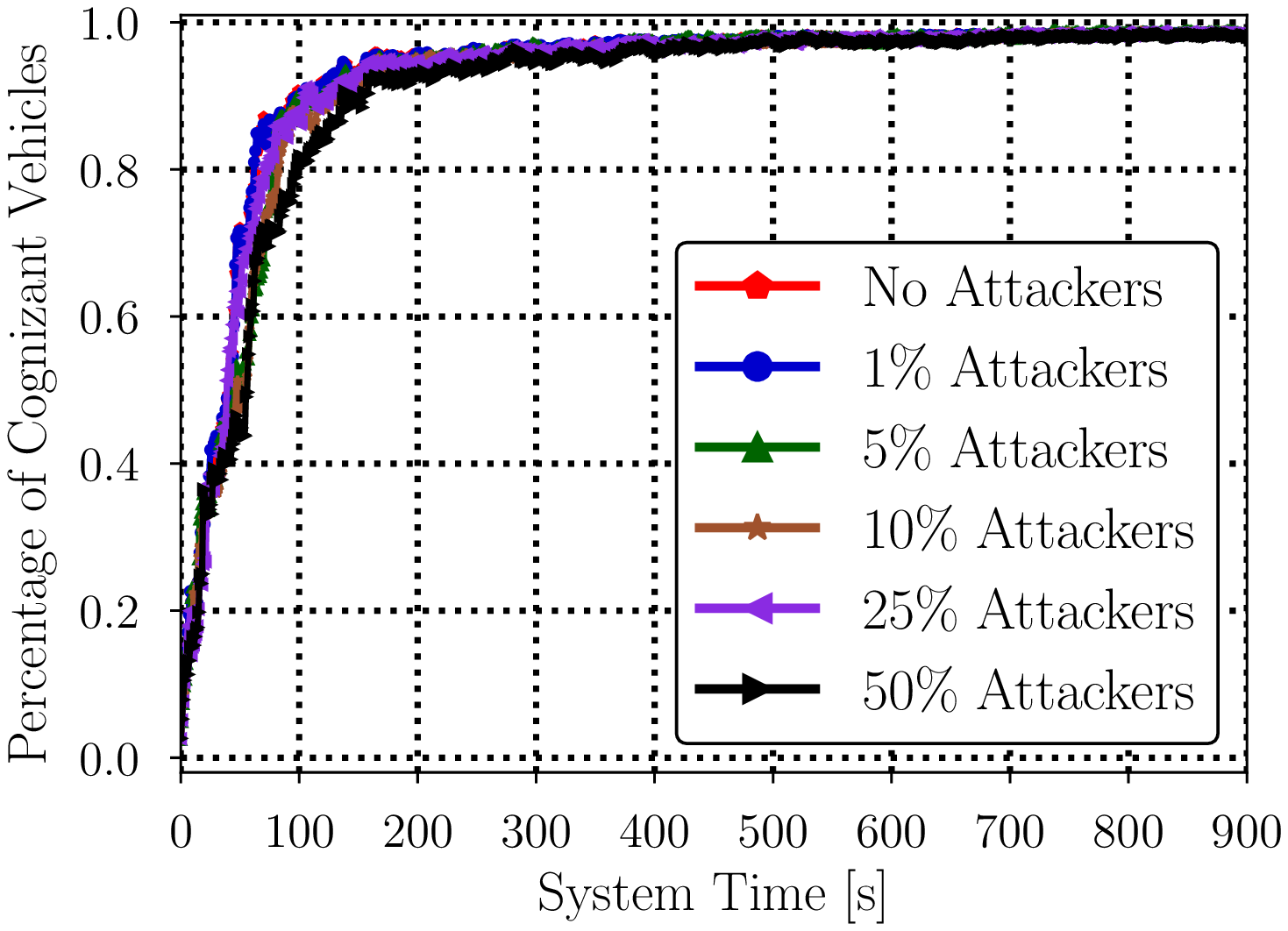}}
		\vspace{-1em}
		\caption{Resilience comparison against pollution and \ac{DDoS} attacks with different number of attackers in the system.}
		\label{fig:crl-dis-dos-attack-resilience-comparison-percentage-of-cognizant-nodes-with-attackers}
	\end{center}
	\vspace{-0em}
\end{figure}

Fig.~\ref{fig:crl-dis-dos-attack-resilience-comparison-percentage-of-cognizant-nodes-with-attackers} shows the percentage of cognizant vehicles when attackers perform pollution and \ac{DDoS} attacks by periodically broadcasting fake \ac{CRL} pieces once every 0.5s. Fig.~\ref{fig:crl-dis-dos-attack-resilience-comparison-percentage-of-cognizant-nodes-with-attackers}.a shows that the baseline scheme is adversely affected once the number of attackers in the system is more than 10\% of the vehicles. In contrast, Fig.~\ref{fig:crl-dis-dos-attack-resilience-comparison-percentage-of-cognizant-nodes-with-attackers}.b illustrates the percentage of cognizant vehicles for our scheme: even if 50\% of the \acp{OBU} are compromised and misbehave in this way, the percentage of cognizant vehicles is not considerably affected and it still converges within a reasonable delay. Again, such resiliency stems from intelligent partitioning of the \ac{CRL}, yielding a huge reduction in the \ac{CRL} size. By integrating the \ac{BF} of a \ac{CRL} in the pseudonyms, we achieve an efficient verification of \ac{CRL} pieces.

\begin{figure} [!t]
	\centering
	\vspace{-2em}
	\begin{center}
		\centering
		\hspace{-1.85em} 
		\subfloat[Computation comparison]{
			\includegraphics[trim=0.3cm 0.25cm 1.3cm 1.28cm, clip=true, width=0.25\textwidth,height=0.25\textheight,keepaspectratio]{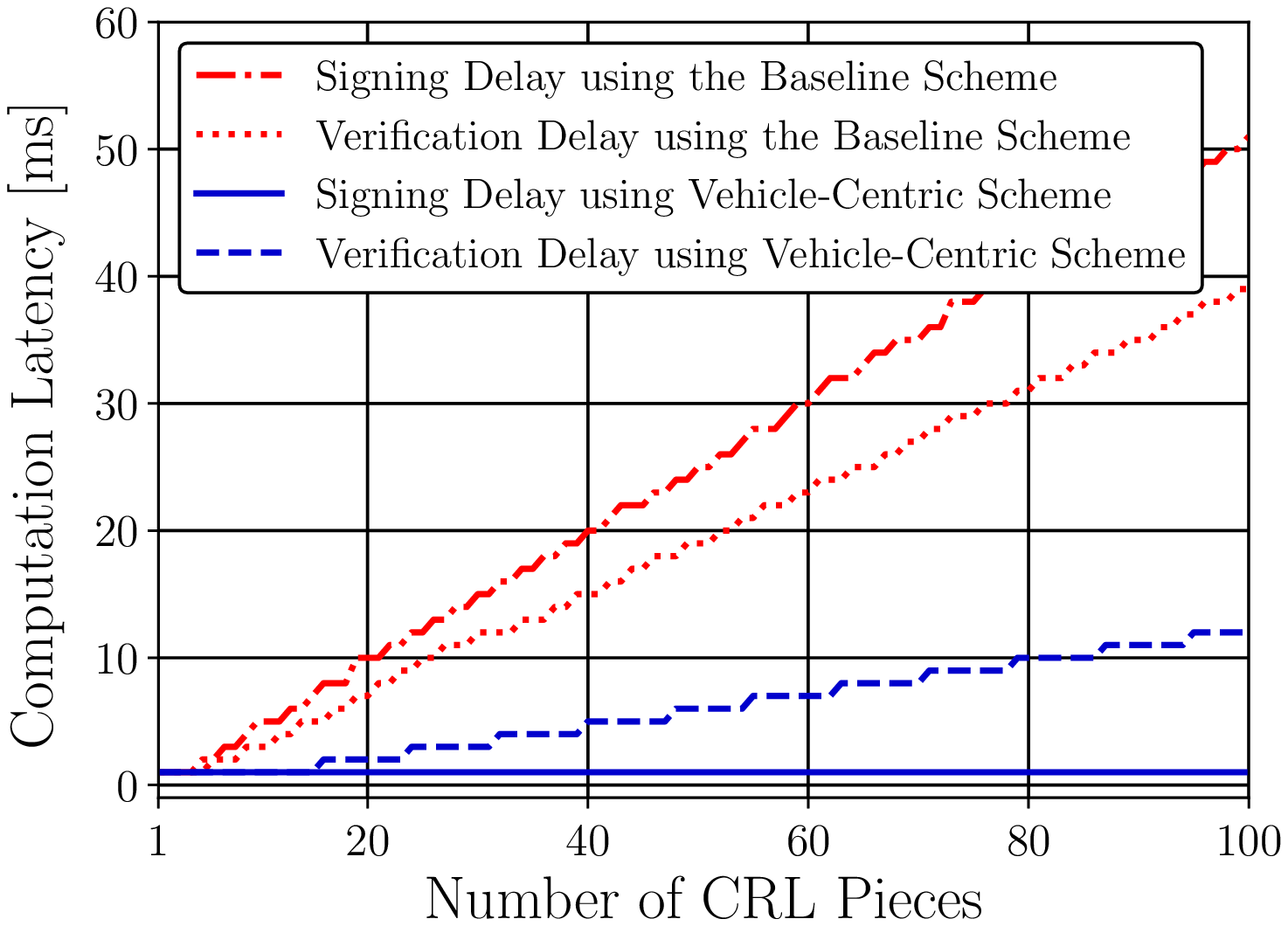}}
		\hspace{-0.45em} 
		\subfloat[Communication comparison]{
			\includegraphics[trim=0.3cm 0.01cm 1.1cm 1.35cm, clip=true, width=0.25\textwidth,height=0.25\textheight,keepaspectratio]{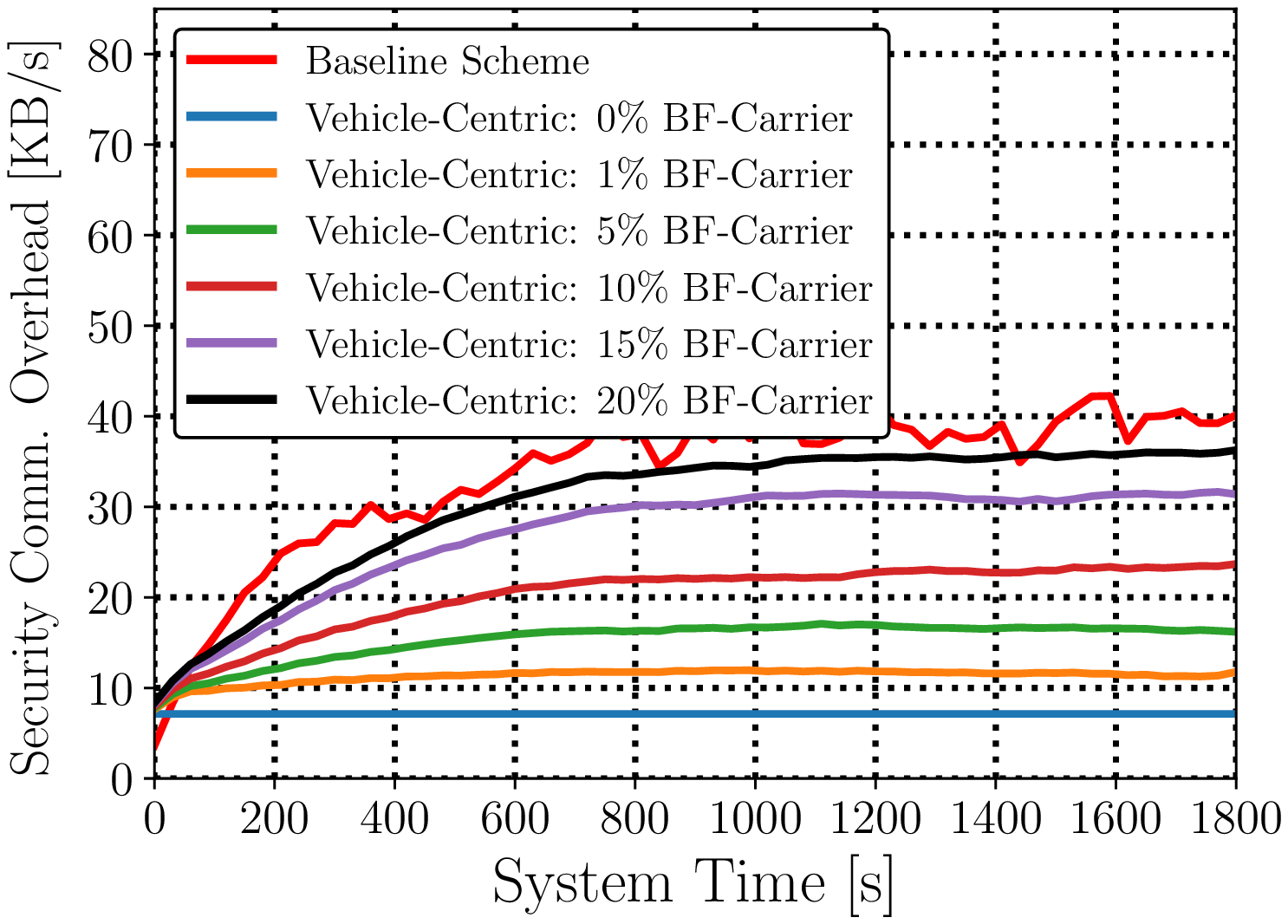}}
		\vspace{-0.75em}
		\caption{(a) Computation latency comparison. (b) Security overhead comparison, averaged every 30s ({\small $\mathbb{R}$=1\%, $\mathbb{B}$ = 50KB/s}).}
		\label{fig:crl-dis-computation-communication-latency-comparison}
	\end{center}
	\vspace{-0em}
\end{figure}

Fig.~\ref{fig:crl-dis-computation-communication-latency-comparison}.a compares the computation delays for generating and validating \ac{CRL} pieces for the baseline and our schemes. Signing and verification delays for the baseline scheme linearly increase with the number of \ac{CRL} pieces. For example, signing and verifying 100 pieces of \ac{CRL} require 51 ms and 39 ms, respectively. Depending on the frequency of revocation events and the size of a \ac{CRL}, this could incur extra overhead for the \ac{PCA} and the vehicles. But the verification delay for our scheme moderately increases with the number of \ac{CRL} pieces thanks to the lightweight \ac{BF} membership validation. The delay to sign the \ac{CRL} pieces is constant (1 ms), in fact one signature for the \ac{BF} of pieces to be broadcasted via \acp{RSU} and zero additional delay for integrating the fingerprints of \ac{CRL} pieces to a subset of pseudonyms during the pseudonym acquisition phase; overall, a significant computational improvement is achieved.

Fig.~\ref{fig:crl-dis-computation-communication-latency-comparison}.b shows the security overhead due to \emph{signatures} and \emph{fingerprints} for \ac{CRL} pieces, for the baseline and the vehicle-centric scheme respectively. The \ac{ECDSA} signature size for the baseline scheme is 72 bytes per piece; the fingerprint in our scheme, 365 bytes long, is signed and it is broadcasted once every 5s via \acp{RSU},  and also integrated in a subset of pseudonyms, 36 bytes ($p=10^{-30}$). Obviously, attaching a pseudonym to every \ac{CAM} is not practical as the packet overhead increases. To reduce overhead, a pseudonym can be attached to \acp{CAM} every $\alpha$ (certificate period) and if there is a pseudonym update process, the new pseudonym is attached every system parameter $\beta$ (push period)~\cite{calandriello2011performance}. We configure $\alpha=10$ and $\beta=1$ with beaconing frequency $\gamma=0.1$ (10 \ac{CAM}/sec) and $\tau_{P}=60s$. Fig.~\ref{fig:crl-dis-computation-communication-latency-comparison}.b shows that the average security overhead (only the signature field) for the baseline scheme is higher than the one for our scheme, even with 20\% of nodes assumed as fingerprint-carriers. Obviously, the longer the pseudonym lifetime, combined with slow neighborhood change, the lower the need to attach pseudonyms, and thus the lower the communication fingerprints overhead. All in all, our scheme outperforms the baseline scheme in terms of computation and communication overhead.

\section{Conclusion and Future Work}
\label{sec:crl-dis-conclusions}

We proposed a practical framework to effectively distribute \acp{CRL} in \ac{VC} systems. Through extensive experimental evaluation, we demonstrated that our scheme is highly efficient and scalable, resilient against \ac{DoS} attacks, and it is a viable solution towards catalyzing the deployment of the secure and privacy-protecting \ac{VC} systems. As future work, we plan to investigate an optimal interval for $\Gamma_{CRL}$ based on different factors, e.g., the frequency of revocation events, to guarantee a narrower vulnerability window.

\bibliographystyle{ACM-Reference-Format}
\bibliography{references} 


\begin{thebibliography}{79}


\ifx \showCODEN    \undefined \def \showCODEN     #1{\unskip}     \fi
\ifx \showDOI      \undefined \def \showDOI       #1{#1}\fi
\ifx \showISBNx    \undefined \def \showISBNx     #1{\unskip}     \fi
\ifx \showISBNxiii \undefined \def \showISBNxiii  #1{\unskip}     \fi
\ifx \showISSN     \undefined \def \showISSN      #1{\unskip}     \fi
\ifx \showLCCN     \undefined \def \showLCCN      #1{\unskip}     \fi
\ifx \shownote     \undefined \def \shownote      #1{#1}          \fi
\ifx \showarticletitle \undefined \def \showarticletitle #1{#1}   \fi
\ifx \showURL      \undefined \def \showURL       {\relax}        \fi
\providecommand\bibfield[2]{#2}
\providecommand\bibinfo[2]{#2}
\providecommand\natexlab[1]{#1}
\providecommand\showeprint[2][]{arXiv:#2}

\bibitem[\protect\citeauthoryear{??}{DOT}{2014}]%
        {DOTHS812014}
 \bibinfo{year}{2014}\natexlab{}.
\newblock \bibinfo{title}{{V2V} {C}ommunications: {R}eadiness of {V2V}
  {T}echnology for {A}pplication}.
\newblock
\newblock
\newblock
\shownote{{N}ational {H}ighway {T}raffic {S}afety {A}dministration, {DOT HS}
  812 014.}


\bibitem[\protect\citeauthoryear{??}{acs}{2015}]%
        {acs-survey}
 \bibinfo{year}{2015}\natexlab{}.
\newblock \bibinfo{title}{{A}merican {C}ommunity {S}urvey ({ACS})}.
\newblock \bibinfo{howpublished}{\url{http://tiny.cc/hc8qqy}}.
\newblock


\bibitem[\protect\citeauthoryear{??}{IEE}{2016}]%
        {IEEE-WAVE-2016}
 \bibinfo{year}{2016}\natexlab{}.
\newblock \showarticletitle{{IEEE} {S}tandard for {W}ireless {A}ccess in
  {V}ehicular {E}nvironments ({WAVE}) \textendash {N}etworking {S}ervices}.
\newblock \bibinfo{journal}{\emph{{IEEE} Vehicular Technology Society}}
  (\bibinfo{date}{Jan.} \bibinfo{year}{2016}).
\newblock


\bibitem[\protect\citeauthoryear{??}{new}{2016}]%
        {newsroom-how-much-motorists-drive}
 \bibinfo{year}{2016}\natexlab{}.
\newblock \bibinfo{title}{{W}hen, {W}here and {H}ow {M}uch {M}otorists
  {D}rive}.
\newblock \bibinfo{howpublished}{\url{tiny.cc/yinqqy}}.
\newblock


\bibitem[\protect\citeauthoryear{??}{ant}{2017a}]%
        {antpoolchina}
 \bibinfo{year}{2017}\natexlab{a}.
\newblock \bibinfo{title}{{A}ntpool, {T}he {A}dvanced {B}itcoin {M}ining
  {P}ool}.
\newblock \bibinfo{howpublished}{\url{antpool.com/}}.
\newblock


\bibitem[\protect\citeauthoryear{??}{ant}{2017b}]%
        {antminerS9Review}
 \bibinfo{year}{2017}\natexlab{b}.
\newblock \bibinfo{title}{{B}itmain {A}ntminer {S9} {R}eview}.
\newblock \bibinfo{howpublished}{\url{http://tiny.cc/12p4qy}}.
\newblock


\bibitem[\protect\citeauthoryear{??}{let}{2017}]%
        {lets-encrypt}
 \bibinfo{year}{2017}\natexlab{}.
\newblock \bibinfo{title}{{L}et's {E}ncrypt {S}tats}.
\newblock \bibinfo{howpublished}{\url{https://letsencrypt.org/stats/}}.
\newblock


\bibitem[\protect\citeauthoryear{??}{omn}{2017}]%
        {omnetpp}
 \bibinfo{year}{2017}\natexlab{}.
\newblock \bibinfo{title}{OMNeT++}.
\newblock \bibinfo{howpublished}{\url{https://www.omnetpp.org/}}.
\newblock


\bibitem[\protect\citeauthoryear{??}{dsr}{2018}]%
        {dsrc-cheaper-than-cellular}
 \bibinfo{year}{2018}\natexlab{}.
\newblock \bibinfo{title}{{DSRC} {W}orks out {C}heaper than {C}ellular
  {C}ommunications for {V2X}}.
\newblock \bibinfo{howpublished}{\url{http://tiny.cc/qqk4qy}}.
\newblock


\bibitem[\protect\citeauthoryear{1609.2}{1609.2}{2016}]%
        {1609-2016}
\bibfield{author}{\bibinfo{person}{{IEEE} 1609.2}.}
  \bibinfo{year}{2016}\natexlab{}.
\newblock \showarticletitle{{IEEE} {S}tandard for {W}ireless {A}ccess in
  {V}ehicular {E}nvironments - {S}ecurity {S}ervices for {A}pplications and
  {M}anagement {M}essages}.
\newblock  (\bibinfo{date}{Mar.} \bibinfo{year}{2016}).
\newblock


\bibitem[\protect\citeauthoryear{Abboud and et~al}{Abboud and et~al}{2016}]%
        {abboud2016interworking}
\bibfield{author}{\bibinfo{person}{K. Abboud} {and} \bibinfo{person}{et al}.}
  \bibinfo{year}{2016}\natexlab{}.
\newblock \showarticletitle{{I}nterworking of {DSRC} and {C}ellular {N}etwork
  {T}echnologies for {V2X} {C}ommunications:{A} {S}urvey}.
\newblock \bibinfo{journal}{\emph{IEEE TVT}} \bibinfo{volume}{65},
  \bibinfo{number}{12} (\bibinfo{date}{July} \bibinfo{year}{2016}),
  \bibinfo{pages}{9457--9470}.
\newblock


\bibitem[\protect\citeauthoryear{Amoozadeh}{Amoozadeh}{2012}]%
        {amoozadeh2012certificate}
\bibfield{author}{\bibinfo{person}{M. Amoozadeh}.}
  \bibinfo{year}{2012}\natexlab{}.
\newblock \emph{\bibinfo{title}{{C}ertificate {R}evocation {L}ist
  {D}istribution in {V}ehicular {C}ommunication {S}ystems}}.
\newblock \bibinfo{thesistype}{Master's\ thesis}. \bibinfo{school}{KTH},
  \bibinfo{address}{Stockholm, Sweden}.
\newblock


\bibitem[\protect\citeauthoryear{Ardelean and et~al}{Ardelean and
  et~al}{2009}]%
        {ardelean2009CRLImplementation}
\bibfield{author}{\bibinfo{person}{P. Ardelean} {and} \bibinfo{person}{et al}.}
  \bibinfo{year}{2009}\natexlab{}.
\newblock \showarticletitle{{I}mplementation and {E}valuation of {C}ertificate
  {R}evocation {L}ist {D}istribution for {VANETs}}.
\newblock \bibinfo{journal}{\emph{Technical Report}} (\bibinfo{date}{Jan.}
  \bibinfo{year}{2009}).
\newblock


\bibitem[\protect\citeauthoryear{Behrisch and et~al}{Behrisch and
  et~al}{2011}]%
        {behrisch2011sumo}
\bibfield{author}{\bibinfo{person}{M. Behrisch} {and} \bibinfo{person}{et al}.}
  \bibinfo{year}{2011}\natexlab{}.
\newblock \showarticletitle{{SUMO}~\textendash~{S}imulation of {U}rban
  {M}Obility}. In \bibinfo{booktitle}{\emph{The 3rd International Conference on
  Advances in System Simulation}}. \bibinfo{address}{Barcelona, Spain}.
\newblock


\bibitem[\protect\citeauthoryear{Bi{\ss}meyer}{Bi{\ss}meyer}{2014}]%
        {bissmeyer2014misbehavior}
\bibfield{author}{\bibinfo{person}{N. Bi{\ss}meyer}.}
  \bibinfo{year}{2014}\natexlab{}.
\newblock \emph{\bibinfo{title}{{M}isbehavior {D}etection and {A}ttacker
  {I}dentification in {V}ehicular {A}d-{H}oc {N}etworks}}.
\newblock \bibinfo{thesistype}{Ph.D. Dissertation}. \bibinfo{school}{Technische
  Universit{\"a}t}.
\newblock


\bibitem[\protect\citeauthoryear{Bi{\ss}meyer and et~al}{Bi{\ss}meyer and
  et~al}{2013}]%
        {bibmeyer2013copra}
\bibfield{author}{\bibinfo{person}{Norbert Bi{\ss}meyer} {and}
  \bibinfo{person}{et al}.} \bibinfo{year}{2013}\natexlab{}.
\newblock \showarticletitle{{C}o{PRA}: {C}onditional {P}seudonym {R}esolution
  {A}lgorithm in {VANET}s}. In \bibinfo{booktitle}{\emph{IEEE WONS}}.
  \bibinfo{address}{Banff, Canada}, \bibinfo{pages}{9--16}.
\newblock


\bibitem[\protect\citeauthoryear{Bloom}{Bloom}{1970}]%
        {bloom1970space}
\bibfield{author}{\bibinfo{person}{Burton~H Bloom}.}
  \bibinfo{year}{1970}\natexlab{}.
\newblock \showarticletitle{{S}pace/{T}ime {T}rade-offs in {H}ash {C}oding with
  {A}llowable {E}rrors}.
\newblock \bibinfo{journal}{\emph{Commun. ACM}} \bibinfo{volume}{13},
  \bibinfo{number}{7} (\bibinfo{date}{July} \bibinfo{year}{1970}),
  \bibinfo{pages}{422--426}.
\newblock


\bibitem[\protect\citeauthoryear{{C}alandriello and et~al}{{C}alandriello and
  et~al}{2011}]%
        {calandriello2011performance}
\bibfield{author}{\bibinfo{person}{G. {C}alandriello} {and} \bibinfo{person}{et
  al}.} \bibinfo{year}{2011}\natexlab{}.
\newblock \showarticletitle{{O}n the {P}erformance of {S}ecure {V}ehicular
  {C}ommunication {S}ystems}.
\newblock \bibinfo{journal}{\emph{IEEE TDSC}} \bibinfo{volume}{8},
  \bibinfo{number}{6} (\bibinfo{date}{Nov.} \bibinfo{year}{2011}),
  \bibinfo{pages}{898--912}.
\newblock


\bibitem[\protect\citeauthoryear{Chariton and e~al}{Chariton and e~al}{2017}]%
        {charitonccsp}
\bibfield{author}{\bibinfo{person}{A-A. Chariton} {and} \bibinfo{person}{e
  al}.} \bibinfo{year}{2017}\natexlab{}.
\newblock \showarticletitle{{CCSP}: a {C}ompressed {C}ertificate {S}tatus
  {P}rotocol}. In \bibinfo{booktitle}{\emph{IEEE INFOCOM}}.
  \bibinfo{address}{Atlanta, GA, USA}.
\newblock


\bibitem[\protect\citeauthoryear{Clark and et~al}{Clark and et~al}{2013}]%
        {clark2013sok}
\bibfield{author}{\bibinfo{person}{J. Clark} {and} \bibinfo{person}{et al}.}
  \bibinfo{year}{2013}\natexlab{}.
\newblock \showarticletitle{{SoK}: {SSL} and {HTTPS}: {R}evisiting {P}ast
  {C}hallenges and {E}valuating {C}ertificate {T}rust {M}odel {E}nhancements}.
  In \bibinfo{booktitle}{\emph{IEEE SnP}}. \bibinfo{address}{Berkeley, USA}.
\newblock


\bibitem[\protect\citeauthoryear{Codeca and et~al}{Codeca and et~al}{2015}]%
        {codeca2015lust}
\bibfield{author}{\bibinfo{person}{L. Codeca} {and} \bibinfo{person}{et al}.}
  \bibinfo{year}{2015}\natexlab{}.
\newblock \showarticletitle{{L}uxembourg {S}UMO {T}raffic ({LuST}) {S}cenario:
  24 {H}ours of {M}obility for {V}ehicular {N}etworking {R}esearch}. In
  \bibinfo{booktitle}{\emph{IEEE VNC}}. \bibinfo{address}{Kyoto, Japan}.
\newblock


\bibitem[\protect\citeauthoryear{Cooper}{Cooper}{2000}]%
        {cooper2000more}
\bibfield{author}{\bibinfo{person}{D. Cooper}.}
  \bibinfo{year}{2000}\natexlab{}.
\newblock \showarticletitle{{A} {M}ore {E}fficient {U}se of {D}elta-{CRLs}}. In
  \bibinfo{booktitle}{\emph{IEEE S\&P}}. \bibinfo{address}{CA, USA}.
\newblock


\bibitem[\protect\citeauthoryear{Das and et~al}{Das and et~al}{2004}]%
        {das2004spawn}
\bibfield{author}{\bibinfo{person}{S. Das} {and} \bibinfo{person}{et al}.}
  \bibinfo{year}{2004}\natexlab{}.
\newblock \showarticletitle{{SPAWN}: {A} {S}warming {P}rotocol for {V}ehicular
  {A}d-{H}oc {W}ireless {N}etworks}. In \bibinfo{booktitle}{\emph{ACM workshop
  on VANET}}. \bibinfo{address}{Philadelphia, PA, USA}.
\newblock


\bibitem[\protect\citeauthoryear{Dierks}{Dierks}{2008}]%
        {dierks2008transport}
\bibfield{author}{\bibinfo{person}{T. Dierks}.}
  \bibinfo{year}{2008}\natexlab{}.
\newblock \showarticletitle{The transport layer security protocol version 1.2}.
\newblock  (\bibinfo{date}{Aug.} \bibinfo{year}{2008}).
\newblock


\bibitem[\protect\citeauthoryear{Era and Preneel}{Era and Preneel}{2015}]%
        {era2015cryptography}
\bibfield{author}{\bibinfo{person}{Snowden Era} {and} \bibinfo{person}{Bart
  Preneel}.} \bibinfo{year}{2015}\natexlab{}.
\newblock \showarticletitle{{C}ryptography and {I}nformation {S}ecurity in the
  {P}ost-{S}nowden era}.
\newblock  (\bibinfo{date}{May} \bibinfo{year}{2015}).
\newblock


\bibitem[\protect\citeauthoryear{{ETSI}}{{ETSI}}{2009}]%
        {ETSI-102-638}
\bibfield{author}{\bibinfo{person}{{ETSI}}.} \bibinfo{year}{2009}\natexlab{}.
\newblock \bibinfo{title}{{I}ntelligent {T}ransport {S}ystems ({ITS});
  {V}ehicular {C}ommunications; {B}asic {S}et of {A}pplications;
  {D}efinitions}.
\newblock
\newblock


\bibitem[\protect\citeauthoryear{Eugster and et~al}{Eugster and et~al}{2003}]%
        {eugster2003many}
\bibfield{author}{\bibinfo{person}{P-T. Eugster} {and} \bibinfo{person}{et
  al}.} \bibinfo{year}{2003}\natexlab{}.
\newblock \showarticletitle{{T}he {M}any {F}aces of {P}ublish/{S}ubscribe}.
\newblock \bibinfo{journal}{\emph{ACM computing surveys (CSUR)}}
  \bibinfo{volume}{35}, \bibinfo{number}{2} (\bibinfo{date}{June}
  \bibinfo{year}{2003}), \bibinfo{pages}{114--131}.
\newblock


\bibitem[\protect\citeauthoryear{Filippi and et~al}{Filippi and et~al}{2017}]%
        {filippiieee802}
\bibfield{author}{\bibinfo{person}{A. Filippi} {and} \bibinfo{person}{et al}.}
  \bibinfo{year}{2017}\natexlab{}.
\newblock \showarticletitle{{IEEE802.11p} {A}head of {LTE-V2V} for {S}afety
  {A}pplications}.
\newblock  (\bibinfo{date}{Nov.} \bibinfo{year}{2017}).
\newblock
\newblock
\shownote{{W}hite {P}aper. {A}ccessed {D}ate: 20-November-2017.}


\bibitem[\protect\citeauthoryear{Fischer and et~al}{Fischer and et~al}{2006}]%
        {fischer2006secure}
\bibfield{author}{\bibinfo{person}{L. Fischer} {and} \bibinfo{person}{et al}.}
  \bibinfo{year}{2006}\natexlab{}.
\newblock \showarticletitle{{S}ecure {R}evocable {A}nonymous {A}uthenticated
  {I}nter-vehicle {C}ommunication ({SRAAC})}. In
  \bibinfo{booktitle}{\emph{ESCAR}}. \bibinfo{address}{Berlin, Germany}.
\newblock


\bibitem[\protect\citeauthoryear{Forn{\'e} and et~al}{Forn{\'e} and
  et~al}{2009}]%
        {forne2009certificate}
\bibfield{author}{\bibinfo{person}{J. Forn{\'e}} {and} \bibinfo{person}{et
  al}.} \bibinfo{year}{2009}\natexlab{}.
\newblock \showarticletitle{{C}ertificate {S}tatus {V}alidation in {M}obile
  {A}d {H}oc {N}etworks}.
\newblock \bibinfo{journal}{\emph{IEEE Wireless Communications}}
  \bibinfo{volume}{16}, \bibinfo{number}{1} (\bibinfo{date}{Mar.}
  \bibinfo{year}{2009}).
\newblock


\bibitem[\protect\citeauthoryear{F{\"o}rster and et~al}{F{\"o}rster and
  et~al}{2014}]%
        {puca2014}
\bibfield{author}{\bibinfo{person}{D. F{\"o}rster} {and} \bibinfo{person}{et
  al}.} \bibinfo{year}{2014}\natexlab{}.
\newblock \showarticletitle{{PUCA}: {A} {P}seudonym {S}cheme with
  {U}ser-{C}ontrolled {A}nonymity for {V}ehicular {A}d-{H}oc {N}etworks}. In
  \bibinfo{booktitle}{\emph{IEEE VNC}}. \bibinfo{address}{Paderborn, Germany}.
\newblock


\bibitem[\protect\citeauthoryear{F{\"o}rster and et~al}{F{\"o}rster and
  et~al}{2015}]%
        {forster2015rewire}
\bibfield{author}{\bibinfo{person}{D. F{\"o}rster} {and} \bibinfo{person}{et
  al}.} \bibinfo{year}{2015}\natexlab{}.
\newblock \showarticletitle{{REWIRE}\textendash{R}evocation {W}ithout
  {R}esolution: {A} {P}rivacy-{F}riendly {R}evocation {M}echanism for
  {V}ehicular {A}d-{H}oc {N}etworks}.
\newblock In \bibinfo{booktitle}{\emph{Trust and Trustworthy Computing}}.
  \bibinfo{address}{Heraklion, Greece}.
\newblock


\bibitem[\protect\citeauthoryear{Ga{\~n}{\'a}n and et~al}{Ga{\~n}{\'a}n and
  et~al}{2012}]%
        {ganan2012toward}
\bibfield{author}{\bibinfo{person}{C. Ga{\~n}{\'a}n} {and} \bibinfo{person}{et
  al}.} \bibinfo{year}{2012}\natexlab{}.
\newblock \showarticletitle{{T}oward {R}evocation {D}ata {H}andling
  {E}fficiency in {VANET}s}. In \bibinfo{booktitle}{\emph{Springer
  Nets4Cars/Nets4Trains}}. \bibinfo{address}{Vilnius, Lithuania}.
\newblock


\bibitem[\protect\citeauthoryear{Ga{\~n}{\'a}n and et~al}{Ga{\~n}{\'a}n and
  et~al}{2013a}]%
        {ganan2013becsi}
\bibfield{author}{\bibinfo{person}{C. Ga{\~n}{\'a}n} {and} \bibinfo{person}{et
  al}.} \bibinfo{year}{2013}\natexlab{a}.
\newblock \showarticletitle{{BECSI}: {B}andwidth {E}fficient {C}ertificate
  {S}tatus {I}nformation {D}istribution {M}echanism for {VANET}s}.
\newblock \bibinfo{journal}{\emph{Hindawi-MIS}} \bibinfo{volume}{9},
  \bibinfo{number}{4} (\bibinfo{date}{Mar.} \bibinfo{year}{2013}),
  \bibinfo{pages}{347--370}.
\newblock


\bibitem[\protect\citeauthoryear{Ga{\~n}{\'a}n and et~al}{Ga{\~n}{\'a}n and
  et~al}{2013b}]%
        {ganan2013coach}
\bibfield{author}{\bibinfo{person}{C. Ga{\~n}{\'a}n} {and} \bibinfo{person}{et
  al}.} \bibinfo{year}{2013}\natexlab{b}.
\newblock \showarticletitle{{COACH}: {C}ollaborative {C}ertificate {S}tatus
  {C}hecking {M}echanism for {VANET}s}.
\newblock \bibinfo{journal}{\emph{Network and Computer Applications}}
  \bibinfo{volume}{36}, \bibinfo{number}{5} (\bibinfo{date}{Sep.}
  \bibinfo{year}{2013}).
\newblock


\bibitem[\protect\citeauthoryear{Gerlach and et~al}{Gerlach and et~al}{2007}]%
        {gerlach2007security}
\bibfield{author}{\bibinfo{person}{M. Gerlach} {and} \bibinfo{person}{et al}.}
  \bibinfo{year}{2007}\natexlab{}.
\newblock \showarticletitle{{S}ecurity {A}rchitecture for {V}ehicular
  {C}ommunication}. In \bibinfo{booktitle}{\emph{Workshop on Intelligent
  Transportation}}. \bibinfo{address}{Hamburg, Germany}.
\newblock


\bibitem[\protect\citeauthoryear{Greenwald}{Greenwald}{2013}]%
        {nsa}
\bibfield{author}{\bibinfo{person}{Glenn Greenwald}.}
  \bibinfo{year}{2013}\natexlab{}.
\newblock \bibinfo{title}{{NSA} {P}rism {P}rogram {T}aps in to {U}ser {D}ata of
  {A}pple, {G}oogle and {O}thers}.
\newblock \bibinfo{howpublished}{\url{tiny.cc/cj4ary}}.
\newblock


\bibitem[\protect\citeauthoryear{Haas and et~al}{Haas and et~al}{2009}]%
        {haas2009design}
\bibfield{author}{\bibinfo{person}{J-J Haas} {and} \bibinfo{person}{et al}.}
  \bibinfo{year}{2009}\natexlab{}.
\newblock \showarticletitle{{D}esign and {A}nalysis of a {L}ightweight
  {C}ertificate {R}evocation {M}echanism for {VANET}}. In
  \bibinfo{booktitle}{\emph{ACM Vehicular Internetworking}}.
  \bibinfo{address}{NY,$\;\:$USA}.
\newblock


\bibitem[\protect\citeauthoryear{Haas, Hu, and Laberteaux}{Haas
  et~al\mbox{.}}{2011}]%
        {haas2011efficient}
\bibfield{author}{\bibinfo{person}{J-J. Haas}, \bibinfo{person}{Y-C. Hu}, {and}
  \bibinfo{person}{K-P. Laberteaux}.} \bibinfo{year}{2011}\natexlab{}.
\newblock \showarticletitle{{E}fficient {C}ertificate {R}evocation {L}ist
  {O}rganization and {D}istribution}.
\newblock \bibinfo{journal}{\emph{IEEE JSAC}} \bibinfo{volume}{29},
  \bibinfo{number}{3} (\bibinfo{year}{2011}), \bibinfo{pages}{595--604}.
\newblock


\bibitem[\protect\citeauthoryear{Hsiao and et~al}{Hsiao and et~al}{2011}]%
        {hsiao2011flooding}
\bibfield{author}{\bibinfo{person}{H-C Hsiao} {and} \bibinfo{person}{et al}.}
  \bibinfo{year}{2011}\natexlab{}.
\newblock \showarticletitle{{F}looding-{R}esilient {B}roadcast {A}uthentication
  for {VANET}s}. In \bibinfo{booktitle}{\emph{ACM Mobile Computing and
  Networking}}. \bibinfo{address}{Las Vegas, Nevada, USA}.
\newblock


\bibitem[\protect\citeauthoryear{Huang and et~al}{Huang and et~al}{2004}]%
        {huang2004publish}
\bibfield{author}{\bibinfo{person}{Y. Huang} {and} \bibinfo{person}{et al}.}
  \bibinfo{year}{2004}\natexlab{}.
\newblock \showarticletitle{{P}ublish/{S}ubscribe in a {M}obile {E}nvironment}.
\newblock \bibinfo{journal}{\emph{Wireless Networks}} \bibinfo{volume}{10},
  \bibinfo{number}{6} (\bibinfo{date}{Nov.} \bibinfo{year}{2004}),
  \bibinfo{pages}{643--652}.
\newblock


\bibitem[\protect\citeauthoryear{Iliadis and et~al}{Iliadis and et~al}{2003}]%
        {iliadis2003towards}
\bibfield{author}{\bibinfo{person}{J. Iliadis} {and} \bibinfo{person}{et al}.}
  \bibinfo{year}{2003}\natexlab{}.
\newblock \showarticletitle{{T}owards a {F}ramework for {E}valuating
  {C}ertificate {S}tatus {I}nformation {M}echanisms}.
\newblock \bibinfo{journal}{\emph{Elsevier ComCom}} \bibinfo{volume}{26},
  \bibinfo{number}{16} (\bibinfo{date}{Jan.} \bibinfo{year}{2003}),
  \bibinfo{pages}{1839--1850}.
\newblock


\bibitem[\protect\citeauthoryear{Khodaei and et~al}{Khodaei and et~al}{2016}]%
        {khodaei2016evaluating}
\bibfield{author}{\bibinfo{person}{M. Khodaei} {and} \bibinfo{person}{et al}.}
  \bibinfo{year}{2016}\natexlab{}.
\newblock \showarticletitle{{E}valuating {O}n-demand {P}seudonym {A}cquisition
  {P}olicies in {V}ehicular {C}ommunication {S}ystems}. In
  \bibinfo{booktitle}{\emph{IoV/VoI}}. \bibinfo{address}{Paderborn, Germany}.
\newblock


\bibitem[\protect\citeauthoryear{Khodaei, Jin, and Papadimitratos}{Khodaei
  et~al\mbox{.}}{2014}]%
        {khodaei2014ScalableRobustVPKI}
\bibfield{author}{\bibinfo{person}{M. Khodaei}, \bibinfo{person}{H. Jin}, {and}
  \bibinfo{person}{P. Papadimitratos}.} \bibinfo{year}{2014}\natexlab{}.
\newblock \showarticletitle{{T}owards {D}eploying a {S}calable \& {R}obust
  {V}ehicular {I}dentity and {C}redential {M}anagement {I}nfrastructure}. In
  \bibinfo{booktitle}{\emph{IEEE VNC}}. \bibinfo{address}{Paderborn, Germany}.
\newblock


\bibitem[\protect\citeauthoryear{Khodaei, Jin, and Papadimitratos}{Khodaei
  et~al\mbox{.}}{2018}]%
        {khodaei2018Secmace}
\bibfield{author}{\bibinfo{person}{M. Khodaei}, \bibinfo{person}{H. Jin}, {and}
  \bibinfo{person}{P. Papadimitratos}.} \bibinfo{year}{2018}\natexlab{}.
\newblock \showarticletitle{{SECMACE}: {S}calable and {R}obust {I}dentity and
  {C}redential {M}anagement {I}nfrastructure in {V}ehicular {C}ommunication
  {S}ystems}.
\newblock \bibinfo{journal}{\emph{IEEE Transactions on Intelligent
  Transportation Systems}} \bibinfo{volume}{19}, \bibinfo{number}{5}
  (\bibinfo{date}{May} \bibinfo{year}{2018}), \bibinfo{pages}{1430--1444}.
\newblock


\bibitem[\protect\citeauthoryear{Khodaei, Messing, and Papadimitratos}{Khodaei
  et~al\mbox{.}}{2017}]%
        {khodaei2017RHyTHM}
\bibfield{author}{\bibinfo{person}{M. Khodaei}, \bibinfo{person}{A. Messing},
  {and} \bibinfo{person}{P. Papadimitratos}.} \bibinfo{year}{2017}\natexlab{}.
\newblock \showarticletitle{{RHyTHM}: {A} {R}andomized {H}ybrid {S}cheme {T}o
  {H}ide in the {M}obile {C}rowd}. In \bibinfo{booktitle}{\emph{IEEE VNC}}.
  \bibinfo{address}{Torino, Italy}.
\newblock


\bibitem[\protect\citeauthoryear{Khodaei and Papadimitratos}{Khodaei and
  Papadimitratos}{2015}]%
        {khodaei2015VTMagazine}
\bibfield{author}{\bibinfo{person}{M. Khodaei} {and} \bibinfo{person}{P.
  Papadimitratos}.} \bibinfo{year}{2015}\natexlab{}.
\newblock \showarticletitle{{T}he {K}ey to {I}ntelligent {T}ransportation:
  {I}dentity and {C}redential {M}anagement in {V}ehicular {C}ommunication
  {S}ystems}.
\newblock \bibinfo{journal}{\emph{IEEE VT Magazine}} \bibinfo{volume}{10},
  \bibinfo{number}{4} (\bibinfo{date}{Dec.} \bibinfo{year}{2015}),
  \bibinfo{pages}{63--69}.
\newblock


\bibitem[\protect\citeauthoryear{Kumar and et~al}{Kumar and et~al}{2017}]%
        {kumar2017binary}
\bibfield{author}{\bibinfo{person}{V. Kumar} {and} \bibinfo{person}{et al}.}
  \bibinfo{year}{2017}\natexlab{}.
\newblock \showarticletitle{{B}inary {H}ash {T}ree based {C}ertificate {A}ccess
  {M}anagement for {C}onnected {V}ehicles}. In \bibinfo{booktitle}{\emph{ACM
  WiSec}}. \bibinfo{address}{Boston, USA}.
\newblock


\bibitem[\protect\citeauthoryear{Laberteaux and et~al}{Laberteaux and
  et~al}{2008}]%
        {laberteaux2008security}
\bibfield{author}{\bibinfo{person}{K-P. Laberteaux} {and} \bibinfo{person}{et
  al}.} \bibinfo{year}{2008}\natexlab{}.
\newblock \showarticletitle{{S}ecurity {C}ertificate {R}evocation {L}ist
  {D}istribution for {VANET}}. In \bibinfo{booktitle}{\emph{ACM VehiculAr
  Inter-NETworking}}. \bibinfo{address}{New York, NY, USA}.
\newblock


\bibitem[\protect\citeauthoryear{Larisch and et~al}{Larisch and et~al}{2017}]%
        {larisch2017crlite}
\bibfield{author}{\bibinfo{person}{J. Larisch} {and} \bibinfo{person}{et al}.}
  \bibinfo{year}{2017}\natexlab{}.
\newblock \showarticletitle{{CRLite}: {A} {S}calable {S}ystem for {P}ushing
  {A}ll {TLS} {R}evocations to {A}ll {B}rowsers}. In
  \bibinfo{booktitle}{\emph{IEEE Symposium on SnP}}. \bibinfo{address}{San
  Jose, CA, USA}.
\newblock


\bibitem[\protect\citeauthoryear{Liang, Liu, and Rajan}{Liang
  et~al\mbox{.}}{2012}]%
        {liang2012optimal}
\bibfield{author}{\bibinfo{person}{Y. Liang}, \bibinfo{person}{H. Liu}, {and}
  \bibinfo{person}{D. Rajan}.} \bibinfo{year}{2012}\natexlab{}.
\newblock \showarticletitle{{O}ptimal {P}lacement and {C}onfiguration of
  {R}oadside {U}nits in {V}ehicular {N}etworks}. In
  \bibinfo{booktitle}{\emph{IEEE VTC}}. \bibinfo{address}{Yokohama, Japan}.
\newblock


\bibitem[\protect\citeauthoryear{Ma and et~al}{Ma and et~al}{2008}]%
        {ma2008pseudonym}
\bibfield{author}{\bibinfo{person}{Zhendong Ma} {and} \bibinfo{person}{et al}.}
  \bibinfo{year}{2008}\natexlab{}.
\newblock \showarticletitle{{P}seudonym-on-demand: {A} {N}ew {P}seudonym
  {R}efill {S}trategy for {V}ehicular {C}ommunications}. In
  \bibinfo{booktitle}{\emph{IEEE VTC}}. \bibinfo{address}{Calgary, BC}.
\newblock


\bibitem[\protect\citeauthoryear{Marias and et~al}{Marias and et~al}{[n. d.]}]%
        {marias2005adopt}
\bibfield{author}{\bibinfo{person}{GF Marias} {and} \bibinfo{person}{et al}.}
  \bibinfo{year}{[n. d.]}\natexlab{}.
\newblock \showarticletitle{{ADOPT}: {A} {A}istributed {OCSP} for {T}rust
  {E}stablishment in {MANET}s}. In \bibinfo{booktitle}{\emph{European Wireless
  Conference}}. \bibinfo{address}{Nicosia, Cyprus}.
\newblock


\bibitem[\protect\citeauthoryear{{M}emo}{{M}emo}{2011}]%
        {c2c}
\bibfield{author}{\bibinfo{person}{{PKI} {M}emo}.}
  \bibinfo{year}{2011}\natexlab{}.
\newblock \bibinfo{title}{{C2C-CC}}.
\newblock \bibinfo{howpublished}{\url{http://www.car-2-car.org/}}.
\newblock


\bibitem[\protect\citeauthoryear{Micali}{Micali}{1996}]%
        {micali1996efficient}
\bibfield{author}{\bibinfo{person}{S. Micali}.}
  \bibinfo{year}{1996}\natexlab{}.
\newblock \showarticletitle{{E}fficient {C}ertificate {R}evocation}.
  \bibinfo{publisher}{MIT}, \bibinfo{address}{MA, USA}.
\newblock


\bibitem[\protect\citeauthoryear{Micali}{Micali}{2002}]%
        {micali2002scalable}
\bibfield{author}{\bibinfo{person}{S. Micali}.}
  \bibinfo{year}{2002}\natexlab{}.
\newblock \showarticletitle{{S}calable {C}ertificate {V}alidation and
  {S}implified {PKI} {M}anagement}. In \bibinfo{booktitle}{\emph{PKI
  workshop}}, Vol.~\bibinfo{volume}{15}.
\newblock


\bibitem[\protect\citeauthoryear{Mitzenmacher}{Mitzenmacher}{2002}]%
        {mitzenmacher2002compressed}
\bibfield{author}{\bibinfo{person}{M. Mitzenmacher}.}
  \bibinfo{year}{2002}\natexlab{}.
\newblock \showarticletitle{{C}ompressed {B}loom {F}ilters}.
\newblock \bibinfo{journal}{\emph{IEEE transactions on networking}}
  \bibinfo{volume}{10}, \bibinfo{number}{5} (\bibinfo{date}{Dec.}
  \bibinfo{year}{2002}), \bibinfo{pages}{604--612}.
\newblock


\bibitem[\protect\citeauthoryear{Moore and et~al}{Moore and et~al}{2008}]%
        {moore2008fast}
\bibfield{author}{\bibinfo{person}{T. Moore} {and} \bibinfo{person}{et al}.}
  \bibinfo{year}{2008}\natexlab{}.
\newblock \showarticletitle{{F}ast {E}xclusion of {E}rrant {D}evices from
  {V}ehicular {N}etworks}. In \bibinfo{booktitle}{\emph{IEEE SECON}}.
  \bibinfo{address}{San Francisco, CA}.
\newblock


\bibitem[\protect\citeauthoryear{Myers and et~al}{Myers and et~al}{1999}]%
        {myers1999x}
\bibfield{author}{\bibinfo{person}{M. Myers} {and} \bibinfo{person}{et al}.}
  \bibinfo{year}{1999}\natexlab{}.
\newblock \bibinfo{booktitle}{\emph{{X}. 509 {I}nternet {P}ublic {K}ey
  {I}nfrastructure {O}nline {C}ertificate {S}tatus {P}rotocol-{OCSP}}}.
\newblock \bibinfo{type}{{T}echnical {R}eport}. \bibinfo{institution}{RFC
  2560}.
\newblock


\bibitem[\protect\citeauthoryear{Nguyen and et~al}{Nguyen and et~al}{2016}]%
        {nguyen2016secure}
\bibfield{author}{\bibinfo{person}{V-T. Nguyen} {and} \bibinfo{person}{et al}.}
  \bibinfo{year}{2016}\natexlab{}.
\newblock \showarticletitle{{S}ecure {C}ontent {D}istribution in {V}ehicular
  {N}etworks}.
\newblock \bibinfo{journal}{\emph{arXiv preprint arXiv:1601.06181}}
  (\bibinfo{date}{Jan.} \bibinfo{year}{2016}).
\newblock
\newblock
\shownote{{A}ccessed {D}ate: 30-July-2017.}


\bibitem[\protect\citeauthoryear{Nowatkowski and et~al}{Nowatkowski and
  et~al}{2009}]%
        {nowatkowski2009cooperative}
\bibfield{author}{\bibinfo{person}{M. Nowatkowski} {and} \bibinfo{person}{et
  al}.} \bibinfo{year}{2009}\natexlab{}.
\newblock \showarticletitle{{C}ooperative {C}ertificate {R}evocation {L}ist
  {D}istribution {M}ethods in {VANET}s}.
\newblock In \bibinfo{booktitle}{\emph{International Conference on Ad Hoc
  Networks}}.
\newblock


\bibitem[\protect\citeauthoryear{Nowatkowski and et~al}{Nowatkowski and
  et~al}{2010a}]%
        {nowatkowski2010certificate}
\bibfield{author}{\bibinfo{person}{M. Nowatkowski} {and} \bibinfo{person}{et
  al}.} \bibinfo{year}{2010}\natexlab{a}.
\newblock \showarticletitle{{C}ertificate {R}evocation {L}ist {D}istribution in
  {VANET}s {U}sing {M}ost {P}ieces {B}roadcast}. In
  \bibinfo{booktitle}{\emph{IEEE SoutheastCon}}. \bibinfo{address}{Concord, NC,
  USA}.
\newblock


\bibitem[\protect\citeauthoryear{Nowatkowski and et~al}{Nowatkowski and
  et~al}{2010b}]%
        {nowatkowski2010scalable}
\bibfield{author}{\bibinfo{person}{M. Nowatkowski} {and} \bibinfo{person}{et
  al}.} \bibinfo{year}{2010}\natexlab{b}.
\newblock \showarticletitle{{S}calable {C}ertificate {R}evocation {L}ist
  {D}istribution in {V}ehicular {A}d {H}oc {N}etworks}. In
  \bibinfo{booktitle}{\emph{IEEE GLOBECOM Workshops}}.
\newblock


\bibitem[\protect\citeauthoryear{Papadimitratos}{Papadimitratos}{2008}]%
        {Papadi:C:08}
\bibfield{author}{\bibinfo{person}{P. Papadimitratos}.}
  \bibinfo{year}{2008}\natexlab{}.
\newblock \showarticletitle{"{O}n the road" - {R}eflections on the {S}ecurity
  of {V}ehicular {C}ommunication {S}ystems}. In
  \bibinfo{booktitle}{\emph{{IEEE} {ICVES}}}. \bibinfo{address}{Columbus, OH,
  USA}.
\newblock


\bibitem[\protect\citeauthoryear{Papadimitratos and et~al}{Papadimitratos and
  et~al}{2006}]%
        {papadimitratos2006securing}
\bibfield{author}{\bibinfo{person}{P. Papadimitratos} {and} \bibinfo{person}{et
  al}.} \bibinfo{year}{2006}\natexlab{}.
\newblock \showarticletitle{{S}ecuring {V}ehicular
  {C}ommunications-{A}ssumptions, {R}equirements, and {P}rinciples}. In
  \bibinfo{booktitle}{\emph{ESCAR}}. \bibinfo{address}{Berlin, Germany}.
\newblock


\bibitem[\protect\citeauthoryear{Papadimitratos and et~al}{Papadimitratos and
  et~al}{2007}]%
        {papadimitratos2007architecture}
\bibfield{author}{\bibinfo{person}{P. Papadimitratos} {and} \bibinfo{person}{et
  al}.} \bibinfo{year}{2007}\natexlab{}.
\newblock \showarticletitle{{A}rchitecture for {S}ecure and {P}rivate
  {V}ehicular {C}ommunications}. In \bibinfo{booktitle}{\emph{IEEE ITST}}.
  \bibinfo{address}{Sophia Antipolis}, \bibinfo{pages}{1--6}.
\newblock


\bibitem[\protect\citeauthoryear{Papadimitratos and et~al}{Papadimitratos and
  et~al}{2008a}]%
        {papadimitratos2008certificate}
\bibfield{author}{\bibinfo{person}{P. Papadimitratos} {and} \bibinfo{person}{et
  al}.} \bibinfo{year}{2008}\natexlab{a}.
\newblock \showarticletitle{{C}ertificate {R}evocation {L}ist {D}istribution in
  {V}ehicular {C}ommunication {S}ystems}. In \bibinfo{booktitle}{\emph{ACM
  {VANET}}}. \bibinfo{address}{San Francisco, CA}.
\newblock


\bibitem[\protect\citeauthoryear{Papadimitratos and et~al}{Papadimitratos and
  et~al}{2008b}]%
        {papadimitratos2008secure}
\bibfield{author}{\bibinfo{person}{P. Papadimitratos} {and} \bibinfo{person}{et
  al}.} \bibinfo{year}{2008}\natexlab{b}.
\newblock \showarticletitle{{S}ecure {V}ehicular {C}ommunication {S}ystems:
  {D}esign and {A}rchitecture}.
\newblock \bibinfo{journal}{\emph{IEEE Comm. Mag.}} \bibinfo{volume}{46},
  \bibinfo{number}{11} (\bibinfo{date}{Nov.} \bibinfo{year}{2008}),
  \bibinfo{pages}{100--109}.
\newblock


\bibitem[\protect\citeauthoryear{{P}roject}{{P}roject}{2015}]%
        {preserve-url}
\bibfield{author}{\bibinfo{person}{{PRESERVE} {P}roject}.}
  \bibinfo{year}{2015}\natexlab{}.
\newblock \bibinfo{howpublished}{\url{www.preserve-project.eu/}}.
\newblock


\bibitem[\protect\citeauthoryear{Raya and et~al}{Raya and et~al}{2006}]%
        {raya2006certificaterevocation}
\bibfield{author}{\bibinfo{person}{M. Raya} {and} \bibinfo{person}{et al}.}
  \bibinfo{year}{2006}\natexlab{}.
\newblock \showarticletitle{{C}ertificate {R}evocation in {V}ehicular
  {N}etworks}.
\newblock \bibinfo{journal}{\emph{{T}echnical {R}eport, EPFL, Switzerland}}
  (\bibinfo{year}{2006}).
\newblock


\bibitem[\protect\citeauthoryear{Raya and et~al}{Raya and et~al}{2007}]%
        {raya2007eviction}
\bibfield{author}{\bibinfo{person}{M. Raya} {and} \bibinfo{person}{et al}.}
  \bibinfo{year}{2007}\natexlab{}.
\newblock \showarticletitle{{E}viction of {M}isbehaving and {F}aulty {N}odes in
  {V}ehicular {N}etworks}.
\newblock \bibinfo{journal}{\emph{IEEE JSAC}} (\bibinfo{date}{Oct.}
  \bibinfo{year}{2007}), \bibinfo{pages}{1557--1568}.
\newblock


\bibitem[\protect\citeauthoryear{Rescorla and et~al}{Rescorla and
  et~al}{2012}]%
        {rescorla2012datagram}
\bibfield{author}{\bibinfo{person}{E. Rescorla} {and} \bibinfo{person}{et al}.}
  \bibinfo{year}{2012}\natexlab{}.
\newblock \showarticletitle{{D}atagram {T}ransport {L}ayer {S}ecurity V.1.2}.
\newblock  (\bibinfo{date}{Jan.} \bibinfo{year}{2012}).
\newblock


\bibitem[\protect\citeauthoryear{Rigazzi and et~al}{Rigazzi and et~al}{2017}]%
        {rigazzi2017optimized}
\bibfield{author}{\bibinfo{person}{G. Rigazzi} {and} \bibinfo{person}{et al}.}
  \bibinfo{year}{2017}\natexlab{}.
\newblock \showarticletitle{{O}ptimized {C}ertificate {R}evocation {L}ist
  {D}istribution for {S}ecure {V2X} {C}ommunications}.
\newblock  (\bibinfo{year}{2017}).
\newblock
\newblock
\shownote{{A}ccessed {D}ate: 30-June-2017.}


\bibitem[\protect\citeauthoryear{Schaub, Kargl, Ma, and Weber}{Schaub
  et~al\mbox{.}}{2010}]%
        {schaub2010v}
\bibfield{author}{\bibinfo{person}{F. Schaub}, \bibinfo{person}{F. Kargl},
  \bibinfo{person}{Z. Ma}, {and} \bibinfo{person}{M. Weber}.}
  \bibinfo{year}{2010}\natexlab{}.
\newblock \showarticletitle{V-tokens for {C}onditional {P}seudonymity in
  {VANET}s}. In \bibinfo{booktitle}{\emph{IEEE WCNC}}.
  \bibinfo{address}{Sydney, Australia}.
\newblock


\bibitem[\protect\citeauthoryear{Solworth}{Solworth}{2008}]%
        {solworth2008instant}
\bibfield{author}{\bibinfo{person}{Jon~A Solworth}.}
  \bibinfo{year}{2008}\natexlab{}.
\newblock \showarticletitle{{I}nstant {R}evocation}.
\newblock In \bibinfo{booktitle}{\emph{European PKI}}.
  \bibinfo{address}{Trondheim, Norway}.
\newblock


\bibitem[\protect\citeauthoryear{Stumpf and et~al}{Stumpf and et~al}{2007}]%
        {stumpf2007trust}
\bibfield{author}{\bibinfo{person}{F. Stumpf} {and} \bibinfo{person}{et al}.}
  \bibinfo{year}{2007}\natexlab{}.
\newblock \showarticletitle{{T}rust, {S}ecurity and {P}rivacy in
  {VANET}s~\textendash~a {M}ultilayered {S}ecurity {A}rchitecture for
  {C2C}-{C}ommunication}.
\newblock \bibinfo{journal}{\emph{Automotive Security}} (\bibinfo{date}{Nov.}
  \bibinfo{year}{2007}).
\newblock


\bibitem[\protect\citeauthoryear{Tarkoma and et~al}{Tarkoma and et~al}{2011}]%
        {tarkoma2012theory}
\bibfield{author}{\bibinfo{person}{S. Tarkoma} {and} \bibinfo{person}{et al}.}
  \bibinfo{year}{2011}\natexlab{}.
\newblock \showarticletitle{{T}heory and {P}ractice of {B}loom {F}ilters for
  {D}istributed {S}ystems}.
\newblock \bibinfo{journal}{\emph{IEEE Communications Surveys \& Tutorials}}
  \bibinfo{volume}{14}, \bibinfo{number}{1} (\bibinfo{date}{Apr.}
  \bibinfo{year}{2011}), \bibinfo{pages}{131--155}.
\newblock


\bibitem[\protect\citeauthoryear{Wasef and Shen}{Wasef and Shen}{2009}]%
        {wasef2009edr}
\bibfield{author}{\bibinfo{person}{A. Wasef} {and} \bibinfo{person}{X. Shen}.}
  \bibinfo{year}{2009}\natexlab{}.
\newblock \showarticletitle{{EDR}: {E}fficient {D}ecentralized {R}evocation
  {P}rotocol for {V}ehicular {A}d hoc {N}etworks}.
\newblock \bibinfo{journal}{\emph{IEEE TVT}} \bibinfo{volume}{58},
  \bibinfo{number}{9} (\bibinfo{year}{2009}), \bibinfo{pages}{5214--5224}.
\newblock


\bibitem[\protect\citeauthoryear{Whyte, Weimerskirch, Kumar, and Hehn}{Whyte
  et~al\mbox{.}}{2013}]%
        {whyte2013security}
\bibfield{author}{\bibinfo{person}{W. Whyte}, \bibinfo{person}{A Weimerskirch},
  \bibinfo{person}{V. Kumar}, {and} \bibinfo{person}{T. Hehn}.}
  \bibinfo{year}{2013}\natexlab{}.
\newblock \showarticletitle{{A} {S}ecurity {C}redential {M}anagement {S}ystem
  for {V2V} {C}ommunications}. In \bibinfo{booktitle}{\emph{IEEE VNC}}.
  \bibinfo{address}{Boston, MA}.
\newblock


\end{thebibliography}

\end{document}